\documentclass[twocolumn]{bmcart}

\usepackage{amsthm,amsmath, amssymb}
\RequirePackage[authoryear]{natbib}
\usepackage[utf8]{inputenc} 
\usepackage{color}
\usepackage{graphicx}

\startlocaldefs

\def\Rsolar{$R_{\odot}$}
\def\Msolar{$M_{\odot}$}

\newcommand{\peryr}{\,{\rm yr^{-1}}}
\newcommand{\kmpers}{\,{\rm km s^{-1}}}
\newcommand{\Mo}{M_{\odot}}
\newcommand{\Ro}{R_{\odot}}
\newcommand{\code}{\texttt{TrES}\,}

\newcommand{\ve}[1]{\boldsymbol{#1}}

\endlocaldefs

\begin{document}

\begin{frontmatter}

\begin{fmbox}
\dochead{Research}


\title{The evolution of hierarchical triple star-systems}


\author[
   addressref={aff1},                   
   corref={aff1},                       
   email={toonen@strw.leidenuniv.nl}   
]{\inits{S}\fnm{S.} \snm{Toonen}}
\author[
   addressref={aff1},
]{\inits{A}\fnm{A.} \snm{Hamers}}
\author[
   addressref={aff1},
]{\inits{S}\fnm{S.} \snm{Portegies Zwart}}


\address[id=aff1]{
  \orgname{Leiden Observatory, Leiden University}, 
  \postcode{PO Box 9513}                                
  \city{Leiden},                              
  \cny{The Netherlands}                                    
}





\begin{abstractbox}

\begin{abstract} 
Field stars are frequently formed in pairs, and many of these binaries are part of triples or even higher-order systems. 
Even though, the principles of single stellar evolution and binary evolution, have been accepted for a long time, the long-term evolution of stellar triples is poorly understood.
The presence of a third star in an orbit around a binary system can significantly alter the evolution of those stars and the binary system.
The rich dynamical behaviour in three-body systems can give rise to Lidov-Kozai cycles, in which the eccentricity of the inner orbit and the inclination between the inner and outer orbit vary periodically. 
In turn, this can lead to an enhancement of tidal effects (tidal friction), gravitational-wave emission and stellar interactions such as mass transfer and collisions. 
The lack of a self-consistent treatment of triple evolution, including both three-body dynamics as well as stellar evolution, hinders the systematic study and general understanding of the long-term evolution of triple systems. 
In this paper, we aim to address some of these hiatus, 
by discussing the dominant physical processes of hierarchical triple evolution, 
and presenting heuristic recipes for these processes. 
To improve our understanding on hierarchical stellar triples, these descriptions are implemented in a public source code \code\,which combines three-body dynamics (based on the secular approach) with stellar evolution and their mutual influences. 
Note that modelling through a phase of stable mass transfer in an eccentric orbit is currently not implemented in \code, but can be implemented with the appropriate methodology at a later stage. 
\end{abstract}



\end{abstractbox}
\end{fmbox}

\end{frontmatter}



\section{Introduction}
\label{sec:intro}

The majority of stars are members of multiple systems.  These include
binaries, triples, and higher order hierarchies. The evolution of
single stars and binaries have been studied extensively and there is
general consensus over the dominant physical processes
\citep{Pos14, Too14}. Many exotic systems, however, cannot easily be
explained by binary evolution, and these have often been attributed to
the evolution of triples, 
for example
 low-mass X-ray binaries \citep{Egg86} and blue stragglers \citep{Per09}.
Our lack of a clear understanding of triple
evolution hinders the systematic exploration of these curious objects.
At the same time triples are fairly common; Our nearest neighbour
$\alpha$ Cen is a triple star system \citep{Tok14I}, but more importantly
$\sim 10$\% of the low-mass stars are in triples \citep{Tok08, Rag10,
  Tok14II, Moe16} a fraction that gradually increases \citep{Duc13} to $\sim
50$\% for spectral type B stars \citep{Rem11, San14, Moe16}.

The theoretical studies of triples can classically be divided into
three-body dynamics and stellar evolution, which both are often discussed
separately.  Three-body dynamics is generally governed by the
gravitational orbital evolution, whereas the stellar evolution is
governed by the internal nuclear burning processes in the individual
stars and their mutual influence.

Typical examples of studies that focused on the three-body dynamics
include \citet{For00, Fab07, Nao13, Nao14, Liu15}, and stellar
evolution studies include \citet{Egg96, Ibe99,
  Kur01}. Interdisciplinary studies, in which the mutual interaction
between the dynamics and stellar aspects are taken into account are
rare \citep{Kra12, Per12, Ham13,Sha13,Mic14, Nao16}, but demonstrate the
richness of the interacting regime.  The lack of a self consistent
treatment hinders a systematic study of triple systems. This makes it
hard to judge the importance of this interacting regime, or how many
curious evolutionary products can be attributed to triple evolution.
Here we discuss triple evolution in a broader context in order to
address some of these hiatus.

In this paper we discuss the principle complexities of triple
evolution in a broader context (Sect.\,\ref{sec:background}). We start by
presenting an overview of the evolution of single stars and binaries, and how
to extend these to triple evolution. In the second part of this paper
we present heuristic recipes for simulating their evolution
(Sect.\,\ref{sec:methods}). These recipes combine three-body dynamics
with stellar evolution and their mutual influences, such as tidal
interactions and mass transfer.  These descriptions are summarized in
a public source code \code\,with which triple evolution can be
studied.

\section{Background}
\label{sec:background}

We will give a brief overview of isolated binary evolution (Sect.\,\ref{sec:bg_ds}) and isolated triple evolution (Sect.\,\ref{sec:bg_ts}). We discuss in particular under what circumstances triple evolution differs from binary evolution and what the consequences are of these differences. We start with a brief summary of single star evolution with a focus on those aspects that are relevant for binary and triple evolution.

\begin{table}
\caption{Necessary parameters to describe a single star system, a binary and a triple. For stellar parameters, age and metallicity of each star can be added. The table shows that as the multiplicity of a stellar system increases from one to three, the problem becomes significantly more complicated.}
\begin{tabular}{lcc}
\hline \hline
Parameters & Stellar & Orbital \\
\hline
Single star & $m$& -  \\
Binary &$m_1$, $m_2$& $a$, $e$\\
Triple &$m_1$, $m_2$, $m_3$& $i_{\rm mutual}$, $a_{\rm in}$, $e_{in}$, $g_{\rm in}$, $h_{\rm in}$ \\
 &&  $a_{\rm out}$, $e_{out} $, $g_{\rm out}$,  $h_{\rm out}$\\
\hline
\end{tabular}
\label{tbl:simon}
\end{table}

\subsection{Single stellar evolution}
\label{sec:bg_ss}

Hydrostatic and thermal equilibrium in a star give rise to temperatures and pressures that allow for nuclear burning, and consequently the emission of the starlight that we observe. 
Cycles of nuclear burning and exhaustion of fuel regulate the evolution of a star, and sets the various phases during the stellar lifetime. 

The evolution of a star is predominantly determined by a single parameter, namely the stellar mass (Tbl.\,\ref{tbl:simon}). It depends only slightly on the initial chemical composition or the amount of core overshooting\footnote{\label{ftnt:overshooting}
Overshooting refers to a chemically mixed region beyond the boundary of the convective core \citep[e.g.][]{Sto63, Mas79,Mae91}, as predicted by basic stellar evolutionary theory, i.e. the Schwarzschild criterion. A possible mechanism is convection carrying material beyond the boundary due to residual velocity. For the effects of overshooting on stellar evolution, see e.g. \citet{Bre15}.}.

\subsubsection{Timescales}
\label{sec:bg_ss_time}

Fundamental timescales of stellar evolution are the dynamical ($\tau_{\rm dyn}$), thermal ($\tau_{\rm th}$), and nuclear timescale ($ \tau_{\rm nucl} $). The dynamical timescale is the characteristic time that it would take for a star to collapse under its own gravitational attraction without the presence of internal pressure: 

\begin{equation}
\tau_{\rm dyn} = \sqrt{\frac{R^3}{Gm}} ,
\label{eq:t_dynamic}
\end{equation}
where $R$ and $m$ are the radius and mass of the star. It is a measure of the timescale on which a star would expand or contract if the hydrostatic equilibrium of the star is disturbed. This can happen for example because of sudden mass loss.

A related timescale is the time required for the Sun to radiate
all its thermal energy content at its current luminosity: 

\begin{equation}
\tau_{\rm th} = \frac{Gm^2R}{L},
\label{eq:t_thermal}
\end{equation}
where $L$ is the luminosity of the star. In other words, when the thermal equilibrium of a star is disturbed, the star will move to a new equilibrium on a thermal (or Kelvin–Helmholtz) timescale

Finally, the nuclear timescale represents the time required for the star to exhaust its supply of nuclear fuel at its current luminosity:

\begin{equation}
\tau_{\rm nucl} = \frac{\epsilon c^2m_{\rm nucl}}{L},
\label{eq:t_nuclear}
\end{equation}
where $\epsilon$ is the efficiency of nuclear energy production, $c$ is the speed of light, and 
$m_{\rm nucl}$ is the amount of mass available as fuel. For core hydrogen burning,  $\epsilon = 0.007$ and $M_{\rm nucl}\approx 0.1M$. 
Assuming a mass-luminosity relation of $L\propto M^{\alpha}$, with empirically $\alpha \approx 3-4$ \citep[e.g.][]{Sal05, Eke15}, 
it follows that massive stars live shorter and evolve faster than low-mass stars.

For the Sun, $\tau_{\rm dyn} \approx$ 30 min,  
$\tau_{\rm th} \approx$ 30 Myr,
and $\tau_{\rm nucl} \approx$ 10 Gyr. 
Typically, $\tau_{\rm dyn}< \tau_{\rm th} < \tau_{\rm nucl} $, which allows us to quantitatively predict the structure and evolution of stars in broad terms.

\subsubsection{Hertzsprung-Russell diagram}
\label{sec:bg_ss_HR}

\begin{figure}[h!]
\caption{\csentence{Hertzsprung-Russell diagram} Evolutionary tracks for seven stars in the HR-diagram with masses 1, 1.5, 2.5, 4, 6.5, 10, and 15\Msolar\,at solar metallicity. Specific moments in the evolution of the stars are noted by blue circles as explained in the text. The tracks are calculated with \texttt{SeBa} \citep{Por96, Too12}. The dashed lines show lines of constant radii by means of the Stefan–Boltzmann law. }
\centering
\includegraphics[width=\columnwidth]{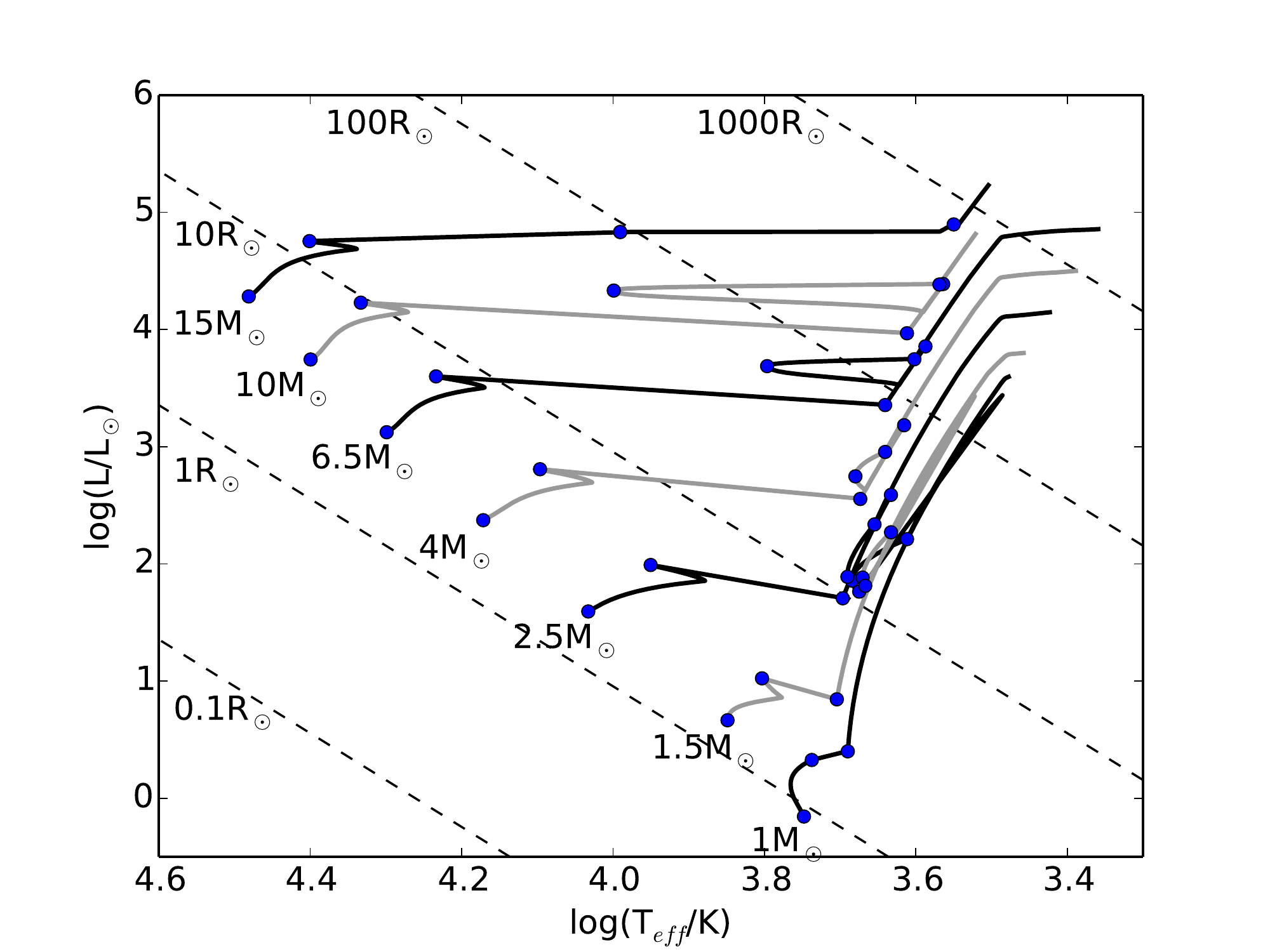} 
\label{fig:HR}
\end{figure}

The Hertzsprung-Russell (HR) diagram in Fig.\ref{fig:HR} shows seven evolutionary tracks for stars of different masses. 
The longest phase of stellar evolution is known as the main-sequence (MS), in which nuclear burning takes place of hydrogen in the stellar core. The MS occupies the region in the HR-diagram between the stellar birth on the zero-age MS (ZAMS, blue circles in Fig.\,\ref{fig:HR}) and the end of the MS-phase (terminal-age MS (TAMS), blue circles in Fig.\,\ref{fig:HR}). 
Stars more massive than 1.2\Msolar\,contract slightly at the end of the MS when the stellar core runs out of hydrogen. This can be seen in Fig.\,\ref{fig:HR} as the hook in the tracks leading up to the TAMS. 

After the TAMS, hydrogen ignites in a shell around the core. Subsequently the outer layers of the star expand rapidly. This expansion at roughly constant luminosity results in a lower effective temperature and a shift to the right in the HR-diagram. 
Stars of less than 13\Msolar\,reach effective temperatures as low as ($10^{3.7}$K) 5000K before helium ignition. At this point (denoted by a blue circle in  Fig.\,\ref{fig:HR}) they start to ascend the red giant branch (RGB) which goes hand in hand with a strong increase in luminosity and radius. On the right of the RGB in the HR-diagram, lies the forbidden region where hydrostatic equilibrium cannot be achieved. 
Any star in this region will rapidly move towards the RGB. 
The red giant star consists of a dense core and an extended envelope up to hundreds of solar radii.  
When the temperature in the core reaches $10^8$K, helium core burning commences and the red giant phase has come to an end. 
For stars less massive than 2\Msolar, helium ignites degenerately in a helium flash. For stars more massive than 13\Msolar, helium ignites before their effective temperature has decreased to a few thousand Kelvin; the shift to the right in the HR-diagram is truncated when helium ignites.

During helium burning the stellar tracks make a loop in the HR-diagram, also known as the horizontal branch. This branch is marked in Fig.\,\ref{fig:HR} by a blue circle at its maximum effective temperature. 
The loop goes hand in hand with a decrease and increase of the stellar radius. As the burning front moves from the core to a shell surrounding the core, the outer layers of the star expand again and the evolutionary track bends back to right in the HR-diagram. 

As the core of the star reaches temperatures of $5\cdot 10^8$K, carbon ignites in the star (denoted by a blue circle in Fig.\,\ref{fig:HR}).

As the core of the star becomes depleted of helium, helium burning continues in a shell surrounding the inert carbon-oxygen core. The star has now has reached the supergiant-phases of its life. The star ascents the asymptotic giant branch (AGB) reaching its maximum size of about a thousand solar radii.

\begin{figure}[h!]
\caption{\csentence{Evolution of stellar radius} 
Radius as a function of stellar age for two stars with masses 4 and 6.5\Msolar\,at solar metallicity. Specific moments in the evolution of the stars are noted by blue circles as for Fig.\,\ref{fig:HR}. The radius evolution is calculated with \texttt{SeBa} \citep{Por96, Too12}. The figure also shows that high-mass stars evolve faster and live shorter than lower-mass stars.  }
\centering
\includegraphics[width=\columnwidth]{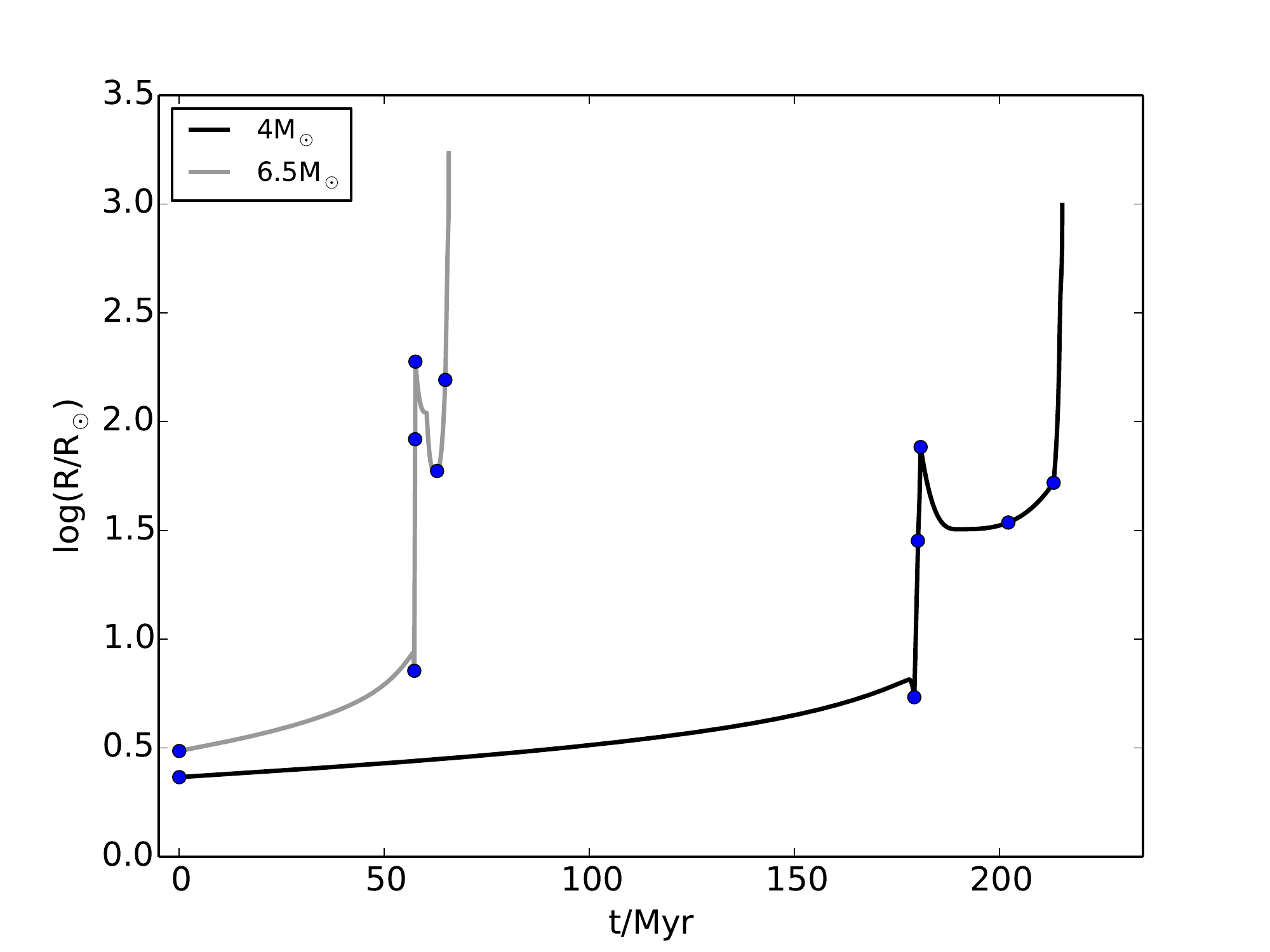} 
\label{fig:rad}
\end{figure}

Fig.\,\ref{fig:rad} shows the variation of the outer radius as the star evolves in its lifetime. It illustrates the dramatic increases in radius during the RGB- and AGB-phases as previously discussed. Shrinkage of star occur after helium ignition, and to a lesser degree at the end of the MS. The radial evolution is of particular interest for binaries and triples, as a star is more likely to initiate mass transfer (i.e. fill its Roche lobe) when its envelope is extended e.g. on the RGB or AGB.

\subsubsection{Stellar winds}
\label{sec:bg_ss_wind}
During the lifetime of a star, a major fraction of the star's mass is lost by means of stellar winds \citep{Lam99, Owo13}. The winds deposit enriched material back into the ISM and can even collide with previously ejected matter to form stellar-wind bubbles and planetary nebulae.  

Stellar winds develop for almost all stars, but the mass losses increases dramatically for more evolved stars and for more massive stars. 
The winds of AGB stars \citep[for a review][]{Hof15} are characterized by extremely high mass-loss rates ( $10^{-7}- 10^{-4}\Mo\peryr$) and low terminal velocities (5-30$\kmpers$). For stars up to 8\Msolar, these 'superwinds' remove the entire stellar envelope. 
AGB-winds are driven by radiation pressure onto molecules and dust grains in the cold outer atmosphere.
The winds are further enhanced by the stellar pulsations that increase the gas density in the extended stellar atmosphere where the dust grains form.

For massive O and B-type stars, strong winds already occur on the MS. These winds \citep[e.g.][]{Pul08, Vin15} are driven by another
 mechanism, i.e. radiation pressure in the continuum and absorption lines of heavy elements. 
The winds are characterized with high mass-loss rates ($10^{-7}-10^{-4}\Mo\peryr$) and high velocities (several 100-1000$\kmpers$) \citep[e.g.][]{Kud00}. 
For stars of more than $\sim$30\Msolar, the mass-loss rate is sufficiently large that the evolution of the star is significantly affected, as the timescale for mass loss is smaller than the nuclear timescale. In turn the uncertainties in our knowledge of the stellar wind mechanisms, introduces considerable uncertainties in the evolution of  massive stars.

\subsubsection{Stellar remnants}
\label{sec:bg_ss_remnants}

The evolution of a star of less than $\sim$6.5\Msolar\,comes to an end as helium burning halts at the end of the AGB. Strong winds strip the core of the remaining envelope and this material forms a planetary nebula surrounding the core. The core cools and contracts to form a white dwarf (WD) consisting of carbon and oxygen (CO). 

Slightly more massive stars up to $\sim$11\Msolar\,experience an additional nuclear burning phase. Carbon burning leads to the formation of a degenerate oxygen-neon (ONe) core. Stars up to $\sim$8\Msolar\,follow a similar evolutionary path discussed above, but they end their lives as oxygen-neon white dwarfs. In the mass range $\sim$8-11\Msolar, the oxygen-neon core reaches the Chandrasekhar mass, and collapses to a neutron star (NS). 

More massive stars than $\sim$11\Msolar\,go through a rapid succession of nuclear burning stages and subsequent fuel exhaustion. The nuclear burning stages are sufficiently short, that the stellar envelope hardly has time to adjust to the hydrodynamical and thermal changes in the core. The position of the star in the HR-diagram remains roughly unchanged. The stellar evolution continues until a iron core is formed after which nuclear burning cannot release further energy. The star then collapses to form a NS or a black hole (BH). 
An overview of the initial mass ranges and the corresponding remnants are given in Tbl.\,\ref{tbl:mimf}. 

\begin{table}
\caption{Initial stellar mass range and the corresponding remnant type and mass. Note that the given masses represent approximate values. They are dependent on the metallicity and on stellar evolutionary processes that are not understood well, such as stellar winds and  core overshooting (footnote\,\ref{ftnt:overshooting}). }
\begin{tabular}{ccc}
\hline \hline
Initial mass (\Msolar)& Remnant type & Remnant mass (\Msolar)\\
\hline
1--6.5 & CO WD& 0.5-1.1\\
6.5--8  & ONe WD  & 1.1--1.44\\
8--$\sim$23& NS & 1.1--2\\
$\gtrsim$23 &  BH& $>$5\\
\hline
\end{tabular}
\label{tbl:mimf}
\end{table}

When a star is part of a compact stellar system, its evolution can be terminated prematurely when the star looses its envelope in a mass-transfer phase. 
After the envelope is lost, the star may form a remnant directly. If on the other hand, the conditions to sustain nuclear burning are fulfilled, the star can evolve further as a  hydrogen-poor helium rich star i.e. helium MS star or helium giant star. 

Due to the mass loss, the initial mass ranges given in Tbl.\,\ref{tbl:mimf} can be somewhat larger.
Furthermore, if a star with a helium core of less than $\sim$0.32\Msolar\,\citep[e.g.][]{Han02}
looses its envelope as a result of mass transfer before helium ignition, the core contracts to form a white dwarf made of helium instead of CO or ONe.

\subsubsection{Supernova explosions}
\label{sec:bg_ss_sn}

When a high-mass star reaches the end of its life and its core collapses to a NS or BH, the outer layers of the star explode in a core-collapse supernova (SN) event. The matter that is blown off the newly formed remnant, enriches the ISM with heavy elements. 
Any asymmetry in the SN, such as in the mass or neutrino loss \citep[e.g.][]{Lai04, Jan12}, can give rise to a natal-kick $\ve{v}_\mathrm{k}$ to the star. Neutron stars are expected to receive a kick at birth of about 400$\kmpers$ \citep[e.g.][]{Cor93, Lyn94, Hob05}, however smaller kick velocities in the range of $\lesssim 50\kmpers$ have been deduced for neutron stars in high-mass X-ray binaries \citep{Pfa02}. Also, whether or not black holes that are formed in core-collapse supernova receive a kick is still under debate \citep[e.g.][]{Gua05, Rep12, Won14, Rep15, Zuo15}.

\subsection{Binary evolution}
\label{sec:bg_ds}

The evolution of a binary can be described by the masses of the stars $m_1$ and $m_2$, the semi-major axis $a$, and the eccentricity $e$. A useful picture for binaries is the Roche model, which describes the effective gravitational potential of the binary. It is generally based on three assumptions: 1) the binary orbit is circular 2) the rotation of the stellar components are synchronized with the orbit 3) the stellar components are small compared to the distance between them. The first two assumptions are expected to hold for binaries that are close to mass transfer because of tidal forces (Sect.\,\ref{sec:bg_ds_tides}). Under the three assumptions given above, the stars are static in a corotating frame of reference. The equipotential surface around a star in which material is gravitationally bound to that star is called the Roche lobe. The Roche radius is defined as the radius of a sphere with the same volume of the nearly spherical Roche lobe, and is often approximated \citep{Egg83} by:
\begin{equation}
R_{\rm L1} \approx  \frac{0.49q^{2/3}}{0.6q^{2/3} + \text{ln}(1+q^{1/3})}
\approx 0.44a\frac{q^{0.33}}{(1+q)^{0.2}},
\label{eq:rl}
\end{equation}
where the mass ratio $q=m_1/m_2$. 
If one of the stars in the binary overflows its Roche lobe, matter from the outer layers of the star can freely move through the first Lagrangian point L1 to the companion star. 
Binaries with initial periods less than several years (depending on the stellar masses) will experience at least one phase of mass transfer, if the stars have enough time to evolve.

If the stars do not get close to Roche lobe overflow (RLOF), the stars in a binary evolve effectively as single stars, slowly decreasing in mass and increasing in radius and luminosity until the remnant stage. The binary orbit can be affected by stellar winds, tides and angular momentum losses such as gravitational wave emission and magnetic braking. These processes are discussed in the following three sections. In the last three sections of this chapter we describe how RLOF affects a binary. 

\subsubsection{Stellar winds in binaries}
\label{sec:bg_ds_wind}
Wind mass loss affects a binary orbit through mass and angular momentum loss. 
Often the assumption is made that the wind is spherically symmetric and fast with respect to the orbit. 
In this approximation, the wind does not interact with the binary orbit directly, such that the process is adiabatic. Furthermore, the orbital eccentricity remains constant \citep{Hua56, Hua63}.

If none of the wind-matter is accreted, the wind causes the orbit to widen. From angular momentum conservation, it follows as:
\begin{equation}
\frac{
\dot{a}_{\rm wind, no-acc}}{a} =\frac{-\dot{m}_1}{m_1 + m_2},
\label{eq:wind_cons_a}
\end{equation}  
where $m_1$ and $m_2$ are the masses of the stars, $\dot{m}_1$ is the mass lost in the wind of the star with mass $m_1$ ($\dot{m}_1 \leqslant 0$), $a$ is the semi-major axis of the orbit, and $\dot{a}_{\rm wind, no-acc}$ the change in the orbital separation with no wind accretion. Eq.\,\ref{eq:wind_cons_a} can be rewritten to:
\begin{equation}
\frac{a_\mathrm{f}}{a_\mathrm{i}} =\frac{m_1+m_2}{m_1 + m_2-\Delta m_{\rm wind}},
\label{eq:wind_cons_a2}
\end{equation}  
where $a_\mathrm{f}$ and $a_\mathrm{i}$ are the semi-major axis of the orbit before and after the wind mass loss, and $\Delta m_{\rm wind}$ is the amount of matter lost in the wind ($\Delta m_{\rm wind}\geqslant 0$).

While the two stars in the binary are in orbit around each other, the stars can accrete some of the wind material of the other star. Including wind accretion, the orbit changes as:
\begin{equation}
\frac{
\dot{a}_{\rm wind}}{a} = \frac{-\dot{m_1}}{m_1} \left( 
-2\beta + 2\beta\frac{m_1}{m_2} - (1-\beta)\frac{m_1}{m_1+m_2}
\right),
\label{eq:wind_a}
\end{equation}  
where the star with mass $m_2$ accretes at a rate of $\dot{m}_2=-\beta\dot{m}_1$. Note that Eq.\,\ref{eq:wind_a} reduces to Eq.\,\ref{eq:wind_cons_a} for complete non-conservative mass transfer i.e. $\beta=0$.
Wind accretion is often modelled by Bondi-Hoyle accretion \citep{Bon44, Liv84}. This model considers a spherical accretion onto a point mass that moves through a uniform medium. 
Wind accretion is an important process known to operate in high-mass X-ray binaries \citep{Tau06, Cha11} and symbiotic stars \citep{Mik02, Sok03}.

The assumptions of a fast and spherically symmetric wind are not always valid. The former is not strictly true for all binary stars i.e. an evolved AGB-star has a wind of 5-30$\kmpers$ \citep[e.g.][]{Hof15}, which is comparable to the velocity of stars in a binary of $a\approx 10^3\Ro$. 
Hydrodynamical simulations of such binaries suggest that the wind of the
donor star is gravitationally confined to the Roche lobe of the donor star \citep{Moh07, deV09,Moh11}. The wind can be focused towards the orbital plane and in particular towards the companion star. This scenario (often called wind Roche-lobe overflow (wRLOF) or gravitational focusing) allows for an accretion efficiency of
up to 50\%, which is significantly higher than for Bondi-Hoyle accretion. 
A requirement for wRLOF to work is that the Roche lobe of the donor star is comparable or smaller than the radius where the wind is accelerated beyond the escape velocity. 
wRLOF is supported by observations of detached binaries with very efficient mass transfer \citep{Kar05, Bli11}

Furthermore, the assumption of adiabatic mass loss is inconsistent with binaries in which the orbital timescale is longer than the mass-loss timescale. The effects of instantaneous mass loss has been studied in the context of SN explosions, and can even lead to the disruption of the binary system (see also Sect.\,\ref{sec:bg_ds_sn}). 
However, also wind mass-loss can have a non-adiabatic effect on the binary orbit \citep[e.g.][]{Had66, Rah09,Ver11} if the mass-loss rate is high and the orbit is wide. 
Under the assumption that mass-loss proceeds isotropically, the wind causes the orbit to widen, as in the case for adiabatic mass loss. However, the eccentricity may decrease or increase, and may even lead to the disruption of the binary system \citep[see e.g.][for a detailed analysis of the effects of winds on sub-stellar binaries in which an exoplanet orbits a host star]{Ver11}. Toonen, Hollands, Gaensicke, \& Boekholt, in prep. show that also (intermediate-mass) stellar binaries can be disrupted during the AGB-phases when the mass loss rates are high ($10^{-7}- 10^{-4}\Mo\peryr$) for orbital separations approximately larger then 
$10^6\Ro$ ($P\approx 10^6$yr where $P$ is the orbital period).

Lastly, anisotropic mass-loss might occur in fast-rotating stars or  systems that harbour bipolar outflows. 
Rotation modifies the structure and evolution of a star, and as such the surface properties of the star where the wind originates \citep[see][for a review]{Mae12}. For an increasing rate of rotation until critical rotation, the stellar winds increasingly depart from a spherical distribution \citep[see e.g.][]{Geo11}. 
Additionally, the bipolar outflows or jets are associated with protostars, evolved post-AGB stars and binaries containing compact objects. Their origin is most likely linked to the central object or the accretion disk \citep[e.g.][]{Bla90}.

The effect of anisotropic mass loss on the orbit of a binary system is important primarily for wide binaries \citep[e.g.][]{Par98, Ver13}. Specifically, \citet{Ver13} show that the relative contribution of the anisotropic terms to the overall motion scale as $\sqrt{a}$. If the mass loss is symmetric about the stellar equator, the mass loss does not affect the orbital motion in another way than for the isotropic case. \citet{Ver13} conclude that the isotropic mass-loss approximation can be used safely to model the orbital evolution of a planet around a host star until orbital separations of hundreds of AU. For a fixed total mass of the system, the effects of anisotropic mass loss are further diminished with decreasing mass ratio (i.e. for systems with more equal masses), such that the assumption of isotropic mass-loss is robust until even larger orbital separations for stellar binaries.

\subsubsection{Angular momentum losses}
\label{sec:bg_ds_aml}

Angular momentum loss from gravitational waves (GW) and magnetic braking act to shrink the binary orbit \citep[e.g.][]{Pet64, Ver81}. Ultimately this can lead to RLOF of one or both components and drive mass transfer.

The strength of GW emission depends strongly on the semi-major axis, and 
to lesser degree on the eccentricity. 
It affects the orbits as:
\begin{equation}
\dot a_{\rm gr} = -\frac{64}{5}\ \frac{G^3 m_1m_2 (m_1+m_2)}{c^5a^3(1-e^2)^{7/2}}\ \left (1 + \frac{73}{24}e^2 + \frac{37}{96}e^4 \right ), 
\label{eq:gw_a}
\end{equation}
and 
\begin{equation}
\dot e_{\rm gr} = -\dfrac{304}{15} e\dfrac{G^3m_1m_2(m_1+m_2)}{c^5a^4(1-e^2)^{5/2}} \left(1+\frac{121}{304}e^2\right ),
\label{eq:gw_e}
\end{equation}
where $\dot{a}_{\rm gr}$ and $\dot{e}_{\rm gr}$ are the change in orbital separation and eccentricity averaged over a full orbit \citep{Pet64}.
Accordingly, GW emission affects most strongly the compact binaries. These binaries are a very interesting and the only known source of GWs for GW interferometers such as LIGO, VIRGO and eLISA. 

Magnetic braking extracts angular momentum from a rotating magnetic star by means of an ionized stellar wind \citep{Sch62, Hua66, Sku72}. Even when little mass is lost from the star, the wind matter can exert a significant spin-down torque on the star. This happens when the wind matter  is forced to co-rotate with the magnetic field.
If the star is in a compact binary and forced to co-rotate with the orbit due to tidal forces, angular momentum is essentially removed from the binary orbit as well \citep{Ver81}. This drain of angular momentum results in a contraction of the orbit.

Magnetic braking plays an important role in the orbital evolution of interacting binaries with low-mass donor stars, such as cataclysmic variables and low-mass X-ray binaries \citep{Kni11, Tau06}. 
For magnetic braking to take place, the donor star is expected to have a mass between 0.2-1.2\Msolar, such that the star has a radiative core and convective envelope to sustain the magnetic field. 
The strength of magnetic braking is still under debate and several prescriptions exist \citep[see][for a review]{Kni11}.

\subsubsection{Tides}
\label{sec:bg_ds_tides}

The presence of a companion star introduces tidal forces in the binary system that act on the surface of the star and lead to tidal deformation of the star.  
If the stellar rotation is not synchronized or aligned with the binary orbit, the tidal bulges are misaligned with the line connecting the centres of mass of the two stars. 
This produces a tidal torque that allows for the transfer of angular momentum between the stars and the orbit. 
Additionally, energy is dissipated in the tides, which drains energy from the orbit and rotation.
Tidal interaction drives the binary to a configuration of lowest energy e.g. it strives to circularize the orbit, synchronize the rotation of the stars with the orbital period and align the stellar spin with respect to the orbital spin. See \citet{Zah08} and \citet{Zah13} for recent reviews.

For binaries with extreme mass ratios, a stable solution does not exist \citep{Dar1879, Hut80}. 
In this scenario  a star is unable
to extract sufficient angular momentum from the orbit to
remain in synchronized rotation. Tidal forces will cause the orbit to decay and the companion to spiral into the envelope of the donor star.
This tidal instability occurs when the angular momentum of the star $J_{\star} > \frac{1}{3} J_{\rm b}$, with $J_{\rm b}$ the orbital angular momentum and $J_{\star}=I\Omega$, where $I$ is the moment of inertia and $\Omega$ the spin angular frequency.

\citet{Hut81} derives a general qualitative picture of tidal evolution and its effect on the orbital evolution of a binary system. 
\citet{Hut81} considers a model in which the tides assume their equilibrium shape, and with very small deviations in position and amplitude with respect to the equipotential surfaces of the stars. If a companion star with mass $m_2$ raises tides on a star with mass $m_1$, the change of binary parameters due to tidal friction is:

\begin{eqnarray}
\begin{array}{lcl}
\dot{a}_{\rm TF}& =& -6 \dfrac{k_{\rm am}}{\tau_{\rm TF}} \tilde{q}(1+\tilde{q}) \left( \dfrac{R}{a}\right)^8 \dfrac{a}{(1-e^2)^{15/2}}  \\
&&\\
&&  \left( f_1(e^2)-(1-e^2)^{3/2} f_2(e^2) \dfrac{\Omega}{\Omega_b}\right),
\end{array}
\label{eq:a_dot_TF}
\end{eqnarray}

\begin{eqnarray}
\begin{array}{lcl}
\dot{e}_{\rm TF} = & =& -27 \dfrac{k_{\rm am}}{\tau_{\rm TF}} \tilde{q}(1+\tilde{q}) \left( \dfrac{R}{a}\right)^8 \dfrac{e}{(1-e^2)^{13/2}}  \\
&&\\
&&  \left( f_3(e^2)-\frac{11}{18}(1-e^2)^{3/2} f_4(e^2) \dfrac{\Omega}{\Omega_b}\right),
\end{array}
\label{eq:e_dot_TF}
\end{eqnarray}

\begin{eqnarray}
\begin{array}{lcl}
\dot{\Omega}_{\rm TF} & =& 3 \dfrac{k_{\rm am}}{\tau_{\rm TF}} \dfrac{\tilde{q}^2}{k^2} \left( \dfrac{R}{a}\right)^6 \dfrac{\Omega_b}{(1-e^2)^{6}}\\
&&\\
&&  \left( f_2(e^2)-(1-e^2)^{3/2} f_5(e^2) \dfrac{\Omega}{\Omega_b}\right),
\end{array}
\label{eq:omega_dot_TF}
\end{eqnarray}

where $\tilde{q}\equiv m_2/m_1$, and $\Omega_b=2\pi/P$ is the mean orbital angular velocity. The star with mass $m_1$ has an apsidal motion constant $k_{\rm am}$, gyration radius $k$, and spin angular frequency $\Omega$. 
$\tau_{\rm TF}$ represents the typical timescale on which significant changes in the orbit take place through tidal evolution. 
The parameters $f_n(e^2)$ are polynomial expressions given by \citep{Hut81}:
\begin{eqnarray}
\left\{
\begin{array}{l c l}
 f_1(e^2) &=& 1+\frac{31}{2}e^2+\frac{255}{8}e^4+\frac{185}{16}e^6+\frac{25}{64}e^8\\
 f_2(e^2) &=& 1+\frac{15}{2}e^2+\frac{45}{8}e^4+\frac{5}{16}e^6\\
 f_3(e^2) &=& 1+\frac{15}{4}e^2+\frac{15}{8}e^4+\frac{5}{64}e^6\\
 f_4(e^2) &=& 1+\frac{3}{2}e^2+\frac{1}{8}e^4\\
 f_5(e^2) &=& 1+3e^2+\frac{3}{8}e^4\\
 \end{array} \right.
\label{eq:fne}
\end{eqnarray}
The degree of tidal interaction strongly increases with the ratio of the stellar radii to the semi-major axis of the orbit (Eq.\,\ref{eq:a_dot_TF} \ref{eq:e_dot_TF}~and~\ref{eq:omega_dot_TF}). Therefore, tidal interaction mostly affect the orbits of relatively close binaries, unless the eccentricities are high and/or the stellar radii are large.

The tidal timescale $T$ ( Eq.\,\ref{eq:a_dot_TF}-\ref{eq:omega_dot_TF}) is subject to debate due to quantitative uncertainties in tidal dissipation mechanisms \citep{Wit99, Wil03, Mei05}. Tidal dissipation causes the misalignment of the tidal bulges with the line connecting the centres of mass of the two stars.
For stars (or planets) with
an outer convection zone, the dissipation is often attributed\footnote{For alternative mechanisms, see \citep{Goo98, Tas00, Sav02}.}
to turbulent friction in the convective regions of the star \citep{Zah77, Zah89, Gol91}. 
For stars with
an outer radiation zone, 
the dominant dissipation mechanism identified so far is radiative damping of stellar oscillations that are exited by the tidal field i.e. dynamical tides \citep{Zah75, Zah77}.  
Despite the uncertainties in tidal dissipation mechanisms, it is generally assumed that circularization and synchronization is achieved before RLOF in a binary.

\subsubsection{Mass transfer}
\label{sec:bg_ds_mt}

Whether or not mass transfer is stable depends on the response of the donor star upon mass loss, and the reaction of the Roche lobe upon the re-arrangement of mass and angular momentum within the binary \citep[e.g.][]{Web85, Hje87, Pol94, Sob97}. 
If the donor star stays approximately within its Roche lobe, mass transfer is dynamically stable. When this is not the case, the donor star will overflow its Roche lobe even further as mass is removed. This leads to a runaway situation that  progresses into a common-envelope \citep[CE, ][]{Pac76}. During the CE-phase, the envelope of the donor star engulfs both stars causing them to spiral inwards until both stars merge or the CE is expelled.

Due to the mass loss, the donor star falls out of hydrostatic and thermal equilibrium. The radius of the star changes as the star settles to a new hydrostatic equilibrium, and subsequently thermal equilibrium. 
The stellar response upon mass loss depends critically on the structure of the stellar envelope i.e. the thermal gradient and entropy of the envelope. In response to mass loss, stars with a deep surface convective zone tend to expand, whereas stars with a radiative envelope tend to shrink rapidly. 
Therefore, giant donor stars with convective envelopes favour CE-evolution upon RLOF 
\footnote{Unless $q \lesssim 0.7$, such that the orbit and the Roche lobe expand significantly upon mass transfer (e.g. Eqs.\,\ref{eq:a_cons_mt}-\ref{eq:a_mt})}. As giants have radii of several hundreds to thousands of Solar radii, the orbit at the onset of mass transfer is of the same order of magnitude. On the other hand, donor stars on the MS with radiative envelope often lead to dynamically stable mass transfer in binaries with short orbital periods \citep[e.g.][]{Too14}.

\subsubsection{Common-envelope evolution}
\label{sec:ce}
During the CE-phase, the core of the donor star and the companion are contained within a CE. Friction between these objects and the slow-rotating envelope is expected to cause the objects to spiral-in. If this process does not release enough energy and angular momentum to drive off the entire envelope, the binary coalesces. On the other hand if a merger can be avoided, a close binary remains in which one or both stars have lost their envelopes. The evolution of such a star is significantly shortened, or even terminated prematurely if it directly evolves to a remnant star.

The systems that avoid a merger lose a significant amount of mass and angular momentum during the CE-phase. The orbital separation of these systems generally decreases by two orders of magnitude, which affects the further evolution of the binary drastically.
 The CE-phase plays an essential role in the formation of short-period systems with compact objects, such as X-ray binaries, and cataclysmic variables. In these systems the current orbital separation is much smaller than the size of the progenitor of the donor star, which had giant-like dimensions at the onset of the CE-phase. 

Despite of the importance of the CE-phase and the enormous efforts of the community, the CE-phase is not understood in detail \cite[see][for a review]{Iva13}. 
The CE-phase involves a complex mix of physical processes, such as energy dissipation, angular momentum transport, and tides, over a large range in time- and length-scales. 
A complete simulation of the CE-phase is still beyond our reach, but great progress has been made with hydrodynamical simulations in the last few years \citep{Ric12, Pas12, Nan15}. 
The uncertainty in the CE-phase is one of the aspects of the theory of binary evolution that affects our understanding of the evolutionary history of a specific binary population most \citep[e.g.][]{Too13, Too14}.

The classical way to treat the orbital evolution due to the CE-phase, is the $\alpha$-formalism. This formalism considers the energy budget of the initial and final configuration \citep{Tut79}; 
\begin{equation}
E_{\rm gr} = \alpha (E_{\rm orb,i}-E_{\rm orb,f}),
\label{eq:alpha-ce}
\end{equation}
where $E_{\rm gr}$ is the binding energy of the envelope, $E_{\rm orb, i}$ and $E_{\rm orb, f}$ are the orbital energy of the pre- and post-mass transfer binary. The $\alpha$-parameter describes the efficiency with which orbital energy is consumed to unbind the CE. When both stars have loosely bound envelopes, such as for giants, both envelopes can be lost simultaneously \citep[hereafter double-CE, see][]{Bro95, Nel01}. In Eq.\,\ref{eq:alpha-ce} $E_{\rm gr}$ is then replaced by the sum of the binding energy of each envelope to its host star: 
\begin{equation}
E_{\rm gr,1}+E_{\rm gr,2} = \alpha (E_{\rm orb,i}-E_{\rm orb,f}).
\label{eq:double-alpha-ce}
\end{equation}

The binding energy of the envelope of the donor star in Eq.\,\ref{eq:alpha-ce}~and~\ref{eq:double-alpha-ce} is given by:
\begin{equation}
E_{\rm gr} = \frac{Gm_{\rm d} m_{\rm d,env}}{\lambda_{\rm ce} R},
\label{eq:Egr_web}
\end{equation} 
where $R$ is the radius of the donor star, $M_{\rm d,env}$ is the envelope mass of the donor and $\lambda_{\rm ce}$ depends on the structure of the donor \citep{deK87, Dew00, Xu10, Lov11}.
The parameters $\lambda_{\rm ce}$ and $\alpha$ are often combined in one parameter $\alpha\lambda_{\rm ce}$.

According to the alternative $\gamma$-formalism \citep{Nel00}, angular momentum is used to expel the envelope of the donor star, according to:
\begin{equation}
\frac{J_{\rm b, i}-J_{\rm b,  f}}{J_{\rm b,i}} = \gamma \frac{\Delta m_{\rm d}}{m_{\rm d}+ m_{\rm a}},
\label{eq:gamma-ce}
\end{equation} 
where $J_{\rm b,i}$ and $J_{\rm b,f}$ are the orbital angular momentum of the pre- and post-mass transfer binary respectively. The parameters $m_d$ and $m_a$ represent the mass of the donor and accretor star, respectively, and $\Delta m_{\rm d}$ is the mass lost by the donor star.
The $\gamma$-parameter describes the efficiency with which orbital angular momentum is used to blow away the CE.  

Valuable constraints on CE-evolution have come from evolutionary reconstruction studies of observed samples of close binaries and from comparing those samples with the results of binary population synthesis studies. 
The emerging picture is that for binaries with low mass ratios, the CE-phase leads to a shrinkage of the orbit.
For the formation of compact WD-MS binaries with low-mass MS companions, the orbit shrinks strongly \citep[$\alpha\lambda_{\rm ce} \approx 0.25 $, see][]{Zor10, Too13, Por13, Cam14}. However, for the formation of the second WD in double WDs, the orbit only shrinks moderately \citep[$\alpha\lambda_{\rm ce} \approx 2 $, see][]{Nel00, Nel01, Van06}. When binaries with approximately equal masses come in contact, mass transfer leads to a modest widening of the orbit, alike the $\gamma$-formalism \citep{Nel00, Nel01}. The last result is based on a study of the first phase of mass transfer for double WDs, in which the first WD is formed. 
\citet{Woo12} suggested that this mass transfer episode can occur stably and non-conservatively even with donor star (early) on the red giant branch. 
Further research is needed to see if this evolutionary path suffices to create a significant number of double WDs.

\subsubsection{Stable mass transfer}
\label{sec:stable_mt}

Whereas the duration of the CE-phase is likely of the order of $10^3$yr (i.e. the thermal timescale of the envelope), stable mass transfer occurs on much longer timescales. Several driving mechanisms exist for stable mass transfer with their own characteristic mass transfer timescales. 
The donor star can drive Roche lobe overflow due to its nuclear evolution or due to the thermal readjustment from the mass loss. Stable mass transfer can also be driven by the contraction of the Roche lobe due to angular momentum losses in the system caused by gravitational wave radiation or magnetic braking.

When mass transfer proceeds conservatively the change in the orbit is regulated by the masses of the stellar components. For circular orbits, 
\begin{equation}
\frac{a_\mathrm{f}}{a_\mathrm{i}} = \left(\dfrac{m_{\rm d,i}m_{\rm a,i}}{m_{\rm d,f}m_{\rm a,f}} \right) ^2,
\label{eq:a_cons_mt}
\end{equation}
where the subscript $\mathrm{i}$~and~$\mathrm{f}$ denote the pre- and post-mass transfer values. 
In general, the donor star will be the more massive component in the binary and the binary orbit will initially shrink in response to mass transfer. After the mass ratio is approximately reversed, the orbit widens. In comparison with the pre-mass transfer orbit, the post-mass transfer orbit is usually wider with a factor of a few \citep{Too14}. 
 
If  the accretor star is not capable of accreting the matter conservatively, mass and angular momentum are lost from the system. The evolution of the system is then dictated by how much mass and angular momentum is carried away.
Assuming angular momentum conservation and neglecting the stellar rotational angular momentum compared to the orbital angular momentum, 
the orbit evolves as \citep[e.g.][]{Mas75, Pol94, Pos14}:
\begin{equation}
\frac{\dot{a}}{a} = -2\frac{\dot{m_\mathrm{d}}}{m_\mathrm{d}}\left[1-\beta\frac{m_\mathrm{d}}{m_\mathrm{a}}-(1-\beta)(\eta+\frac{1}{2})\frac{m_\mathrm{d}}{m_\mathrm{d}+m_\mathrm{a}}\right],
\label{eq:a_mt}
\end{equation}
where the accretor star captures a fraction $\beta\equiv -\dot{m_\mathrm{a}}/\dot{m_\mathrm{d}}$ of the transferred matter, and the matter that is lost carries specific angular momentum $h$ equal to a multiple $\eta$ of the specific orbital angular momentum of the binary: \begin{equation}
h\equiv \frac{\dot{J}}{\dot{m_\mathrm{d}}+\dot{m_\mathrm{a}}} = \eta \frac{J_\mathrm{b}}{m_\mathrm{a}+m_\mathrm{d}}.
\label{eq:spec_J}
\end{equation}
Different modes of angular momentum loss exist which can lead to a relative expansion or contraction of the orbit compared to the case of conservative mass transfer \citep{Sob97, Too14}. For example, the generic description of orbital evolution of Eq.\,\ref{eq:a_mt} reduces to that of conservative mass transfer (Eq.\,\ref{eq:a_cons_mt}) for $\beta =1$ or $\dot{m_\mathrm{a}}=\dot{m_\mathrm{d}}$.  Also, Eq.\,\ref{eq:a_mt} reduces to Eq.\,\ref{eq:wind_a} describing the effect of stellar winds on the binary orbit, under the assumption of specific angular momentum loss equal to that of the donor star 
($h=J_\mathrm{d}/m_\mathrm{d} = m_\mathrm{a}/m_\mathrm{d} \cdot J_\mathrm{b}/(m_\mathrm{d}+m_\mathrm{a})$ or $\eta=m_\mathrm{a} /m_\mathrm{d}$). 
Depending on which mode of angular momentum loss is applicable, the further orbital evolution and stability of the system varies.

Stable mass transfer influences the stellar evolution of the donor star and possibly that of the companion star. 
The donor star is affected by the mass loss, which leads to a change in the radius on long timescales compared to a situation without mass loss \citep{Hur00}. 
Stable mass transfer tends to terminate when the donor star has lost most of its envelope, and contracts to form a remnant star or to a hydrogen-poor helium rich star. In the latter case the evolution of the donor star is significantly shortened, and in the former it is stopped prematurely, similar to what was discussed previously for the CE-phase.

If the companion star accretes a fraction or all of the transferred mass, evolution of this star is affected as well. Firstly, if due to accretion, the core grows and fresh fuel from the outer layers is mixed into the nuclear-burning zone, the star is 'rejuvenated' \citep[see e.g.][]{Van94}. These stars can appear significantly younger than their co-eval neighbouring stars in a cluster\footnote{Stable mass transfer is one of the proposed evolutionary pathways for the formation of blue stragglers \citep[for a review][]{Dav15}. These are MS stars in open and globular clusters that are more luminous and bluer than other MS stars in the same cluster.}. 
Secondly, the accretor star adjusts its structure to a new equilibrium. 
If the timescale of the mass transfer is shorter than the thermal timescale of the accretor, the star will temporarily fall out of thermal equilibrium.
The radial response of the accretor star will depend on the structure of the envelope (as discussed for donor stars in Sect.\,\ref{sec:bg_ds_mt}). A star with a radiative envelope is expected to expand upon mass accretion, whereas a star with a convective envelope shrinks. In the former case, the accretor may swell up sufficiently to fill its Roche lobe, leading to the formation of a contact binary.

\subsubsection{Supernova explosions in binaries}
\label{sec:bg_ds_sn}

If the collapsing star is part of a binary or triple, natal kick $\ve{v}_\mathrm{k}$ alters the orbit and it can even unbind the system. 
Under the assumption that the SN is instantaneous and the SN-shell does not impact the companion star(s), the binary orbit is affected by the mass loss and velocity kick \citep{Hil83, Kal96, Tau98, Pij12} through:

\begin{eqnarray}
\begin{array}{l}
\dfrac{a_\mathrm{f}}{a_\mathrm{i}} = \left (1-\dfrac{\Delta m}{m_\mathrm{t,i}} \right ) \cdot  \\
\\
\left (1-\dfrac{2a_\mathrm{i} }{r_\mathrm{i}}\dfrac{\Delta m }{ m_\mathrm{t,i}} -\dfrac{2(\ve{v}_\mathrm{i}\cdot \ve{v}_\mathrm{k})}{v_\mathrm{c}^2}  - \dfrac{ v_\mathrm{k}^2}{v_\mathrm{c}^2}\right )^{-1},
\end{array}
\label{eq:a_sn_bin}
\end{eqnarray}

where $a_\mathrm{i}$ and $a_\mathrm{f}$ are the semi-major axis of the pre-SN and post-SN orbit,
$\Delta m$ is the mass lost by the collapsing star, $m_\mathrm{t,i}$ is the total mass of the system pre-SN, $r_\mathrm{i}$ is the pre-SN distance between the two stars, $\ve{v}_\mathrm{i}$ is the pre-SN relative velocity of the collapsing star relative to the companion, and

\begin{equation}
v_\mathrm{c} \equiv \sqrt{\frac{Gm_\mathrm{t,i}}{a_\mathrm{i}}}
\label{eq:v_c_bin}
\end{equation}
is the orbital velocity in a circular orbit. 
A full derivation of this equation and that for the post-SN eccentricity is given in Appendix\,\ref{sec:app_sn}. Note that the equation for the post-SN eccentricity of \citet[their Eq.8a][]{Pij12} is incomplete.

Eq.\,\ref{eq:a_sn_bin} shows that with a negligible natal kick, a binary survives the supernova explosion if less than half of the mass is lost. 
Furthermore, the binary is more likely to survive if the SN occurs at apo-astron. 
With substantial natal kicks compared to the pre-SN orbital velocity, survival of the binary depends on the magnitude ratio  and angle between the two (through $\ve{v}_\mathrm{i}\cdot \ve{v}_\mathrm{k}$ in Eq.\,\ref{eq:a_sn_bin}). Furthermore, the range of angles that lead to survival is larger at peri-astron than apo-astron \citep{Hil83}.
If the direction of the natal kick is opposite to the orbital motion of the collapsing star, the binary is more likely to survive the SN explosion.

\subsection{Triple evolution}
\label{sec:bg_ts}
The structures of observed triples tend to be hierarchical, i.e. the triples consist of an inner binary and a distant star (hereafter outer star) that orbits the centre of mass of the inner binary \citep{Hut83}. 
To define a triple star system, no less than 10 parameters are required (Tbl.\,\ref{tbl:simon}):\\
- the masses of the stars in the inner orbit $m_1$ and $m_2$, and the mass of the outer star in the outer orbit $m_3$; \\
- the semi-major axis $a$, the eccentricity $e$, the argument of pericenter $g$ of both the inner and outer orbits. Parameters for the inner and outer orbit are denoted with a subscript '$\rm in$' and '$\rm out$', respectively;\\
- the mutual inclination $i_{r}$ between the two orbits.\\
The longitudes of ascending nodes $h$ specify the orientation of the triple on the sky, and not the relative orientation. Therefore, they do not affect the intrinsic dynamical evolution. From total angular momentum conservation $h_{\rm in} - h_{\rm out}= \pi$ for a reference frame with the z-axis aligned along the total angular momentum vector \citep{Nao13}.

In some cases, the presence of the outer star has no significant effect on the evolution of the inner binary, such that the evolution of the inner and outer binary can be described separately by the processes described in Sect.\,\ref{sec:bg_ss}~and~\ref{sec:bg_ds}. 
In other cases, there is an interaction between the three stars that is unique to systems with multiplicities of higher orders than binaries. In this way, many new evolutionary pathways open up compared to binary evolution. The additional processes are described in the following sections, such as the dynamical instability and Lidov-Kozai cycles.

\subsubsection{Stability of triples}
\label{sec:bg_ts_stability}

The long-term behaviour of triple systems has fascinated scientists for centuries. Not only stellar triples have been investigated, but also systems with planetary masses, such as the Earth-Moon-Sun system by none other than Isaac Newton. It was soon realised that the three-body problem does not have closed-form solutions as in the case for two-body systems. 
Unstable systems dissolve to a lower order systems on dynamical timescales \citep{Van07}.

It is hard to define the boundary between stable and unstable systems, as stability can occur on a  range of timescales. Therefore, many stability criteria exist \citep{Mar01b,Geo08}, that can be divided in three categories: analytical, numerical integration and chaotic criteria. The commonly used criterion of \citet{Mar99}:

\begin{eqnarray}
\begin{array}{l c l}
\dfrac{a_{\rm out}}{a_{\rm in}}|_{\rm crit} &= &\dfrac{2.8}{1-e_{\rm out}} (1-\dfrac{0.3i}{\pi}) \cdot \\
&&\\
&&\left( \dfrac{(1.0+q_{\rm out})\cdot(1+e_{\rm out})}{\sqrt{1-e_{\rm out}}} \right)^{2/5}, \\
 \end{array} 
\label{eq:stab_crit}
\end{eqnarray}

where systems are unstable if $\dfrac{a_{\rm out}}{a_{\rm in}} < \dfrac{a_{\rm out}}{a_{\rm in}}|_{\rm crit}$ and $q_{\rm out}\equiv \dfrac{m_3}{m_1+m_2}$. 
This criterion is based on the concept of chaos and the consequence of overlapping resonances. The criterion is conservative, as the presence of chaos in some cases is not necessarily the same as an instability. By comparison with numerical integration studies, it was shown that Eq.\,\ref{eq:stab_crit} works well for a wide range of parameters \citep{Aar01, Aar04}.

Most observed triples have hierarchical structures, because democratic triples tend to be unstable and short-lived \citep{Van07}. 
Hierarchical triples that are born in a stable configuration can become unstable as they evolve.
Eq.\,\ref{eq:stab_crit} shows that when the ratio of the semi-major axes of the outer and inner orbit decreases sufficiently, the system enters the instability regime. Physical mechanisms that can lead to such an event, are stellar winds from the inner binary and stable mass transfer in the inner binary \citep{Kis94, Ibe99, Fre11, Por11}. 
Regarding wind mass losses from the inner binary exclusively,  
the fractional mass losses $|\dot{m}|/(m_1 + m_2) > |\dot{m}|/(m_1 + m_2 + m_3)$. Therefore, the fractional orbital increases $\dot{a}_{\rm in}/a_{\rm in} > \dot{a}_{\rm out}/a_{\rm out}$, following Eq.\,\ref{eq:wind_cons_a}. 
\citet{Per12} shows that such a triple evolution dynamical instability (TEDI) lead to close encounters, collisions, and exchanges between the stellar components. They find that the TEDI evolutionary channel caused by stellar winds is responsible for the majority of stellar collisions in the Galactic field.

\subsubsection{Lidov-Kozai mechanism}
\label{sec:bg_ts_KL}

Secular dynamics can play a mayor role in the evolution of triple systems.  The key effect is the Lidov-Kozai mechanism \citep{Lid62,Koz62}, see Sect.\,\ref{sec:ex_gliese} for an example of a triple undergoing Lidov-Kozai cycles. Due to a mutual torque between the inner and outer binary orbit, angular momentum is exchanged between the orbits. The orbital energy is conserved, and therefore the semi-major axes are conserved as well \citep[e.g.][]{Mar01}. As a consequence, the orbital inner eccentricity and mutual inclination vary periodically. 
The maximum eccentricity of the inner binary is reached when the inclination between the two orbits is minimized.
Additionally, the argument of pericenter may rotate periodically (also known as precession or apsidal motion) or librate. 
For a comprehensive review of the Lidov-Kozai effect, see \citet{Nao16b}.

The Lidov-Kozai mechanism is of great importance in several astrophysical phenomena. For example, it can play a mayor role in the eccentricity and obliquity of exoplanets \citep[e.g.][]{Hol97, Ver10, Nao11} including high-eccentricity migration to form hot Jupiters \citep[e.g.][]{Wu03, Cor11, Pet15}, and for accretion onto black holes in the context of tidal disruption events \citep[e.g.][]{Che09, Weg11} or mergers of (stellar and super-massive) black hole binaries \citep[e.g.][]{Bla02, Mil02, Ant14}.
In particular for the evolution of close binaries, the Lidov-Kozai oscillations may play a key role \citep[e.g.][]{Har69, Maz79,Kis98,Fab07, Nao14}, e.g. for black hole X-ray binaries \citep{Iva10}, blue stragglers \citep{Per09}, and supernova type Ia progenitors \citep{Tho11, Ham13}.

When the three-body Hamiltonian is expanded to quadrupole order in $a_{\rm in}/a_{\rm out}$, 
the timescale for the Lidov-Kozai cycles is \citep{Kin99}:
\begin{equation}
t_{\rm kozai} = \alpha \dfrac{P_{\rm out}^2}{P_{\rm in}} \dfrac{m_1+m_2+m_3}{m_3} \left(1-e_{\rm out}^2\right)^{3/2},
\label{eq:t_kozai}
\end{equation}
where $P_{\rm in}$ and $P_{\rm out}$ are the periods of the inner and outer orbit, respectively. The dimensionless quantity $\alpha$ depends weakly on the mutual inclination, and on the eccentricity and argument of periastron of the inner binary, and is of order unity \citep{Ant15}.
The timescales are typically much longer than the periods of the inner and outer binary. 

Within the quadrupole approximation, the maximum eccentricity $e_{\rm max}$ is a function of the initial mutual inclination  $i_{\rm i}$ as \citep{Inn97}:
\begin{equation}
e_{\rm max} = \sqrt{1-\frac{5}{3} \mathrm{cos}^2(i_{\rm i})},
\label{eq:e_max}
\end{equation}
in the test-particle approximation \citep{Nao13}, i.e. nearly circular orbits ($e_{\rm in}=0$, $e_{\rm out} = 0$) with one of the inner two bodies a massless test particle ($m_1 \ll m_0, m_2$) and the inner argument of pericenter $g_{\rm in} = 90^{\circ}$. In this case, the (regular) Lidov-Kozai cycles only take place when the initial inclination is between 39.2-140.8$^{\circ}$. For larger inner eccentricities, the range of initial inclinations expands.

For higher orders of $a_{\rm in}/a_{\rm out}$ i.e. the octupole level of approximation, even richer dynamical behaviour is expected than for the quadrupole approximation \citep[e.g.][]{For00, Bla02,Lit11, Nao13, Sha13, Tey13}.
The octupole term is non-zero when the outer orbit is eccentric or if the stars in the inner binary have unequal masses. Therefore it is often deemed the 'eccentric Lidov-Kozai mechanism'. In this case the z-component of the angular momentum of the inner binary is no longer conserved. It allows for a flip in the inclination such that the inner orbit flips from prograde to retrograde or vice versa (hereafter 'orbital flip'). Another consequence of the eccentric Lidov-Kozai mechanism is that the eccentricity of the inner binary can be excited very close to unity. The octupole parameter $\epsilon_{\rm oct}$ measure the importance of the octupole term compared to the quadropole term, and is defined by:
\begin{equation}
\epsilon_{\rm oct} = \dfrac{m_1-m_2}{m_1+m_2} \dfrac{a_{\rm in}}{a_{\rm out}} \dfrac{e_{\rm out}}{1-e_{\rm out}^2}.
\label{eq:e_oct}
\end{equation}
Generally, when $|\epsilon_{\rm oct}| \gtrsim 0.01$, the eccentric Lidov-Kozai mechanism can be of importance \citep{Nao11, Sha13}.

The dynamical behaviour of a system undergoing regular or eccentric Lidov-Kozai cycles can lead to extreme situations. For example, as the eccentricity of the inner orbit increases, the corresponding pericenter distance decreases.
The Lidov-Kozai mechanism is therefore linked to a possible enhanced rate of grazing interactions, physical collisions, and tidal disruptions events of one of the stellar components \citep{For00, Tho11}, and to the formation of eccentric semi-detached binaries (Sect.\,\ref{sec:bg_ts_mt_inner}).

\subsubsection{Lidov-Kozai mechanism with mass loss}
\label{sec:bg_ts_ml}
Eqs.\,\ref{eq:t_kozai}~and~\ref{eq:e_oct} show that the relevance of the Lidov-Kozai mechanism for a specific triple strongly depends on the masses and mass ratios of the stellar components. If one of the components loses mass, the triple can change from one type of dynamical behaviour to another type. For example, mass loss from one of the stars in the inner binary, can increase $|\epsilon_{\rm oct}|$ significantly. As a result the triple can transfer from a regime with regular Lidov-Kozai cycles to a regime where the eccentric Lidov-Kozai mechanism is active. 
This behaviour is known as mass-loss induced eccentric Kozai (MIEK) \citep{Sha13, Mic14}. See also Sect.\,\ref{sec:ex_miek} for an example of this evolutionary pathway. 

The inverse process (inverse-MIEK), when a triple changes state from the octupole to the quadrupole regime, can also occur. Eq.\,\ref{eq:e_oct} shows this is the case when mass loss in the inner binary happens to create an fairly equal mass binary, or when the semi-major axis of the outer orbit increases. This latter is possible when the outer star loses mass in a stellar wind (Sect.\,\ref{sec:bg_ds_wind}). 

Another example comes from \citet{Mic14}, who studied the secular freeze-out (SEFO). In this scenario mass is lost from the inner binary such that the Lidov-Kozai timescale increases (Eq.\,\ref{eq:t_kozai}). This induces a regime change from the quadrupole regime, to a state where secular evolution is either quenched or operates on excessively long time-scales. 

The three examples given above illustrate that the dynamical evolution of a triple system is intertwined with the stellar evolution of its components.
Thus, in order to gain a clear picture of triple evolution, both three-body dynamics and stellar evolution need to be taken into account simultaneously.

\subsubsection{Precession}
\label{sec:bg_ts_prec}
Besides precession caused by the Lidov-Kozai mechanism, other sources of precession exist in stellar triples.  These include general relativistic effects \citep{Bla02}: 
\begin{equation}
\dot{g}_{\rm GR} = \frac{3a^2\Omega_b^3}{c^2(1-e^2)}  .
\label{eq:g_dot_gr}
\end{equation}

Furthermore, orbital precession can be caused by the distortions of the individual stars by tides \citep{Sme01}:
\begin{equation}
\dot{g}_{\rm tides} = \frac{15k_{\rm am}}{(1-e^2)^5\Omega_b}\left( 1+\frac{3}{2}e^2+\frac{1}{8}e^4\right) \frac{m_\mathrm{a}}{m_\mathrm{d}}  \left( \frac{R}{a} \right )^5,
\label{eq:g_dot_tides}
\end{equation}

and by intrinsic stellar rotation \citep{Fab07}:
\begin{equation}
\dot{g}_{\rm rotate} = \frac{k_{\rm am} \Omega^2}{(1-e^2)^2\Omega_b} \frac{m_\mathrm{d}+m_\mathrm{a}}{m_\mathrm{d}}  \left( \frac{R}{a} \right )^5,
\label{eq:g_dot_rot}
\end{equation}
where $m_d$ is the mass of the distorted star that instigates the precession and $m_a$ the companion star in the two-body orbit. The distorted star has a classical apsidal motion constant $k_{\rm am}$, radius $R$, and a spin frequency $\Omega$.
The precession rates in Eq.\,\ref{eq:g_dot_gr}, as well as Eq.\,\ref{eq:g_dot_tides}~and~Eq.\,\ref{eq:g_dot_rot},
are always positive. This implies that relativistic effects, tides and stellar rotation mutually stimulate  precession in one direction. Note, that precession due to these processes also take place in binaries, which affects the binary orientation, but not the evolution of the system.

If the timescales\footnote{See for example Eq.\,\ref{eq:gw_a}~and~\ref{eq:gw_e}. The timescales for GW emission are shorter for binaries with small separations and large eccentricities.} 
for these processes become comparable or smaller than the Lidov-Kozai timescales, the Lidov-Kozai cycles are suppressed. 
Because Lidov-Kozai cycles are driven by tidal forces between the outer and inner orbit, 
 the additional precession tends to destroy the resonance  \citep{Liu15}. 
As a result of the suppression of the cycles, 
the growth of the eccentricity is limited, and orbital flips are limited to smaller ranges of the mutual inclination \citep{Nao12, Pet15, Liu15}.

\subsubsection{Tides and  gravitational waves}
\label{sec:bg_ts_tides_gws}
As mentioned earlier, the Lidov-Kozai mechanism can lead to very high eccentricities that drives the stars of the inner binary close together during pericenter passage. 
During these passages, tides and GW emission can effectively alter the orbit \citep{Maz79,Kis98}. Both processes are dissipative, and act to circularize the orbit and shrink the orbital separation (Sect.\,\ref{sec:bg_ds_aml}~and~\ref{sec:bg_ds_tides}). The combination of Lidov-Kozai cycles with tides or GW emission can then lead to an enhanced rate of mergers and RLOF. For GW sources, the  merger time of a close binary can be significantly reduced, if an outer star is present that gives rise to  Lidov-Kozai cycles on a short timescale \citep{Tho11}. 
This is important in the context of supernova type Ia and gamma-ray bursts.

The combination of Lidov-Kozai cycles with tidal friction (hereafter LKCTF)  can also lead to an enhanced formation of close binaries \citep{Maz79, Kis98}. 
This occurs when a balance can be reached between the eccentricity excitations of the (regular or eccentric) Lidov-Kozai mechanism and the circularisation due to tides\footnote{A balance is also possible between 
the eccentricity excitations of the Lidov-Kozai mechanism and other sources of precession, see Sect.\ref{sec:bg_ts_prec}). }.
The significance of LKCTF is illustrated by \citet{Fab07}, who show that MS binaries with orbital periods of 0.1--10d are produced from binaries with much longer periods up to $10^5$d. Observationally, 96\% of close low-mass MS binaries are indeed part of a triple system \citep{Tok06}. 
Several studies of LKCTF for low-mass MS stars exist \citep{Maz79,Egg01, Fab07, Kis10, Ham13}, however, a study of the effectiveness of LKCTF for high-mass MS triples or triples with more evolved components is currently lacking. Due to the radiative envelopes of high-mass stars, LKCTF is likely less effective compared to the low-mass MS case. However, evolved stars develop convective envelopes during the giant phases for which tidal friction is expected to be effective. Hence, in order to understand the full significance of LKCTF for triple evolution, it is necessary to model three-body dynamics and stellar evolution consistently.

\subsubsection{Mass transfer initiated in the inner binary...}
\label{sec:bg_ts_mt_inner}

In Sect.\,\ref{sec:bg_ds_mt}, we described the effect of mass transfer on a circularized and synchronized binary. However, as Lidov-Kozai cycles can lead effectively to RLOF in eccentric inner binaries, the simple picture of synchronization and circularisation before RLOF, is no longer generally valid for triples. 
In an eccentric binary, there does not exist a frame in which all the material is corotating, and the binary potential becomes time-dependent. 
Studies of the Roche lobe for eccentric and/or asynchronous binaries, show that the Roche lobe can be substantially altered \citep{Pla58, Reg05, Sep07}. In an eccentric orbit, the Roche lobe of a star at periastron may be significantly smaller than that in a binary that is circularized at the same distance $r_p=a(1-e)$. 
The Roche lobe is smaller for stars that rotate super-synchronously at periastron compared to the classical Roche lobe (Eq.\,\ref{eq:rl}), and larger for sub-synchronous stars. 
It is even possible that the Roche lobe around the accretor star opens up. When mass is transferred from the donor star through L1, it is not necessarily captured by the accretor star, and mass and angular momentum may be lost from the binary system.

The modification of the Roche lobe affects the evolution of the mass transfer phase, e.g. the duration and the mass loss rate. Mass transfer in eccentric orbits of isolated binaries has been studied in recent years with SPH techniques \citep{Lay98, Reg05, Chu09, Laj11, Van16} as well as analytical approaches \citep{Sep07b, Sep09, Sep10, Dav13, Dou16, Dou16b}. 
These studies have shown that (initially) the mass transfer is episodic. The mass transfer rate peaks just after periastron, and its evolution during the orbit shows a Gaussian-like shape with a FWHM of about 10\% of the orbital period. 

The long-term evolution of eccentric binaries undergoing mass transfer can be quite different compared to circular binaries. 
The long-term evolution has been studied with analytics adopting a delta-function for the mass transfer centred at periastron \citep{Sep07b, Sep09, Sep10, Dou16, Dou16b}. 
Under these assumptions, the semi-major axis and eccentricity can increase as well as decrease depending on the properties of the binary at the onset of mass transfer. In other words, the secular effects of mass transfer can enhance and compete with the orbital effects from tides. Therefore, rapid circularization of eccentric binaries during the early stages of mass transfer is not generally justified.

The current theory of mass transfer in eccentric binaries predicts that some binaries can remain eccentric for long periods of time. The possibility of mass transfer in eccentric binaries is supported by observations. For example, the catalogue of eccentric binaries of \citet{Pet99} contains 19 semi-detached and 6 contact systems out of 128 systems. That circularisation is not always achieved before RLOF commences, is supported by observations of some detached, but close binaries e.g. ellipsoidal variables \citep{Nic12} and Be X-ray binaries, which are a subclass of high-mass X-ray binaries \citep{Rag05}.

\subsubsection{... and its effect on the outer binary}
\label{sec:bg_ts_mt_inner_onto_outer}
Mass transfer in the inner binary can affect the triple as a whole. The most simple case is during conservative stable mass transfer, when the outer orbit remains unchanged. If the inner orbit is circularized and synchronised, mass transfer generally leads to an increase in the inner semi-major axis by a factor of a few (Eq.\,\ref{eq:a_cons_mt}). When the ratio $a_{\rm out}/a_{\rm in}$ decreases, the triple approaches and possibly crosses into the dynamically unstable regime (Sect.\,\ref{sec:bg_ts_stability} and Eq.\,\ref{eq:stab_crit}). 

If the mass transfer in the inner binary occurs non-conservatively, the effect on the outer binary is completely determined by the details of the mass loss from the inner binary. We conceive three scenarios for this to take place. First, if during stable mass transfer to a hydrogen-rich star, matter escapes from the inner binary, it is likely the matter will escape from L2 in the direction of the orbital plane of the inner binary. Second, during mass transfer to a compact object, a bipolar outflow or jet may develop. 
Thirdly, matter may be lost from the inner binary as a result of a common-envelop phase,  which we will discuss in Sect.\,\ref{sec:bg_ts_ce_inner_onto_outer}.

If matter escapes the inner binary, its velocity must exceed the escape velocity:
\begin{equation}
v_{\rm esc, in} = \sqrt{\frac{2G(m_1+m_2)}{a_{\rm in} (1+e_{\rm in})}},
\label{eq:v_esc_in}
\end{equation}
and analogously, to escape from the outer binary, and the triple as a whole: 
\begin{equation}
v_{\rm esc, out} = \sqrt{\frac{2G(m_1+m_2+m_3)}{a_{\rm out} (1+e_{\rm out})}}. 
\label{eq:v_esc_out}
\end{equation}
For stable triples, such that $a_{\rm out}/a_{\rm in} \gtrsim 3$, $v_{\rm esc, in} > v_{\rm esc, out}$ unless $m_3\gtrsim f(m_1+m_2)$. The factor $f$ is of the order of one, e.g. for circular orbits $f=2$. In the catalogue of \citet{Tok14II} it is uncommon that $m_3\gtrsim (m_1+m_2)$. Out of 199 systems, there are 3 systems with $(m_1+m_2) <m_3< 2(m_1+m_2)$ and none with $m_3\gtrsim 2(m_1+m_2)$. 
Therefore, if the inner binary matter is energetic enough to escape from the inner binary, it is likely to escape from the triple as a whole as well.

For isolated binary evolution, it is unclear if the matter that leaves both Roche lobes is energetic enough to become unbound from the system, e.g. when mass is lost through L2. Instead a circumbinary disk may form that gives rise to a tidal torque between the disk and the binary. This torque can efficiently extract angular momentum from the binary i.e. Eq.\,\ref{eq:a_mt} with 
$\eta=\sqrt{\frac{a_{\rm ring}}{a}}\frac{(m_1+m_2)^2}{m_1 m_2}$, where $a_{\rm ring}$ is the radius of the circumbinary ring \citep{Sob97}. For example, for a binary with $q=2$, angular momentum is extracted more than 10 faster if the escaping matter forms a circumbinary disk compared to a fast stellar wind \citep{Sob97, Too14}. Hence, the formation of a circumbinary disk leads to a stronger reduction of the binary orbit, and possibly a merger.

If a circumbinary disk forms around the inner binary of a triple, we envision two scenarios. Firstly, the outer star may interact with the disk directly if its orbit crosses the disk. Secondly, the disk gives rise to two additional tidal torques, with the inner binary and the outer star. It has been shown that the presence of a fourth body can lead to a suppression of the Lidov-Kozai cycles, however, bodies less massive than a Jupiter mass have a low chance of shielding \citep{Ham15,Ham16, Mar15,  Mun15}.

\subsubsection{The effect of common-envelope on the outer binary}
\label{sec:bg_ts_ce_inner_onto_outer}

Another scenario exists in which material is lost from the inner binary; in stead of a stable mass transfer phase, mass is expelled through a CE-phase.
For isolated binaries the CE-phase and its effect on the orbit is an unsolved problem, and the situation becomes even more complicated for triples. 
In the inner binary the friction between the stars and the material is expected to cause a spiral-in. If the outer star in the triple is sufficiently close to the inner binary, the matter may interact with the outer orbit such that a second spiral-in takes place. If on the other hand, the outer star is in a wide orbit, and the CE-matter is lost in a fast and isotropic manner, the effect on the outer orbit would be like a stellar wind (Sect.\,\ref{sec:bg_ds_wind}). 
\citet{Ver12} study the effect of a CE in a binary with a planet on a wider orbit. Assuming that the CE affects the planetary orbit as an isotropic wind, they find that planetary orbits of a few $10^4$\Rsolar\,are readily dissolved. 
In this scenario the CE-phase operates virtually as a instantaneous mass loss event, and therefore the maximum orbital separation for the outer orbit to remain bound is strongly dependent on the uncertain timescale of the CE-event (Sect.\,\ref{sec:bg_ds_wind}). Disruption of a triple due to a CE-event may also apply to stellar triples, however, the effect is likely less dramatic, as the relative mass lost in the CE-event to the total system mass is lower.

Note that most hydrodynamical simulations of common-envelope evolution show that matter is predominantly lost in the orbital plane of the inner binary \citep[e.g.][]{Ric12, Pas12}, however, these simulations are not been able to unbind the majority of the envelope. In contrast, in the recent work of \citet{Nan15}, the envelope is expelled successfully due to the inclusion of recombination energy in the equation of state. These simulations show a more spherical mass loss. Roughly 60\% of the envelope mass is ejected during the spiral-in phase in the orbital plane, while the rest of the mass is ejected after the spiral-in phase in a closely spherical way (priv. comm. Jose Nandez).

 The first scenario of friction onto the outer orbit has been proposed to explain the formation of two low-mass X-ray binaries with triple components (4U 2129+47 (V1727 Cyg) \citep{Por11} and PSR J0337+1715  \citep{Tau14}), however, it could not be ruled out that the desired decrease in the outer orbital period did not happen during the SN explosion in which the compact object was formed. Currently, it is unclear if the CE-matter is dense enough at the orbit of the outer star to cause significant spiral-in. \citet{Sab15} suggests that if there is enough matter to bring the outer star closer, the CE-phase would lead to a merger in the inner binary.

\subsubsection{Mass loss from the outer star}
\label{sec:bg_ts_mt_outer}
In about 20\% of the multiples in the Tokovinin catalogue of multiple star systems in the Solar neighbourhood, the outer star is more massive than the inner two stars. For these systems the outer star evolves faster than the other stars (Sect.\,\ref{sec:bg_ss}). In about 1\% of the triples in the Tokovinin catalogue, the outer orbit is significantly small that the outer star is expected to fill its Roche lobe at some point in its evolution \citep{DeV14}. 

What happens next has not been studied to great extent, i.e. the long-term evolution of a triple system with a mass-transferring outer star. It is an inherently complicated problem where the dynamics of the orbits, the hydrodynamics of the accretion stream and the stellar evolution of the donor star and its companion stars need to be taken into account consistently. 

Such a phase of mass transfer has been invoked to explain the triple system PSR J0337+1715, consisting of a millisecond pulsar with two WD companions \citep{Tau14, Sab15}. 
\citet{Tau14} note one of the major uncertainties in their modelling of the evolution of PSR J0337+1715 comes from the lack of understanding of the accretion onto the inner binary system and the poorly
known specific orbital angular momentum of the ejected mass during the outer mass transfer phase. \citet{Sab15} proposes that if the inner binary spirals-in to the envelope of the expanding outer star, the binary can break apart from tidal interactions. 

To the best of our knowledge, only \citet{DeV14} have performed detailed simulations of mass transfer initiated by the outer star in a triple. They use the same software framework (\texttt{AMUSE}, Sect.\,\ref{sec:methods}) as we use for our code \code. \citet{DeV14} simulate the  mass transfer phase initiated by the outer star for two triples in the Tokovinin catalogue, $\xi$ Tau and HD97131. For both systems, they find that the matter lost by the outer star does not form an accretion disk or circumbinary disk, but instead the accretion stream intersects with the orbit of the inner binary. The transferred matter forms a gaseous cloud-like structure and interacts with the inner binary, similar to a CE-phase. The majority of the matter is ejected from the inner binary, and the inner binary shrinks moderately to weakly with $\alpha \lambda_{\rm ce} \gtrsim 3$ depending on the mutual inclination of the system. 
In the case of HD 97131, this contraction leads to RLOF in the inner binary. 
The vast majority of the mass lost by the donor star is funnelled through L1, and eventually ejected from the system by the inner binary through the L3 Lagrangian point\footnote{Here the L3 Lagrangian point is located behind the inner binary on the line connecting the centres of mass of the outer star and inner binary.} of the outer orbit. As a consequence of the mass and angular momentum loss, the outer orbit shrinks with\footnote{There is an inconsistency in the meaning of $\eta$ between Eqs.4~and~5 in \citet{DeV14}, denoted as $\beta$ in their equations. In their fits $\eta$ represents the ratio $\frac{\Delta J}{J_b}$, where $\Delta J$ is the amount of angular momentum that is lost from the system, and $J_b$ the orbital angular momentum. It does not represent $\frac{\Delta J}{J_a}$ where $J_a$ is the angular momentum of the accretor star.} $\eta \approx$ 3--4 in Eq.\,\ref{eq:a_mt}. 
During the small number of outer periods that are modelled, the inner and outer orbits approaches contraction at the same fractional rate. Therefore the systems remain dynamically stable.

Systems that are sufficiently wide that the outer star does not fill its Roche lobe, might still be affected by mass loss from the outer star in the form of stellar winds.                                    
\citet{Sok04} has studied this scenario for systems where the outer star is on the AGB, such that the wind mass loss rates are high. Assuming Bondi-Hoyle-Littleton accretion and 
\begin{equation}
R_1 \ll R_{\rm acc, column} \lesssim a_{\rm in} \ll R_{\rm acc, B-H} \ll a_{\rm out},
\end{equation}
where $R_{\rm acc, column}$ is the width of the accretion column at the binary location, and $R_{\rm acc, B-H}$ the Bondi-Hoyle accretion radius, 
\citet{Sok04} finds that a large fraction of triples the stars in the inner binary may accrete from an accretion disk around the stars. 
The formation of an accretion disk depends strongly on the orientation of the inner and outer orbit. 
When the inner and outer orbits are parallel to each other, no accretion disk forms. On the other hand, when the inner orbit is orientated perpendicular to the outer orbit, an accretion disk forms if 
$q \lesssim 3.3$. In this case the accretion is in a steady-state and mainly towards the most massive star of the inner binary.

\subsubsection{Triples and planetary nebulae}
\label{sec:bg_ts_pn}

Interesting to mention in the context of triple evolution are planetary nebulae (PNe), in particular those with non-spherical structures. 
The formation of these PNe is not well understood, but maybe attributed to interactions between an AGB-star and a companion \citep[e.g.][]{Bon90, Bon00, DeM15, Zij15} or multiple companions \citep[e.g.][]{Bon02, Ext10, Sok16, Bea16}. Where a binary companion can impose a non-spherical symmetry on the resulting PN, and even a non-axisymmetry \citep[see e.g. ][for eccentric binaries]{Sok01}, triple evolution can impose structures that are not axisymmetric, mirrorsymmetric, nor pointsymmetric \citep{Bon02, Ext10, Sok16}.

Since the centers of many elliptical and bipolar PNe host close binaries, the systems are expected to have undergone a CE-phase. 
In the context of triples, PNe formation channels have been proposed that concern outer stars on the AGB whose envelope matter just reaches or completely engulfs a tight binary system, e.g. the PN SuWt\,2 \citep{Bon02, Ext10}. Another proposed channel involves systems with a very wide outer orbit of tens to thousands of AU and in which the outer star interacts with the material lost by the progenitor-star of the PN \citep{Sok92, Sok94}. For a detailed review of such evolutionary channels, see \citet{Sok16}. Under the assumptions that PN from triple evolutionary channels give rise to irregular PNe, \citet{Sok16} and \citet{Bea16} find that about 1 in 6-8 PNe might have been shaped by an interaction with an outer companion in a triple system.

\subsubsection{Supernova explosions in triples}
\label{sec:bg_ts_sn}

\citet{Pij12} study the effect of a supernova explosion in a triple star system under the same assumptions as \citet{Hil83}. 
The authors show that for a hierarchical triple in which the outer star collapses in a SN event, the inner binary is not affected, and the effect on the outer orbit can be approximated by that of an isolated binary. 
For a SN taking place in the inner binary, the inner binary itself is modified similar to an isolated binary (Sect.\,\ref{sec:bg_ds_sn}). The effect on the outer binary can be 
viewed as that of an isolated binary in which the inner binary is replaced by an effective star at the center of mass of the inner binary. The effective star changes mass and position (as the center of mass changes) instantaneously in the SN event. The semi-major axis of the outer orbit is affected by the SN in the inner binary as:

\begin{eqnarray}
\begin{array}{l}
\dfrac{a_\mathrm{f}}{a_\mathrm{i}} = \left(1-\dfrac{\Delta m}{m_\mathrm{t,i}}\right) \cdot \biggl( 1-\dfrac{2a_\mathrm{i}\Delta m}{r_\mathrm{f}m_\mathrm{t,i}}  \\
\\
  -\dfrac{2a(\ve{v}_\mathrm{i}\cdot \ve{v}_{\rm sys})}{v_\mathrm{c}^2}  - \dfrac{v_{\rm sys}\,^2 }{v_\mathrm{c}^2} + 2a_i\dfrac{r_\mathrm{i}-r_\mathrm{f}}{r_\mathrm{i}r_\mathrm{f}} \biggr)^{-1},
\end{array}
\label{eq:a_sn_tri}
\end{eqnarray}

where $m_\mathrm{t,i}$ is the total mass of the pre-SN triple,  
$r_\mathrm{i}$ and $r_\mathrm{f}$ are the pre-SN and post-SN distance between the star in the outer orbit and the center of mass of the inner binary, 
$\ve{v}_\mathrm{i}$ is the pre-SN relative velocity of the center of mass of the inner binary relative to the outer star, $\ve{v}_{\rm sys}$ is the systemic velocity the inner binary due to the SN, and 
\begin{equation}
v_\mathrm{c} \equiv \sqrt{\frac{Gm_\mathrm{t,i}}{a_\mathrm{i}}}
\label{eq:v_c_tri}
\end{equation}
the orbital velocity in a circular orbit. Note that in comparison with the circular velocity in binaries (Eq.\,\ref{eq:v_c_bin}), here $m_\mathrm{t,i}$ refers to $m_1+m_2+m_3$ and $a=a_{\rm out}$.

A full derivation of the change in semi-major axis (Eq.\,\ref{eq:a_sn_tri}) and eccentricity (Eq.\,\ref{eq_app:e_out}) of the outer orbit due to a SN in the inner orbit, are given in Appendix\,\ref{sec:app_sn}. Note that the equation for the post-SN eccentricity of the outer orbit in \citet[][their Eq.27]{Pij12} is incomplete (see Appendix\,\ref{sec:app_sn}).

\subsection{Quadruples and higher-order hierarchical systems}

Although quadruples star systems are less common than triple systems, hierarchical quadruples still comprise about 1\% of F/G dwarf systems in the field \citep{Tok14I, Tok14II}. 
While for triples, there is one type of hierarchy that is stable on long-terms (compared to stellar lifetimes), quadruples can be arranged in two distinct long-term stable configurations: the `2+2’ or `binary-binary’ configuration, and the `3+1’ or `triple-single’ configuration.
In the first case, two binaries orbit each others barycentre, and in the latter case a hierarchical triple is orbited by a fourth body. In the sample of F/G dwarfs of \citet{Tok14I, Tok14II}, the `2+2’ systems comprise about 2/3 of quadruples, and the `3+1’ about 1/3.

The secular dynamics of the `2+2’ systems were investigated by \citet{Pej13} using N-body methods. \citet{Pej13} find that in these systems, orbital flips and associated high eccentricities are more likely compared to equivalent triple systems (i.e. with the companion binary viewed as a point mass). 
The `3+1’ configuration was studied by \citet{Ham15} \footnote{\citet{Ham15} derive the orbit-averaged Hamiltonian expressions for the `3+1’ as well as the `2+2’ configuration.}. 
For highly hierarchical systems, i.e. in which the three binaries are widely separated, the global dynamics can be qualitatively described in terms of the (initial) ratio of the Lidov-Kozai time-scales of the two inner most binaries compared to that of the outer two binaries. 
This was applied to the `3+1’ F/G systems of \citet{Tok14I, Tok14II}, and most (90\%) of these systems were found to be in a regime in which the inner three stars are effectively an isolated triple, i.e. the fourth body does not affect the secular dynamical evolution of the inner triple. 

We note that in the case of `3+1’ quadruples and with a low-mass third body (in particular, a planet), the third body can affect the Lidov-Kozai cycles that would otherwise have been induced by the fourth body. In particular, under specific conditions the third body can `shield’ the inner binary from the Lidov-Kozai oscillations, possibly preventing the inner binary from shrinking due to tidal dissipation, and explaining the currently observed lack of circumbinary planets around short-period binaries \citep{Mar15, Mun15, Ham16}. A similar process could apply to more massive third bodies, e.g. low-mass MS stars.

It is currently largely unexplored how non-secular effects such as stellar evolution, tidal evolution and mass transfer affect the evolution of hierarchical quadruple systems, or, more generally, in higher-order multiple systems. The secular dynamics of the latter could be efficiently modelled using the recent formalism of \citet{Ham16b}.

\section{Methods}
\label{sec:methods}

In the previous section, we gave an overview of the most important ingredients of the evolution of stars in single systems, binaries and triples. For example, nuclear evolution of a star leads to wind mass loss, that affects the dynamics of binaries and triples, and can even lead to a dynamical instability in multiple systems. 
Three-body dynamics can give rise to oscillations in the eccentricity of the inner binary system of the triple, which can lead to an amplified tidal effect and an enhanced rate of stellar mass transfer, collisions, and mergers. 
Additionally, a triple system can transition from one to another dynamical regime (i.e. without Lidov-Kozai cycles, regular and eccentric Lidov-Kozai cycles) due to stellar evolution, e.g. wind mass loss or an enhancement of tides as the stellar radius increases in time. 
These examples illustrate that for the evolution of triple stars, stellar evolution and dynamics are intertwined.
Therefore, in order to study the evolution of triple star systems consistently, three-body dynamics and stellar evolution need to be taken into account simultaneously.

In this paper, we present a public source code \code\,to simulate the evolution of wide and close, interacting and non-interacting triples consistently. The code is designed for the study of coeval, dynamically stable, hierarchical, stellar triples. The code is based on heuristic recipes that combine three-body dynamics with stellar evolution and their mutual influences. These recipes are described here.

The code can be used to evaluate the distinct evolutionary channels of a specific population of triples or the importance of different physical processes in triple evolution. 
As an example, it can be used to assess the occurrence rate of stable and unstable mass transfer initiated in circular and eccentric inner orbits of triple systems (Toonen, Hamers, \& Portegies Zwart in prep.).  
We stress that modelling though a phase of stable mass transfer in an eccentric orbit is currently not implemented in \code, but we aim to add this to the capabilities of \code\,in a later version of the code.

The code \code\,is based on the secular approach to solve the dynamics (Sect.\,\ref{sec:orbital_evolution}) and stellar evolution is included in a parametrized way through the fast stellar evolution code \texttt{SeBa} (Sect.\,\ref{sec:stellar_evolution}). 
\code\,is written in the Astrophysics Multipurpose
Software Environment, or \texttt{AMUSE} \citep{Por09, Por13}. This is a component library
with a homogeneous interface structure based on Python. \texttt{AMUSE} can be downloaded for free at amusecode.org and github.com/amusecode/amuse.
In the AMUSE framework new and existing code from different domains (e.g. stellar dynamics, stellar evolution, hydrodynamics and radiative transfer) can be easily used and coupled. 
As a result of the easy coupling, the triple code can be easily extended to include a detailed stellar evolution code (i.e. that solve the stellar structure equations) or a direct N-body code to solve the dynamics of triples that are unstable or in the semi-secular regime (Sect.\,\ref{sec:secular_approach}).

\subsection{Structure of \code}
\label{sec:three_parts}
The code consist of three parts:\\
step 1) stellar evolution,\\
step 2) stellar interaction, \\
step 3) orbital evolution. \\
At the beginning of each timestep we estimate an appropriate timestep $dt_{\rm trial}$ and evolve the stars as single stars for this timestep (step 1).
The trial timestep is estimated with:
\begin{equation}
dt_{\rm trial} = min(dt_{\rm star}, dt_{\rm wind}, dt_{\rm R}, f_{\rm prev} dt_{\rm prev}),
\label{eq:dt}
\end{equation}
where $dt_{\rm star}$, $dt_{\rm wind}$, $dt_{\rm R}$ and $f_{\rm prev} dt_{\rm prev}$ are the minimum timesteps due to stellar evolution, stellar wind mass losses, stellar radius changes and the previous timestep. Each star gives rise to a single value for $dt_{\rm trial}$, where the minimum is adopted as a trial timestep in \code.
The timestep $dt_{\rm star}$ is determined internally by the stellar evolution code (\texttt{SeBa}, Sect.\ref{sec:stellar_evolution}). It is the maximum attainable timestep for the next iteration of this code and is mainly chosen such that the stellar masses that evolve due to winds, are not significantly affected by the timesteps. Furthermore, when a star changes its stellar type (e.g. from a horizontal branch star to an AGB star), the timestep is minimized to ensure a smooth transition. 
For \code, we require a more strict constraint on the wind mass losses, such that 
$dt_{\rm wind} = f_{\rm wind} m/\dot{m}_{\rm wind}$, 
where $f_{\rm wind}=0.01$ and $\dot{m}_{\rm wind}$ is the wind mass loss rate given by the stellar evolution code. The numerical factor $f_{\rm wind}$ establishes a maximum average of 1\% mass loss from stellar winds per timestep.
Furthermore, we ensure that the stellar radii change by less then a percent per timestep through 
$dt_{\rm R} = f_{R}f'_{R}\cdot R/\dot{R}$, 
where $f_{R}$ and $f'_{R}$ are numerical factors. We take $f_{R} = 0.005$ and 

\begin{equation}
f'_{R} =
\left\{
	\begin{array}{ll}
		0.1  & \mbox{for } \dot{R}_{\rm prev}  = 0~\Ro/\rm{ yr} \\
		0.01  & \mbox{for } \dot{R} \cdot \dot{R}_{\rm prev} < 0 ~(\Ro/\rm{ yr})^2\\
		
		1  & \mbox{for } \dot{R} < \dot{R}_{\rm prev} ~\&~\rm{not~MS}\\
		\dot{R}/\dot{R}_{\rm prev}  & \mbox{for } \dot{R} < \dot{R}_{\rm prev} ~\& ~\rm{MS}\\
		\dot{R}_{\rm prev}/\dot{R}  & \mbox{for } \dot{R} > \dot{R}_{\rm prev}\\
	\end{array}
\right.
\label{eq:f'_R}
\end{equation}
where $\dot{R}$ and $\dot{R}_{\rm prev}$ represent the time derivative of the radius of the current and previous timestep, respectively. This limit is particularly important since the degree of tidal interaction strongly depends on the stellar radius. Lastly, we require that 
$dt_{\rm trial} < f_{\rm prev} dt_{\rm prev}$,
where $dt_{\rm prev}$ is the previous successfully accomplished timestep and $f_{\rm prev}$ a numerical factor with a value of 100.

The trial timestep is accepted and the code continues to step 2, only if:\\
case a) no star has started to fill its Roche lobe,\\ 
case b) stellar radii  have changed by less than 1\%, \\
case c) stellar masses have changed by less than 5\% \\
within the trial timestep. 
We have tested that these percentages give accurate results with respect to the orbital evolution. 
Condition b is not applied at moments when the stellar radius changes discontinuously, such as during the helium flash or at white dwarf formation. 

Conditions b \& c are not applied, when a massive star collapses to a neutron star or black hole. Note that when a star undergoes such a supernova explosion, the timestep is minimized through $dt_{\rm star}$. Additionally step 2 and 3 are skipped and the triple is adjusted according to Sect.~\ref{sec:si_SN}.
 
If conditions a, b and c are not met, the timestep is reverted and step 1 is tried again with a smaller timestep $dt'_{\rm trial}$. This process is done iteratively until the conditions are met or until the timestep is sufficiently small, $dt_{\rm min} = 10^{-3}$yr. 
If the change in the stellar parameters is too large (i.e. case b \& c), the new trail timestep is taken to be: 
\begin{equation}
dt'_{\rm trial} = {\rm min}[0.9dt_{\rm trial}, dt'_{\rm trial}(\dot{R}')],
\label{eq:dt_trial_r}
\end{equation}
where $0.9 dt_{\rm trial}$ represents 90\% of the previous timestep and $dt'_{\rm trial}(\dot{R}')$ is a newly calculated timestep according to Eq.~\ref{eq:dt} for which the time derivative of the radius $\dot{R}'$ from the last trial timestep is used.

During mass transfer, the timestep is estimated by: 
\begin{equation}
dt_{\rm trial, MT} = {\rm min}(dt_{\rm trial}, f_{\rm MT} \frac{m}{\dot{m}_{\rm MT}}, f_{\rm MT, prev} dt_{\rm prev}),
\label{eq:dt_mt}
\end{equation}
where $\dot{m}_{\rm MT}$ is the mass loss rate from mass transfer (Sect.\,\ref{sec:si_stable_mt}), and the numerical factors $f_{\rm MT}=0.01$ and $f_{\rm MT, prev}=5$.
If a star starts filling its Roche lobe in a timestep, $dt'_{\rm trial} = 0.5dt_{\rm trial}$. 
If $dt'_{\rm trial} < dt_{\rm trial, MT}$, mass transfer is allowed to commence.

Step 2 in our procedure regards the modelling of the stellar interactions such as stable mass transfer, contact evolution and common-envelope evolution (Sect.~\ref{sec:stellar_interaction}).   

The last step involves the simulation of the orbital evolution of the system by solving a system of differential equations (Sect.\,\ref{sec:orbital_evolution}). 
If the evolution leads to the initiation of RLOF during the trial timestep, both the orbit and stellar evolution are reverted to the beginning of the timestep. If the time until RLOF is shorter than 1\% of $dt_{\rm MT}$, the latter is taken to be the new trial timestep and mass transfer is allowed to commence. If not, the timestep is taken to be the time until RLOF that was found during the last trial timestep. If during the orbital evolution the system becomes dynamically unstable, the simulation is terminated. The stability criterion of \citet{Mar01} is used (Eq.\ref{eq:stab_crit}). 
In all other cases, the trial timestep is accepted, and the next iteration begins.

\subsection{Stellar evolution}
\label{sec:stellar_evolution}
Single stellar evolution\footnote{Note that \texttt{SeBa} is not used to model binary evolution in \code.} is included through the fast stellar evolution code \texttt{SeBa} \citep{Por96, Nel01, Too12, Too13}. \texttt{SeBa} is a parametrized stellar evolution code providing parameters such as radius, luminosity and core mass as a function of initial mass and time. \texttt{SeBa} is based on the stellar evolution tracks from \citet{Hur00}. These tracks are fitted to the results of a detailed stellar evolution code \citep[based on][]{Egg71, Egg72} that solves the stellar structure equations.

\subsection{Orbital evolution}
\label{sec:orbital_evolution}
\code\,solves the orbital evolution through a system of first-order ordinary differential equations (ODE):

\begin{eqnarray}
\left\{
\begin{array}{l c l}
 \dot{a}_{\rm in} &=& \dot{a}_{\rm in, GR} +\dot{a}_{\rm in, TF} +\dot{a}_{\rm in, wind} +\dot{a}_{\rm in, MT} \\
 \dot{a}_{\rm out} &=& \dot{a}_{\rm out, GR} +\dot{a}_{\rm out, TF} +\dot{a}_{\rm out, wind} +\\
 &&\dot{a}_{\rm out, MT} \\
 \dot{e}_{\rm in} &=& \dot{e}_{\rm in,3b} + \dot{e}_{\rm in,GR} +\dot{e}_{\rm in,TF} \\
 \dot{e}_{\rm out} &=& \dot{e}_{\rm out,3b} +\dot{e}_{\rm out,GR} + \dot{e}_{\rm out,TF} \\
 \dot{g}_{\rm in} &=& \dot{g}_{\rm in,3b} +  \dot{g}_{\rm in,GR} + \dot{g}_{\rm in,tides} + \dot{g}_{\rm in,rotate}\\
 \dot{g}_{\rm out} &=& \dot{g}_{\rm out, 3b} + \dot{g}_{\rm out,GR} + \dot{g}_{\rm out,tides} +\\ &&\dot{g}_{\rm out,rotate}\\
 \dot{h}_{\rm in} &=& \dot{h}_{\rm in, 3b}\\
 \dot{\theta} &=& \frac{-1}{J_{\rm b, in}J_{\rm b, out}} [\dot{J}_{\rm b, in}(J_{\rm b, in}+J_{\rm b, out}\theta) + \\
&& \dot{J}_{\rm b, out}(J_{\rm b, out}+ J_{\rm b, in}\theta)]\\
 \dot{\Omega}_{1} &=& \dot{\Omega}_{1, TF}  +\dot{\Omega}_{1, I} +\dot{\Omega}_{\rm 1, wind} \\
 \dot{\Omega}_{2} &=& \dot{\Omega}_{2, TF}  +\dot{\Omega}_{2, I}  +\dot{\Omega}_{\rm 2, wind}\\
 \dot{\Omega}_{3} &=& \dot{\Omega}_{3, TF}  +\dot{\Omega}_{3, I}  +\dot{\Omega}_{\rm 3, wind},\\
 \end{array} \right.
\label{eq:ODE}
\end{eqnarray}

where $\theta \equiv {\rm cos}(i)$, $J_{\rm b, in}$ and $J_{\rm b,  out}$ are the orbital angular momentum of the inner and outer orbit,  and $\Omega_1$, 
$\Omega_2$~and~$\Omega_3$ the spin frequency of the star with mass $m_1$, $m_2$~and~$m_3$, respectively, and $I$ the moment of inertia of the corresponding star. $\dot x$ represents the time derivative of parameter $x$.  
Eq.\,\ref{eq:ODE} includes secular three-body dynamics (with subscript 3b), general relativistic effects (GR), tidal friction (TF), precession, stellar wind effects and mass transfer (MT). 
The quadrupole terms of the three-body dynamics are based on \citet{Har68}, and the octupole terms on \citet{For00} with the modification of \citet{Nao13}. 
Gravitational wave emission is included as in Eq.\,\ref{eq:gw_a}~and~\ref{eq:gw_e} \citep{Pet64}. Our treatment of tidal friction and precession is explained in Sect.\,\ref{sec:si_tides}. Magnetic braking is currently not included.
The treatment of stellar winds and mass transfer is described in Sect.\,\ref{sec:si_wind}-\ref{sec:si_stable_mt}.  

The ODE solver routine uses adaptive timesteps to simulate the desired timestep $dt_{\rm trial}$. Within the ODE solver, parameters that are not given in Eq.\,\ref{eq:ODE} (e.g. gyration radius), are assumed to be constant during $dt_{\rm trial}$. An exception to this is the stellar radius, mass and moment of inertia. Even though $dt_{\rm trial}$ is chosen such that the parameters do not change significantly within this timestep (Sect.\,\ref{sec:three_parts}), there is a cumulative effect that can violate angular momentum conservation on longer timescales if $\dot{\Omega}_{I}$ is not taken into account. As a non-interacting star evolves and the mass, radius and moment of inertia change, the spin frequency of the star evolves accordingly due to conservation of spin angular momentum. The change in the spin frequency is:
\begin{equation}
\dot{\Omega}_{I} = \frac{-\dot{I}\Omega}{I},
\label{eq:spin_dot_I}
\end{equation} 
where 
\begin{equation}
I = k_2(m-m_c)R^2 + k_3m_cR_c^2,
\label{eq:I}
\end{equation}
where $m_c$ and $R_c$ are the mass and radius of the core, $k_2=0.1$ and $k_3=0.21$ \citep{Hur00}. 
Thus we approximate the moment of inertia with a component for the core and for the envelope of the star. This method works well for evolved stars that have developed dense cores, as well as for MS stars with $m_c\equiv 0$\Msolar, and compact objects for which $m-m_c\equiv 0$\Msolar.

The initial spin periods of the stellar components of the triple are assumed to be similar to that of ZAMS stars. Based on observed rotational velocities of MS stars from \citet{Lan92}, \citet{Hur00} proposed the fit:
\begin{equation}
\Omega = \frac{2058}{R} \frac{330M^{3.3}}{15.0+M^{3.45}} \, \rm{yr^{-1}}. 
\label{eq:rot}
\end{equation}
As in \citet{Ham13}, we make the simplifying assumption that the stellar spin axes are aligned with the orbital axis of the corresponding star. For the vast majority of stellar triples the magnitude of the spin angular momenta are small compared to that of the orbital angular momenta. A consequence of this assumption, is that tidal friction from a spin-orbit misalignment is absent. The change in the mutual inclination in Eq.\,\ref{eq:ODE} is based on the conservation of total angular momentum \citep{Ham13}.

The set of orbital equations of Eq.\,\ref{eq:ODE} are solved by a routine based on the ODE solver routine presented in \citet{Ham13}. 
This routine uses the CVODE library, which is designed to integrate stiff ODEs \citep{Coh96}. It has been verified by comparing integrations with example systems presented in \citet{For00}, \citet{Bla02} and \citet{Nao11}, and comparing with analytical solutions at the quadrupole-order level of approximation assuming $J_{\rm b, out}\gg J_{\rm b, in}$, given by \citet{Kin99}.

The ODE consists of 
a combination of prescriptions for the main physical processes for triple evolution (e.g. three-body dynamics and tides) which are described in Sect.\,\ref{sec:background}. In order to set up the ODE, we have assumed  that the physical processes are independent of one another, such that their analytical treatments can be added linearly. 
\citet{Mic14} show that the dynamics of a hierarchical triple including mass loss and transfer can be well modelled with this approach. They find that the secular approach shows excellent agreement with full N-body simulations. Additionally, we note that the ODE of Eq.\,\ref{eq:ODE} is valid as long as the included processes occur on timescales longer than the dynamical timescale. In the next section we discuss when this criterion is violated, and describe the alternative treatments in \code.

\subsubsection{The secular approach }
\label{sec:secular_approach}

The components in Eq.\,\ref{eq:ODE} ($\dot{e}_{\rm 3b}$, $\dot{g}_{\rm 3b}$, and $\dot{h}_{\rm 3b}$) 
that describe the secular three-body dynamics, including the regular Lidov-Kozai cycles, are derived using the orbital-averaging technique \citep[e.g.][]{Mic14, Nao16b, Luo16}. 
With this method, the masses of the components of the triple system are distributed over the inner and outer orbit. The three-body dynamics is then approximated by the interaction between these ellipses. 
Furthermore, the Hamiltonian is expanded up to third order in $a_{\rm in}/a_{\rm out}$ (i.e. the octopole term). The quadrupole level of approximation refers to the second-order expansion, which is the lowest non-trivial expansion order. 
The quadrupole level is sufficient for systems in which the octupole parameter (Eq.\,\ref{eq:e_oct}) is sufficiently low, i.e. $|\epsilon_{\rm oct}| < 10^{-4}$. For example, this includes systems with $m_1 \approx m_2$.   
To accurately model the dynamics of a population of systems with a wide range of mass ratios, eccentricities and orbital separations, one needs to include the octupole term as well.

However, whereas the orbit-averaged equations of motion are valid for strongly hierarchical systems, they become inaccurate for triples with weaker hierarchies \citep[e.g.][]{Ant12, Kat12, Nao13, Ant14, Ant14b, Bod14, Luo16}. For these systems, the timescale of the perturbations due to the outer star can become comparable to or shorter than the dynamical timescales of the system.  
This is problematic as the orbit-average treatment neglects any 
modulations on short orbital timescales per definition. 
For moderately hierarchical systems, the outer star can significantly change the angular momentum of the inner binary between two successive pericenter passages causing rapid oscillations in the corresponding eccentricity. 
The orbit-average treatment is valid when: 
\begin{equation}
\sqrt{1-e_{\rm in}} \gtrsim 5\pi \frac{m_3}{m_1+m_2} \left ( \frac{a_{\rm in}}{a_{\rm out}(1-e_{\rm out})} \right)^3,
\label{eq:orbit_average}
\end{equation}
as derived by \citet{Ant14}.

In the non-secular regime, the inner binary can be driven to much higher eccentricities than the secular approximation predicts, and subsequently lead to more collisions of e.g. black holes \citep{Ant14}, neutron stars \citep{Set13} or white dwarfs \citep{Kat12}. These are interesting in the context of gravitational wave emission and type Ia supernovae. Due to the very short timescale of the eccentricity oscillations, and therefore rapid changes of the periapse distance, tidal or general relativistic effects do not play a role \citep{Kat12}. With the secular approach, as in \code, the maximum inner eccentricity and therefore the number of collisions is probably underestimated for moderately hierarchical systems \citep[see also][]{Nao16}.

Furthermore, \citet{Luo16} showed recently that the rapid oscillations accumulate over time and alter the long-term evolution of the triple systems (e.g. whether or not an orbital flip occurs).  
The non-secular behaviour discussed by \citet{Luo16} occurs in systems in which the mass of the outer star
is comparable or larger than that of the inner binary, and in which the octupole term is important ($|\epsilon_{\rm oct}| \gtrsim 0.05$).

\subsection{Stellar interaction}
\label{sec:stellar_interaction}

\subsubsection{Stellar winds in \code}
\label{sec:si_wind}
The mass loss rate of each star depends on the evolutionary stage and is determined by the stellar evolution code. We make the common assumption that the wind is fast and spherically symmetric with respect to the orbit, as discussed in Sect.\,\ref{sec:bg_ds_wind}. 
For the inner binary, we assume a fraction $\beta_{1\rightarrow2}$ of the wind mass lost from $m_1$ at a rate $\dot{m}_1$ can be accreted by $m_2$, and $\beta_{2\rightarrow1}$ for the mass flowing in the other direction. 
Following Eq.\,\ref{eq:wind_a}, the effect on the inner orbit is then:
\begin{equation}
\dot{a}_{\rm in, wind} = 
\dot{a}_{wind}(\dot{m}_1, \beta_{1\rightarrow2}) +
\dot{a}_{wind}(\dot{m}_2, \beta_{2\rightarrow1}).
\label{eq:wind_a_in}
\end{equation}

The wind mass lost from the inner binary is $\dot{m}_{\rm in}=(1-\beta_{1\rightarrow2})\dot{m}_1+(1-\beta_{2\rightarrow1})\dot{m}_2$, of which the outer star can accrete a fraction $\beta_{in\rightarrow3}$. We do not allow the inner binary to accrete mass from the outer star. The winds widen the orbit according to:
\begin{equation}
\dot{a}_{\rm out, wind} =
\dot{a}_{wind}(\dot{m}_{\rm in}, \beta_{in\rightarrow3}) +
\dot{a}_{wind, no-acc}(\dot{m}_3),
\label{eq:wind_a_out}
\end{equation}  
see Eq.\,\ref{eq:wind_cons_a}~and~Eq.\,\ref{eq:wind_a}.

The wind matter carries away an amount of angular momentum which affects the spin of the star. Under the assumption that the wind mass decouples from the star as a spherical shell: 
\begin{equation}
\dot{\Omega}_{\rm wind} = \frac{-2/3\dot{m}_{\rm wind}R^2\Omega}{I}.
\label{eq:wind_spin}
\end{equation}
If wind matter is accreted by a star, we assume the accretor star spins up i.e. the stellar spin angular momentum increases with the specific angular momentum of the wind matter. For example for $m_1$, the total change in the spin due to winds is:
\begin{equation}
\dot{\Omega}_{\rm1,  wind} = \frac{-2/3\dot{m}_{\rm 1, wind}R_1^2\Omega_1}{I_1} + \frac{2/3\beta_{2\rightarrow 1}\dot{m}_{\rm 2, wind}R_2^2\Omega_2}{I_1}
\label{eq:wind_spin2}
\end{equation}

\subsubsection{Tides and precession}
\label{sec:si_tides}

Tidal friction is included in \code\,as described in Eq.\,\ref{eq:a_dot_TF}-\ref{eq:omega_dot_TF}. The dominant tidal dissipation mechanism is linked with the type of energy transport in the outer zones of the star. 
We follow \citet{Hur02}, and distinguish three types: damping in stars with convective envelopes, radiative envelopes (i.e. dynamical tide), and degenerate stars. 
The quantity $k_{\rm am}/\tau_{\rm TF}$ of Eq.\,\ref{eq:a_dot_TF}-\ref{eq:omega_dot_TF} is given for these three types of stars in their Eq.30, 42\footnote{
Note that there is an error in Eq.42 of \citet{Hur02}. The factor $MR^2/a^5$ should be raised to the power 1/2, which means that $k_{\rm am}/\tau_{\rm TF} \propto R $ instead of $k_{\rm am}/\tau_{\rm TF} \propto R^2 $.
}~and~47, respectively. We assume that  
radiative damping takes place in MS stars with $M>1.2$\Msolar, in helium-MS stars and horizontal branch stars. Excluding compact objects, all other stars are assumed to have convective envelopes. 
For the mass and radius of the convective part of the stellar envelope, we follow \citet{Hur00} (their Sec.7.2) and \citet{Hur02} (their Eq.36-40), respectively, with the modification that MS stars to have convective envelopes in the mass range 0.3-1.2\Msolar.
Regarding the gyration radius $k$, for stars with convective or radiative envelopes we assume $k=k_2$, for compact objects $k=k_3$ (see Eq.\,\ref{eq:I}).

We include precession from three-body dynamics, general relativistic effects \citep[Eq.\,\ref{eq:g_dot_gr},][]{Bla02}, tides \citep[Eq.\,\ref{eq:g_dot_tides},][]{Sme01}, and stellar rotation \citep[Eq.\,\ref{eq:g_dot_rot},][]{Fab07}. 
The latter two equations 
require an expression for the apsidal motion constant $k_{\rm am}$ (instead of $k_{\rm am}/\tau_{\rm TF}$ as required for the tidal equations of Eq.\,\ref{eq:a_dot_TF}-\ref{eq:omega_dot_TF}). For MS, helium-MS, and WDs, we assume $k_{\rm am}=0.0144$ \citep{Bro55}, for neutron stars $k_{\rm am}=0.260$ \citep{Bro55}, for black holes $k_{\rm am}=0$, and for other stars $k_{\rm am}=0.05$ \citep{Cla92}. For low-mass ($m<0.7$\Msolar) MS stars that are fully or deeply convective, we take $k_{\rm am}=0.05$ \citep{Cla92}.

\subsubsection{Stability of mass transfer initiated in the inner binary}
\label{sec:si_stability}
When one of the inner stars fills its Roche lobe, we test for the stability of the mass transfer:
\begin{itemize}
\item Tidal instability;\\
Tidal friction can lead to an instability in the binary system and subsequent orbital decay (see Sect.\,\ref{sec:bg_ds_tides}). The tidal instability takes place in compact binaries with extreme mass ratios. It occurs when there is insufficient angular momentum to keep the star in synchronization i.e.  $J_{\star} > \frac{1}{3} J_{\rm b}$, with $J_{\star}=I\Omega$. 
When RLOF occurs due to a tidal instability, we assume that a CE develops around the inner binary. This will lead further orbital decay, and finally either a merger or ejection of the envelope.

\item RLOF instability;\\
The stability of the mass transfer depends  on the response of the radius and the Roche lobe to the imposed mass loss. 
In the fundamental work of \citet{Hje87}, theoretical stability criteria are derived for polytropes. Stability criteria have been improved with the use of more realistic stellar models, see \citep[e.g.][]{Ge10, Woo10_Myk, Pas12b, Woo12, Ge15}. Our incomplete understanding of the stability of mass transfer leads to differences between synthetic binary populations \citep{Too14}.

The response of the Roche lobe is strongly dependent on the envelope of the donor star and the mass ratio of the system\footnote{and to a lesser degree also the accretion efficiency and the corresponding angular momentum loss mode \citep[e.g.][]{Sob97, Too14}.}. Therefore the stability of mass transfer is often described by a critical mass ratio $q_{\rm crit}< q \equiv m_d/m_a $ for different types of stars. 
For unevolved stars with radiative envelopes, mass transfer can proceed in a stable manner for relatively large mass ratios. 
We assume $q_{\rm crit} = 3$, unless the star is on the MS for which we take $q_{\rm crit} = 1.6$ \citep{DeM07b, Cla14}.
Stars with convective envelopes
are typically unstable to mass transfer, unless the donor is considerably less massive than the companion. 
For giants, we adopt $q_{\rm crit} = 0.362 + [3(1-M_c/M)]^{-1},$ where $M_c$ is the core mass of the donor star \citep{Hje87}. 
For naked helium giants, low-mass MS stars ($M<0.7\Mo$), and white dwarfs, we follow \citet{Hur02} and adopt $q_{\rm crit} = 0.784$,  $q_{\rm crit} = 0.695$ and $q_{\rm crit} = 0.628$, respectively.

\end{itemize}

\subsubsection{Common-envelope evolution in the inner binary}
\label{sec:si_ce}
As CE-evolution is a fast, hydrodynamic process, the ODE solver routine is disabled during the modelling of the CE-phase. 
If the donor star is a star without a clear distinction of the core and the envelope (i.e. MS stars, helium MS stars and remnants), we assume the phase of unstable mass transfer leads  to a merger. 
For other types of donor stars, the treatment of the CE-phase consists of three different models that are based on combinations of the formalisms described in Sect.\,\ref{sec:ce}. In model 1 and model 2, the $\alpha$-formalism (Eq.\,\ref{eq:alpha-ce}) and $\gamma$-formalism (Eq.\,\ref{eq:gamma-ce}) are used to determine the outcome of the CE-phase, respectively. When two giants are involved, the double-CE is applied (Eq.\,\ref{eq:double-alpha-ce}). 
In the standard model, the $\gamma$-formalism is applied unless the CE is triggered by a tidal instability or the binary contains a remnant star. This is based on modelling the evolution of double white dwarfs \citep{Nel00, Nel01, Too12}. The standard values of $\alpha\lambda_{\rm ce}$ and $\gamma$ are taken to be 2.0 and 1.75 \citep{Nel00}.

The companion star in the inner binary is probably not able to accrete from the overflowing material of the CE-phase, because of its relatively long thermal timescale compared to the short timescale on which the CE is expected to place. Therefore, we assume that the CE occurs completely non-conservatively. 

The effect of a CE-phase on the outer star of a triple is poorly studied or constrained (Sect.\,\ref{sec:bg_ts_ce_inner_onto_outer}). 
For stable, hierarchical systems, if the CE-material is energetic enough to escape from the inner binary, the matter is likely energetic enough to escape from the triple as well (Sect.\ref{sec:bg_ts_mt_inner_onto_outer}). We assume that the escaping CE-matter has expanded and diluted sufficiently to avoid a second CE-phase with the outer star, as we only consider stable triples with $a_{\rm out}/a_{\rm in} \gtrsim 3$. 
We allow the matter to escape as a fast wind in a non-conservative way i.e. according to Eq.\,\ref{eq:wind_a_out} with $\dot{m}_3=0$ and $\beta_{in\rightarrow3} = 0$, i.e. 
\begin{equation}
\dot{a}_{\rm out, wind} =
\dot{a}_{wind, no-acc}(\dot{m}_{\rm in}).
\label{eq:ce_a_out}
\end{equation}
There may be friction between the CE-matter and the outer star, if the CE-matter is primarily expelled in the orbital plane and the inner and outer orbital planes are parallel. Therefore we multiply Eq.\,\ref{eq:ce_a_out} with a factor 
\begin{equation}
f_{\rm fric} = \mathrm{min}\left(1,\dfrac{|\mathrm{sin}(i)|}{|\mathrm{sin}(i_{\rm crit})|}\right), 
\label{eq:f_fric}
\end{equation}
where $i_{\rm crit}$ is a minimum inclination necessary for the friction to take place.

\subsubsection{Stable mass transfer in a circular inner binary}
\label{sec:si_stable_mt}
We assume mass transfer in the inner binary takes place on either the thermal or nuclear timescale of the donor star. Mass transfer driven by angular momentum loss is currently not implemented in the code. The mass transfer rate is estimated by $\dot{m}_{\rm MT} =  m/\tau_{\rm MT}$, where $\tau_{\rm MT}$ is the timescale of mass transfer.
The thermal timescale of a star is given by Eq.\,\ref{eq:t_thermal}, where $R$ and $L$ are  given by the stellar evolution code used in \code, \texttt{SeBa}. 
 The nuclear timescale of a MS or helium-MS star is estimated by Eq.\,\ref{eq:t_nuclear}.  For other stars we take $\tau_{\rm nucl} = R/\dot{R}$, where $\dot{R}$ is the time derivative of the radius, calculated from the current and previous timestep. If the star is shrinking, which is possible for horizontal branch stars or evolved AGB stars, we estimate the nuclear timescale by 10\% of the stellar age. 
Rejuvenation of the accretor star, and the opposite process for the donor star are taken into account by \texttt{SeBa}. Their method is explained in Appendix A.2.1 of \citet{Too12}.

The orbital evolution of the inner binary is approximated with Eq.\,\ref{eq:a_mt} where $\beta$ and $\eta$ are taken to be constants. 
If the companion star in the inner binary fills its Roche lobe during the mass transfer phase in response to the accretion, a contact binary is formed. We allow the inner binary to go through a CE-phase as described in Sect.\,\ref{sec:si_ce}, which likely leads to a merger of the system.

The degree of conservativeness $\beta$ of the mass transfer is one of the major uncertainties in binary evolutionary calculations. 
The accretor star is expected to spin-up due to the accretion. 
Even if the companion only accretes a few percent of its own mass, the accretor is spin-up to critical rotation \citep{Pac81}. 
This has been invoked to limit the amount of accretion that can take place. 
However, as some binaries have managed to experience a phase of fairly conservative mass transfer \citep[e.g. $\phi$ Per][]{Pol07b}, additional mechanisms of angular momentum loss must play a role during mass transfer \citep{DeM07}. 

The lack of synchronisation can  affect the size of the Roche lobe significantly. For example, for a star that is rotating 100 times faster than synchronization, the Roche lobe is only 5-10\% of that of the classical Roche lobe \citep[based on][]{Sep07}. 
For simplicity, we make the common assumption that any circularized system entering RLOF, is and will remain synchronized during the mass transfer phase.

For stable, hierarchical triples systems, the matter lost by the inner binary is likely energetic enough to escape from the triple (Sect.\,\ref{sec:bg_ts_mt_inner_onto_outer}), and we model this as a fast wind i.e. according to Eq.\,\ref{eq:ce_a_out}. 
To incorporate the effect of friction between the matter and the outer star, we multiply Eq.\,\ref{eq:ce_a_out} with a factor $0<f_{\rm crit}<1$ (Eq.\,\ref{eq:f_fric}), similar to the case of a CE-phase in the inner binary (Sect.\,\ref{sec:si_ce}. 
During mass transfer, the effect of wind mass losses on the inner and outer orbit (Eqs.\,\ref{eq:wind_a_in}~and~\ref{eq:wind_a_out}) are taken into account simultaneously.

\subsubsection{Mass transfer initiated by the outer star}
\label{sec:si_outer_mt}
Mass transfer from an outer star onto a binary is an intriguing new evolutionary pathway opened up by stellar triples. Even though it is relatively common for triples (about 1\%), it is a complex process that has not been studied in much detail. The study of \citep{DeV14} focuses on two triples undergoing mass transfer initiated in the outer star, however, the study is limited in parameter space (as they study two triples), and in time (as the hydrodynamical simulations they perform are expensive). 
For this reason, we have not implemented the process of outer mass transfer, and currently the code is stopped when the outer star fills its Roche lobe.

\subsubsection{Supernova explosions in \code}
\label{sec:si_SN}

During a SN event, the star collapses on a dynamical timescale, for which the secular approach is not valid. The ODE solver routine is therefore disabled, and the orbital evolution due to the SN event is solved for in a separate function as detailed below.

The amount of mass-loss in the SN-event, and the type or remnant that is left behind are determined by the stellar evolution code \texttt{SeBa}. The effect of the SN ejecta on the companion stars (e.g. compositions and velocities) is usually small \citep[e.g.][]{Kal96,Hir14, Liu15b,Rim15}, unless the pre-supernova separation between the stars is smaller than a few solar radii. For this reason, we assume the dynamics of the companion stars are not affected by the expanding shell of material, and the companion stars neither accrete nor are stripped of mass.

We make the common assumption that the SN takes place instantaneously. 
As a result, the positions of the stars just before and after the SN are not changed. As \code\,is based on orbit-averaged techniques, we do not follow the position of the stars along the orbit as a function of time. In order to obtain the position at the moment of the SN, we randomly sample the mean anomaly from a uniform distribution. 
The natal kick is randomly drawn from either of three distributions (\citet{Pac90}, \citet{Han97} or \citet{Hob05}) in a random direction. 
Our method simply consists of two coordinate transformations (thus we do not use Eq.\,\ref{eq:a_sn_bin}, 
\ref{eq:a_sn_tri}, 
\ref{eq_app:a_in}, 
\ref{eq_app:e_in}, 
\ref{eq_app:a_out}, 
nor Eq.\ref{eq_app:e_out} directly). We convert from our standard orbital parameters of $i$ and $a,e,g,h$ for the inner and outer orbit to orbital vectors i.e. eccentricity $\hat{e}$ and angular momentum vector $\hat{J_{\rm b}}$ for both orbits. After the mass of the dying star is reduced and the natal kick is added to it, we convert back to the orbital elements. 
The reason for performing two coordinate transformation, to orbital vectors and back, is that the  orbital elements in the code are defined with respect to the `invariable’ plane, i.e. in a frame defined by the total angular momentum. In the case of a SN, however, the total orbital angular momentum vector is not generally conserved, which implies that the coordinate frame changes after the SN. In contrast, the orbital vectors are defined with respect to an arbitrary inertial frame that is not affected by the SN. The post-SN orbital vectors are transformed to the orbital elements in the new `invariable’ plane, i.e. defined with respect to the new total angular momentum vector.
An additional advantage of the double coordinate transformation is that the pre-supernova orbit can be circular as well as have an arbitrary eccentricity.

If the post-supernova eccentricity of an orbit is larger than one, the orbit is unbound. We distinguish four situations:
\begin{itemize}
\item Both the inner as the outer orbit remain bound, and the system remains a triple. The simulation of the evolution of the triple is continued.
\item When the inner orbit remains bound, and the outer orbit becomes unbound, the outer star and inner binary remain as separated systems. We assume the outer star does not dynamically affect the inner binary. 
With the default options in \code\,the simulation is stopped here unless the user specifies otherwise. 
\item When both the inner as the outer orbit become unbound, the stars evolve further as isolated stars. As in the previous scenario, by default the simulation is stopped unless the user specifies otherwise.

\item The inner orbit becomes unbound, but at the moment just after the SN the outer star remains bound the inner system. In this case, \code\,cannot simulate the evolution of this system further. The evolution of these systems should be followed up with an N-body code. 
\end{itemize}

\section{Examples}
\label{sec:ex}

In Sect.\,\ref{sec:bg_ds}~and~Sect.\,\ref{sec:bg_ts}, we discussed several physical processes and how they affect the long-term evolution of inner binaries and triple systems. 
Here we illustrate those processes by simulating the evolution of a few realistic triple star systems. For example, 
the evolution of Gliese 667 displays the Lidov-Kozai cycles, and the evolution of Eta Carinae illustrates the effect of precession and stellar winds.  The evolutionary pathways of the triple systems are simulated with the new triple code \code, such that the examples below also demonstrate the capabilities of \code.

\begin{table}
\caption{Initial conditions for the three triple systems discussed in Sect.\,\ref{sec:ex}.}
\begin{tabular}{lccc}
\hline \hline
Parameters & Gliese 667 & Eta Carinae & MIEK\\
\hline
$m_1$ (\Msolar) & 0.73& 110 & 7\\
$m_2$ (\Msolar) & 0.69& 30 & 6.5\\
$m_3$ (\Msolar) & 0.37& 30 & 6\\
$a_{\rm in}$ (AU) & 12.6& 1 & 10\\
$a_{\rm out}$ (AU) & 250& 25 & 250\\
$e_{\rm in}$  & 0.58& 0.1 & 0.1\\
$e_{\rm out}$ & 0.5& 0.2 & 0.7\\
$i_{\rm mutual}$ ($^{\circ}$) & 90& 90 & 60\\
$g_{\rm in}$ (rad) & 0.1&0.1  & 0\\
$g_{\rm out}$ (rad) & 0.5& 0.5 & $\pi$\\
\hline
\end{tabular}
\label{tbl:ex}
\end{table}

\subsection{Gliese 667}
\label{sec:ex_gliese}

Gliese 667 is a nearby triple system in the constellation of Scorpius. 
The orbital parameters of the system are described in Tbl.\,\ref{tbl:ex} based on \citep{Tok08}. The outer star is in an orbit of $a_{\rm out}>230$AU, but for simplicity, we will assume $a_{\rm out}=250$AU in the following. 
The outer star is also a planetary host-star; up to five planets have been claimed, of which two have been confirmed so far \citep{Fer14}. The orbit of the planet Gliese 667 Cb lies just within the habitable zone, which makes this planet a prime candidate in the search for liquid water and life on other planets \citep{Ang12}. In the following, we will neglect the dynamical effect of the presence of planets on the  evolution of the triple. 

\begin{figure}[h!]
\caption{\csentence{Inner eccentricity evolution} The evolution of the inner eccentricity $e_{\rm in}$ as a function of time for the first 3Myr of the evolution of Gliese 667. The figure shows that Gliese 667 is susceptible for Lidov-Kozai cycles. The initial conditions are given in Tbl.\,\ref{tbl:ex}. }
\centering
\includegraphics[width=\columnwidth]{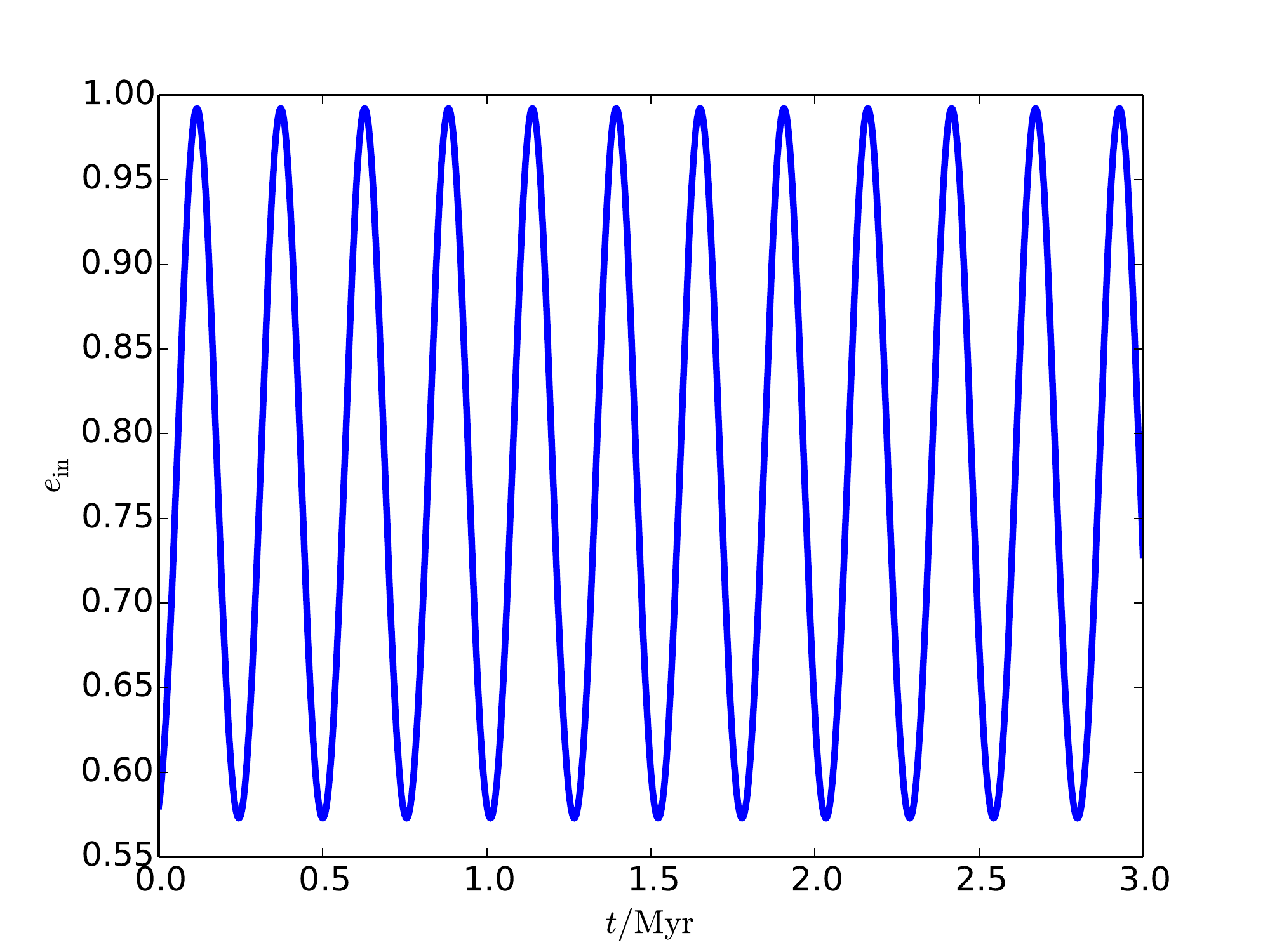} 
\label{fig:ex_gliese_ein_zoom}
\end{figure}

\begin{figure}[h!]
\caption{\csentence{Mutual inclination evolution}
The evolution of the mutual inclination $i$ as a function of time for the first 3Myr of the evolution of Gliese 667. This triple shows the cyclic behaviour in inclination and inner eccentricity (Fig.\,\ref{fig:ex_gliese_ein_zoom}) related to Lidov-Kozai cycles. 
}
\centering
\includegraphics[width=\columnwidth]{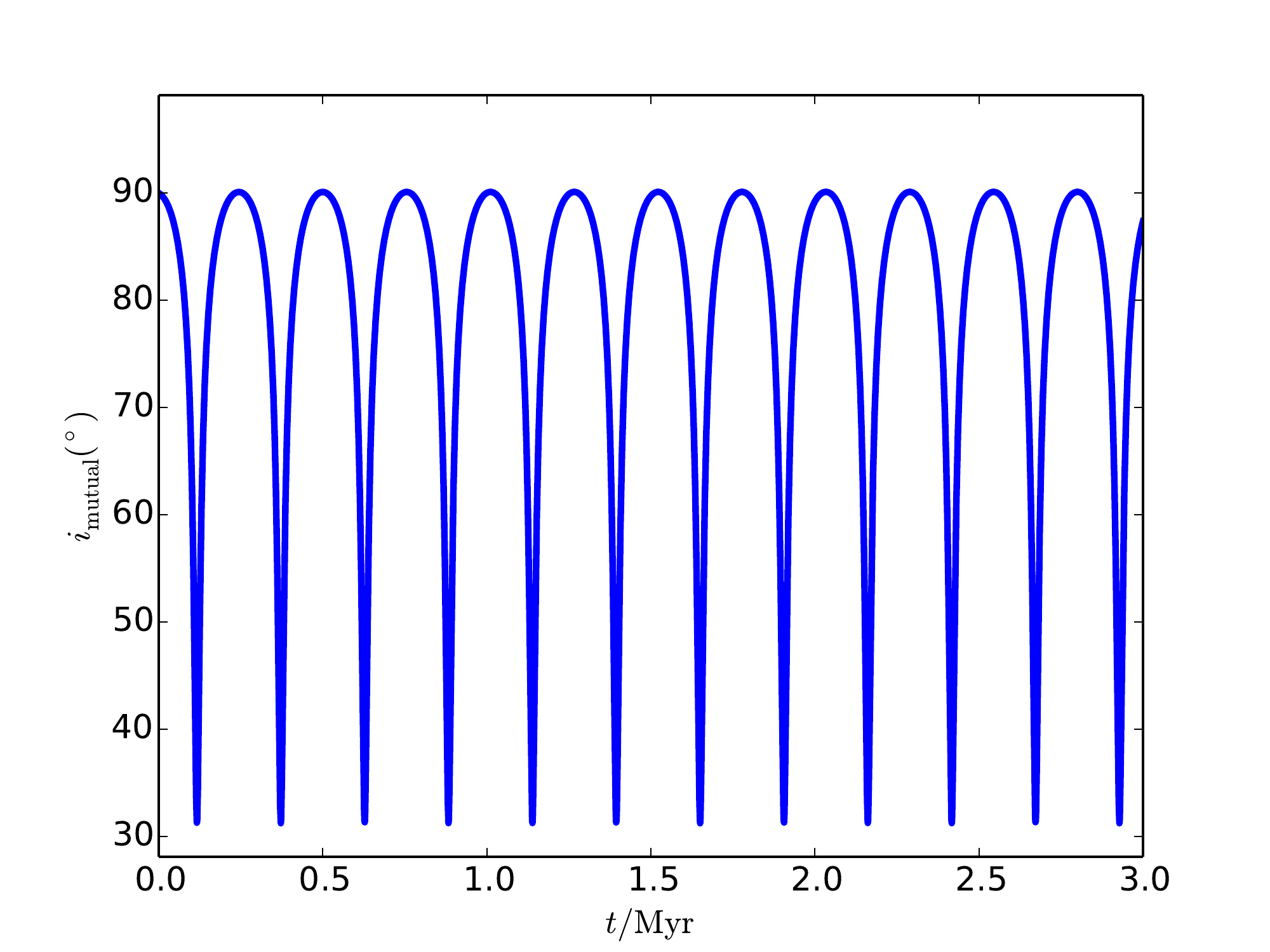} 
\label{fig:ex_gliese_i_zoom}
\end{figure}

\begin{figure}[h!]
\caption{\csentence{Inner semi-major axis evolution}
 The evolution of the inner semi-major axis as a function of time for the first 3Myr of the evolution of Gliese 667. The figure shows a decreasing semi-major axis due to the combination of Lidov-Kozai cycles with tidal friction, i.e. LKCTF. Note the small scale on the y-axis, where $a_0=12.599999999$. 
}
\centering
\includegraphics[width=\columnwidth]{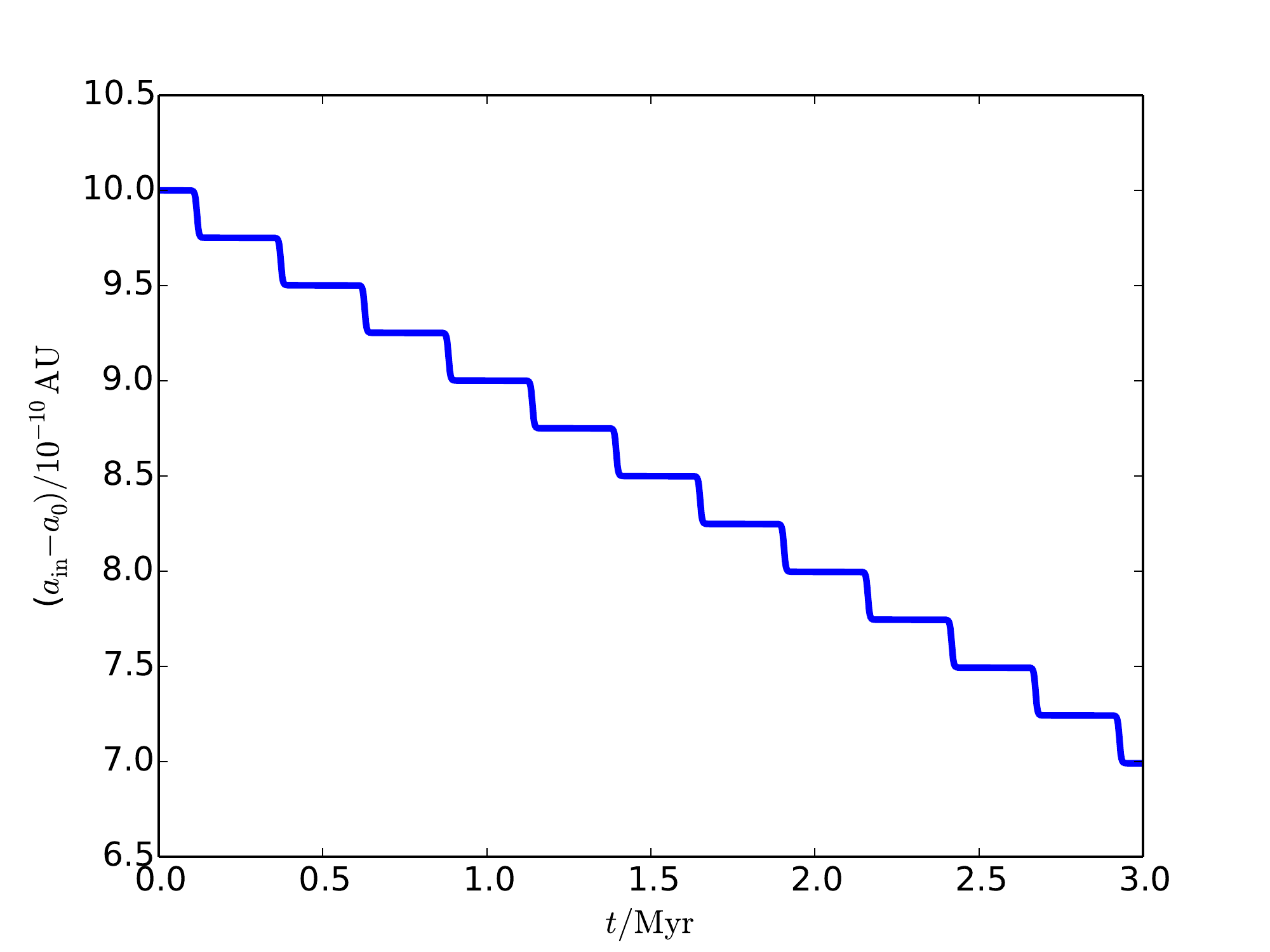} 
\label{fig:ex_gliese_ain_zoom}
\end{figure}

Gliese 667 is a prime example of a triple system undergoing Lidov-Kozai cycles.  Figs.\,\ref{fig:ex_gliese_ein_zoom}~and~\ref{fig:ex_gliese_i_zoom} show the evolution of the inner eccentricity and mutual inclination for the first 3Myr after the birth of the system. under the assumption of $e_{\rm out} = 0.5$, $i=90^{\circ}$, $g_{\rm in}=0.1$ and $g_{\rm out}=0.5$. 
For different values for the outer eccentricity, arguments of pericenters, and mutual inclination the general behaviour of Figs.\,\ref{fig:ex_gliese_ein_zoom}-\ref{fig:ex_gliese_ain_zoom} remains the same, but the timescale and amplitude of the Lidov-Kozai cycles varies (to the point where the cycles are not notable). 
Figs.\ref{fig:ex_gliese_ein_zoom}~and~\ref{fig:ex_gliese_i_zoom} show the cyclic behaviour of eccentricity and inclination in Gliese 667. When the eccentricity is at its maximum, the inclination between the orbits is minimal. 
The timescale of the oscillations is a few 0.1Myr, which is consistent with the order of magnitude approximation of 0.4Myr of Eq.\,\ref{eq:t_kozai}. 
The octupole parameter $\epsilon_{\rm oct} <0.001 $, which indicates that the eccentric Lidov-Kozai mechanism is not of much importance here.

For the same timescale as Fig.\,\ref{fig:ex_gliese_ein_zoom}~and~\ref{fig:ex_gliese_i_zoom}, 
Fig.\,\ref{fig:ex_gliese_ain_zoom} shows the evolution of the inner-orbital semi-major axis $a_{\rm in}$. The change in the inner semi-major axis of Gliese 667 is negligibly small, however, the figure illustrates the effect of Lidov-Kozai cycles with tidal friction or LKCTF. When the inner eccentricity is at its maximum, and the inner stars are at their closest approach during pericenter passage, the inner semi-major axis decreases due to tidal forces. In this way, LKCTF could lead to RLOF in or a merger of the inner system.

Regarding the long-term evolution of this system,
it is analogous to that discussed for the first 3 Gyr, 
 After 10 Gyr \citep[approximately the age of the Galactic thin disk, e.g.][]{Osw96, Del05, Sal09}, 
 the system is still detached. The orbital separation has decreased by only $\sim$7km. The stellar masses are sufficiently low, that the stars do not evolve off the MS within 10 Gyr, and as such do not experience a significant growth in radius that could lead to RLOF. 
Even taking into account the low metallicity of Gliese A \citep{Cay01}, and the corresponding speed-up of the evolutionary timescales, a 0.73\Msolar\,star is not massive enough to evolve of the MS within 10 Gyr. Furthermore, the stars are not massive enough to lose a considerable amount of matter in stellar winds, such that the triple is not affected dynamically by wind mass losses.

\subsection{Eta Carinae}
\label{sec:ex_eta_carinae}

Eta Carinae is a binary system with two massive stars ($m_1\sim90\Mo$ and $m_2\sim30\Mo$) in a highly eccentric orbit ($e=0.9$) with a period of 5.5yr \citep{Dam97}. Both stars are expected to explode as supernovae at the end of their stellar lives. 
Eta Carinae is infamous for its 'Great Eruption'. From 1837 to 1857, it brightened considerably, and in 1843 it even became the second brightest star in the sky \citep{deV52}. 
The system is surrounded by the Homunculus Nebulae, that was formed during the Great Eruption, and heavily obscures the binary stars \citep{Hum99}. The kinetic energy of the Homunculus Nebulae is large i.e. $10^{49.7}$erg \citep{Smi03} and comes close to that of normal supernovae. However, as both stars have survived the Great Eruption, Eta Carinae is often referred to as a 'supernova imposter'. 

The cause of the Great Eruption remains unexplained. A massive outflow, as during the Great Eruption, can be driven by a strong interaction between two stars \citep[e.g.][]{Har09, Smi11} or a merger of two stars \citep[e.g.][]{Soker03}. Recently, \citet{Por16} tested the hypothesis that the Eta Carinae system is formed from the merger of a massive inner binary of a triple system.
 According to their model, the merger was triggered by the gravitational interaction with a massive third companion star, which is the current $\sim$30\Msolar\,companion star in Eta Carinae. Here, we simulate the evolution of their favourite model with the initial conditions as given by Tbl.\,\ref{tbl:ex}. 
Furthermore, \citet{Por16} assume that the argument of periastron does not affect the tidal evolution, and therefore we arbitrarily set $g_{\rm in} = 0.1 $ and $g_{\rm out} = 0.5$.

\begin{figure}[h!]
\caption{\csentence{Inner eccentricity evolution} The evolution of the inner eccentricity $e_{\rm in}$ as a function of time for the first 20000yr of the evolution of triple progenitor of Eta Carinae. The figure shows that the system undergoes Lidov-Kozai cycles. 
 The initial conditions of the triple progenitor are given in Tbl.\,\ref{tbl:ex}.  }
\centering
\includegraphics[width=\columnwidth]{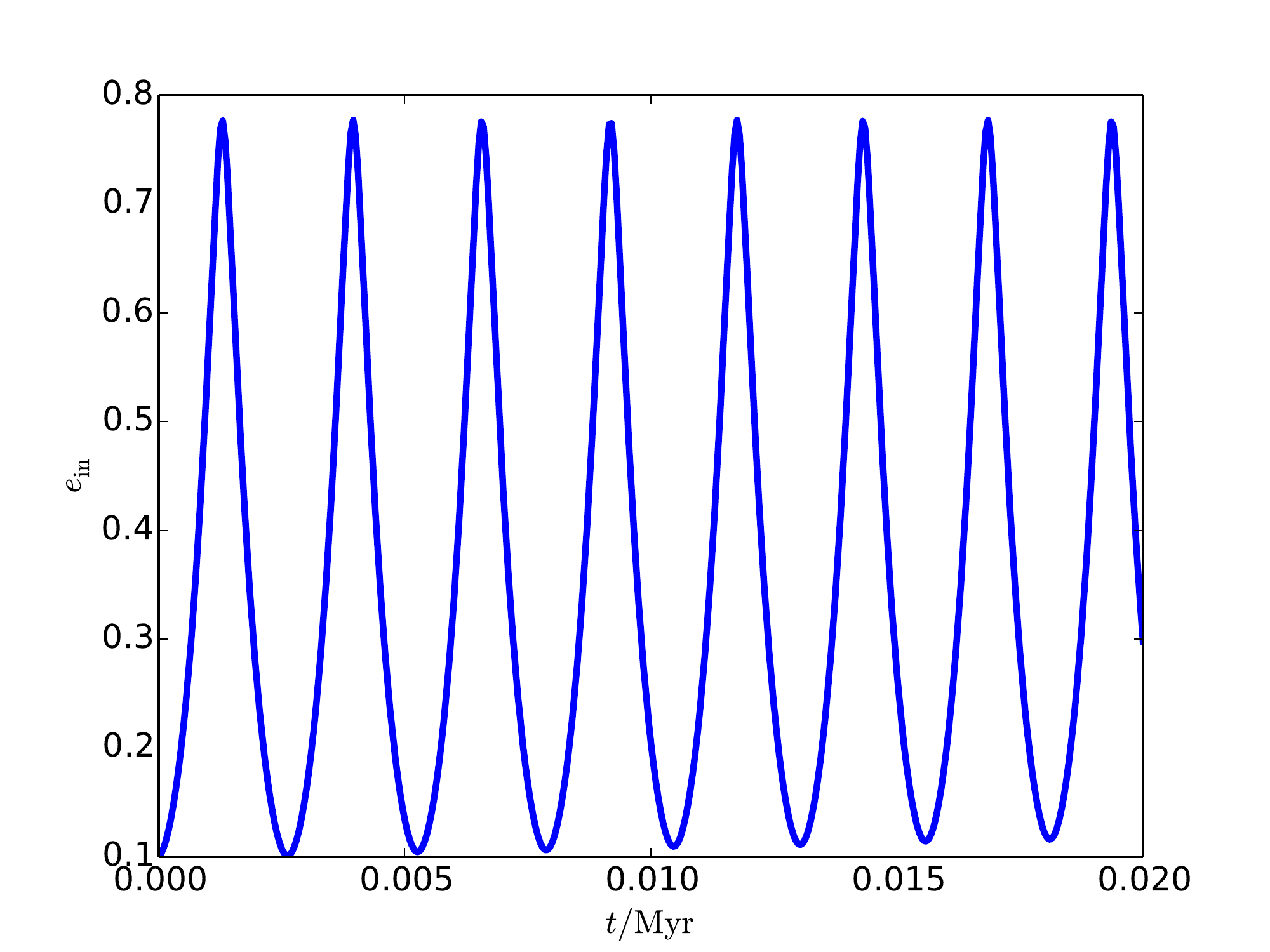}
\label{fig:ex_eta_carinae_ein_zoom}
\end{figure}

During the early evolution of the triple, the system  experiences Lidov-Kozai cycles with a timescale of a few kyr, see Fig.\,\ref{fig:ex_eta_carinae_ein_zoom}. 
The octupole parameter $\epsilon_{\rm oct} = 0.068 >0.01$, which indicates that the system is in the eccentric Lidov-Kozai regime with a timescale of order $t_{\rm oct}\sim t_{\rm kozai}/\epsilon_{\rm oct} \sim $few tens of kyr.

\begin{figure}[h!]
\caption{\csentence{Semi-major axis evolution} The evolution of the inner eccentricity $e_{\rm in}$ as a function of time for the first 20000yr of the evolution of triple progenitor of Eta Carinae. The figure shows the characteristic increase in semimajor-axes due to stellar wind mass loss, and the periodic decrease in inner semimajor-axis  when the eccentricity is high due to LKCTF. }
\centering
\includegraphics[width=\columnwidth]{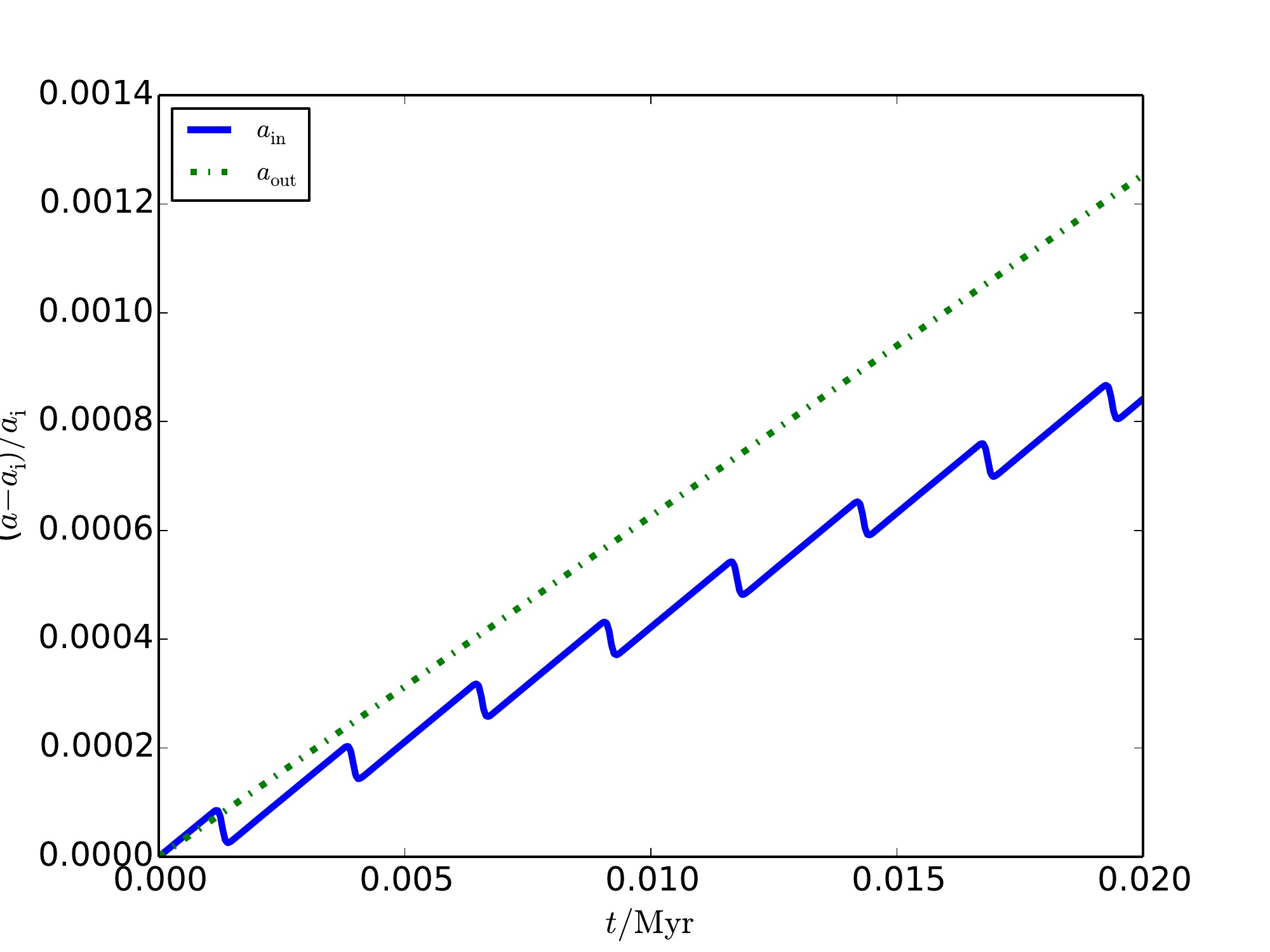} 
\label{fig:ex_eta_carinae_ain_zoom}
\end{figure}

The evolution of the semi-major axis shows two characteristics in Fig.\,\ref{fig:ex_eta_carinae_ain_zoom}. Firstly, as for Gliese 667, the system is affected by LKCTF i.e. the semi-major axis shrinks periodically, due to strong tides at pericenter when the inner eccentricity is at its maximum. Secondly, as the primary star is very massive, strong winds remove large amounts of mass while the star is still on the MS (Sect.\,\ref{sec:bg_ss_wind}). The dynamical effect of such a fast wind is that the inner and outer orbits expand (Sect.\,\ref{sec:bg_ds_wind}). Initially the inner orbit expands faster than the outer orbit, as expected for a strong wind from the inner orbit (Sect.\,\ref{sec:bg_ts_stability}). However, due to the combination of stellar winds with LKCTF for the Eta Carinae progenitor, its outer orbit expands faster, and the triple becomes more dynamically stable.

\begin{figure}[h!]
\caption{\csentence{Radius and mass evolution} The evolution of the radii (dashed line) and mass (dash-dotted line) as a function of time for the stars in the triple progenitor of Eta Carinae. The primary star of initial mass 110\Msolar\, is shown in blue. The secondary and tertiary star, both of initial mass 30\Msolar, are shown in red. The Roche lobe of the primary and secondary are overplotted (blue solid line and red solid line respectively). The Roche lobe of the tertiary is about 2000-3000\Rsolar. After about 3Myr, the primary star fills its Roche lobe. 
}
\centering
\includegraphics[width=0.95\columnwidth]{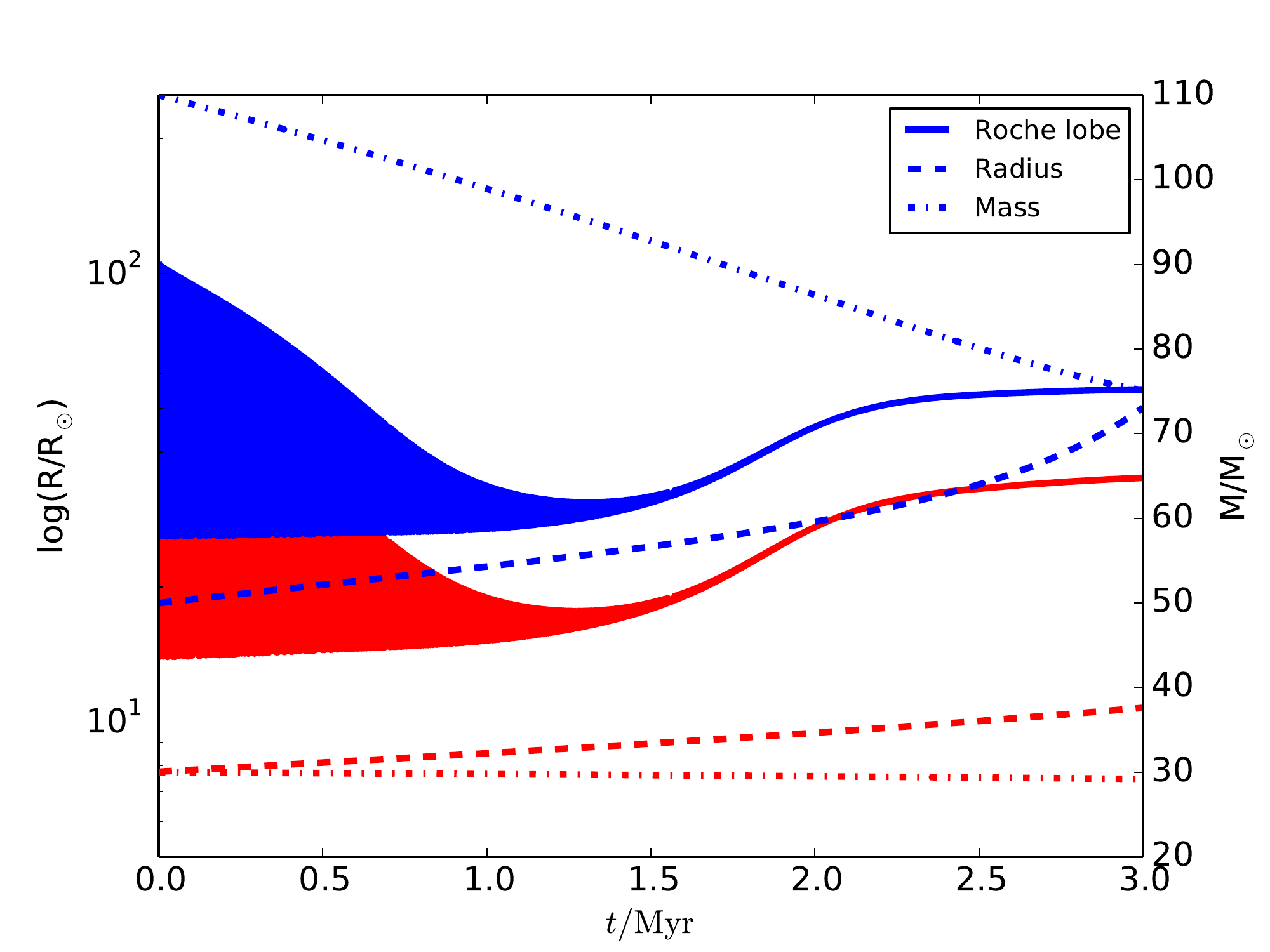}
\label{fig:ex_eta_carinae_mr}
\end{figure}

On a longer timescale, the triple moves from the eccentric Lidov-Kozai regime to the regular regime ($|\epsilon_{\rm oct}| \lesssim 0.01$) as the inner binary loses matter and angular momentum in the stellar winds (Fig.\,\ref{fig:ex_eta_carinae_mr}). After 3Myr, the octupole parameter has decreased from $\epsilon_{\rm oct} = 0.068$ initially, to $\epsilon_{\rm oct} = 0.001$.

\begin{figure}[h!]
\caption{\csentence{Inner eccentricity evolution} The evolution of the inner eccentricity on a timescale of 3Myr. The timescale of the Lidov-Kozai cycles is a few kyr, such that the lines overlap in Fig.\,\ref{fig:ex_eta_carinae_ein}. As the system evolves, the amplitude of the cycles reduces, until the system circularizes.  }
\centering
\includegraphics[width=\columnwidth]{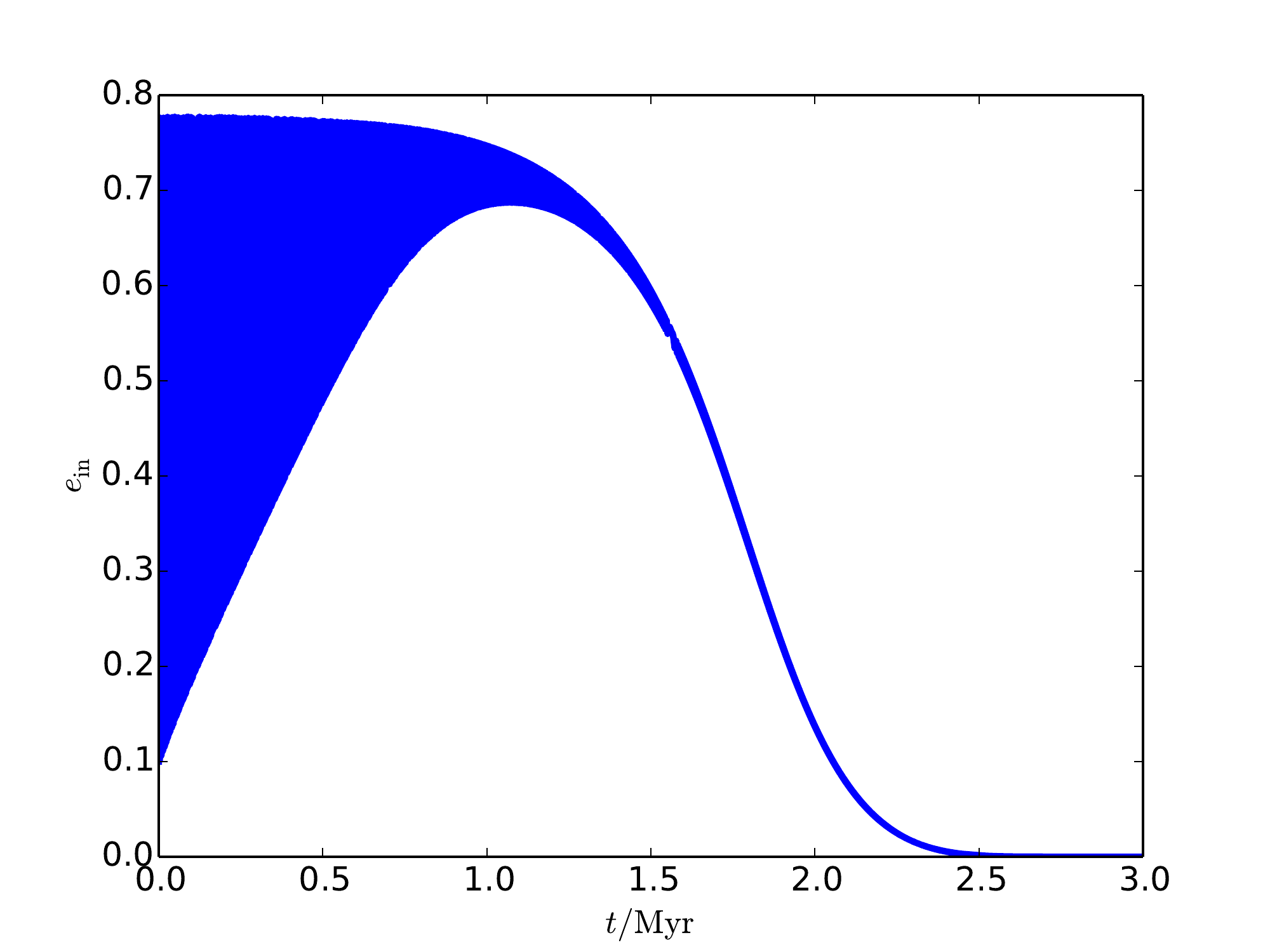} 
\label{fig:ex_eta_carinae_ein}
\end{figure}

\begin{figure}[h!]
\caption{\csentence{Mutual inclination evolution} 
The evolution of the mutual inclination on the same timescale as Fig.\,\ref{fig:ex_eta_carinae_ein}. The variation in the inclination decreases with time; as the stars evolve, their radii increase, and tidal effects become stronger (Eq.\,\ref{eq:a_dot_TF}-\ref{eq:omega_dot_TF}). 
 }
\centering
\includegraphics[width=\columnwidth]{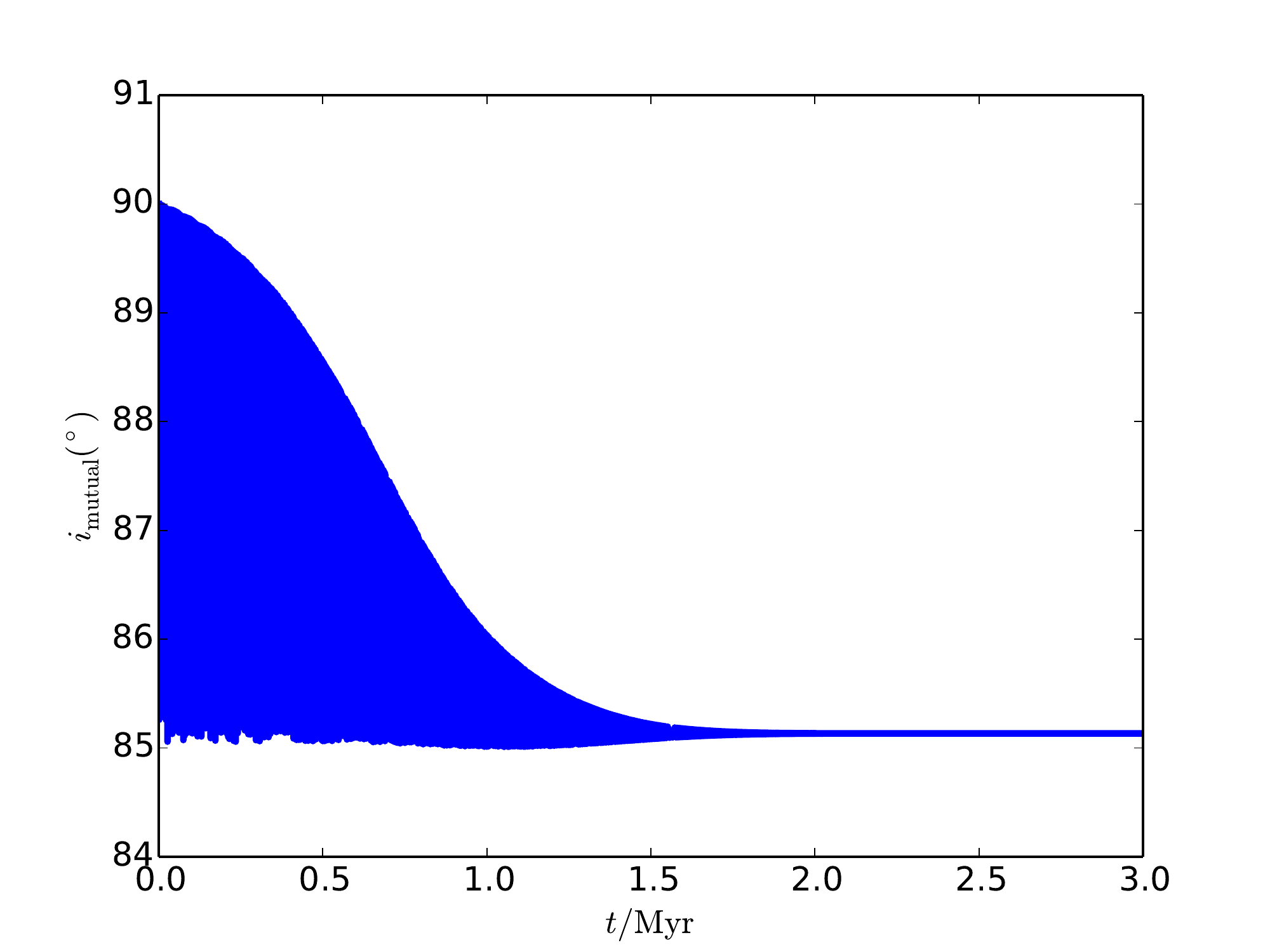} 
\label{fig:ex_eta_carinae_i}
\end{figure}

\begin{figure}[h!]
\caption{\csentence{Inner semi-major axis evolution} The evolution of the inner semimajor-axis on the same timescale as Fig.\,\ref{fig:ex_eta_carinae_ein}~and~\ref{fig:ex_eta_carinae_i}. In the first Myr, the evolution of the system is dominated by the Lidov-Kozai mechanism and the inner semimajor-axis remains more or less constant. In the following Myr, the system circularizes and the semimajor-axis decreases by a factor 2. After synchronisation and circularisation has been reached, the inner semimajor-axis increases due to the ejection of stellar winds. }
\centering
\includegraphics[width=\columnwidth]{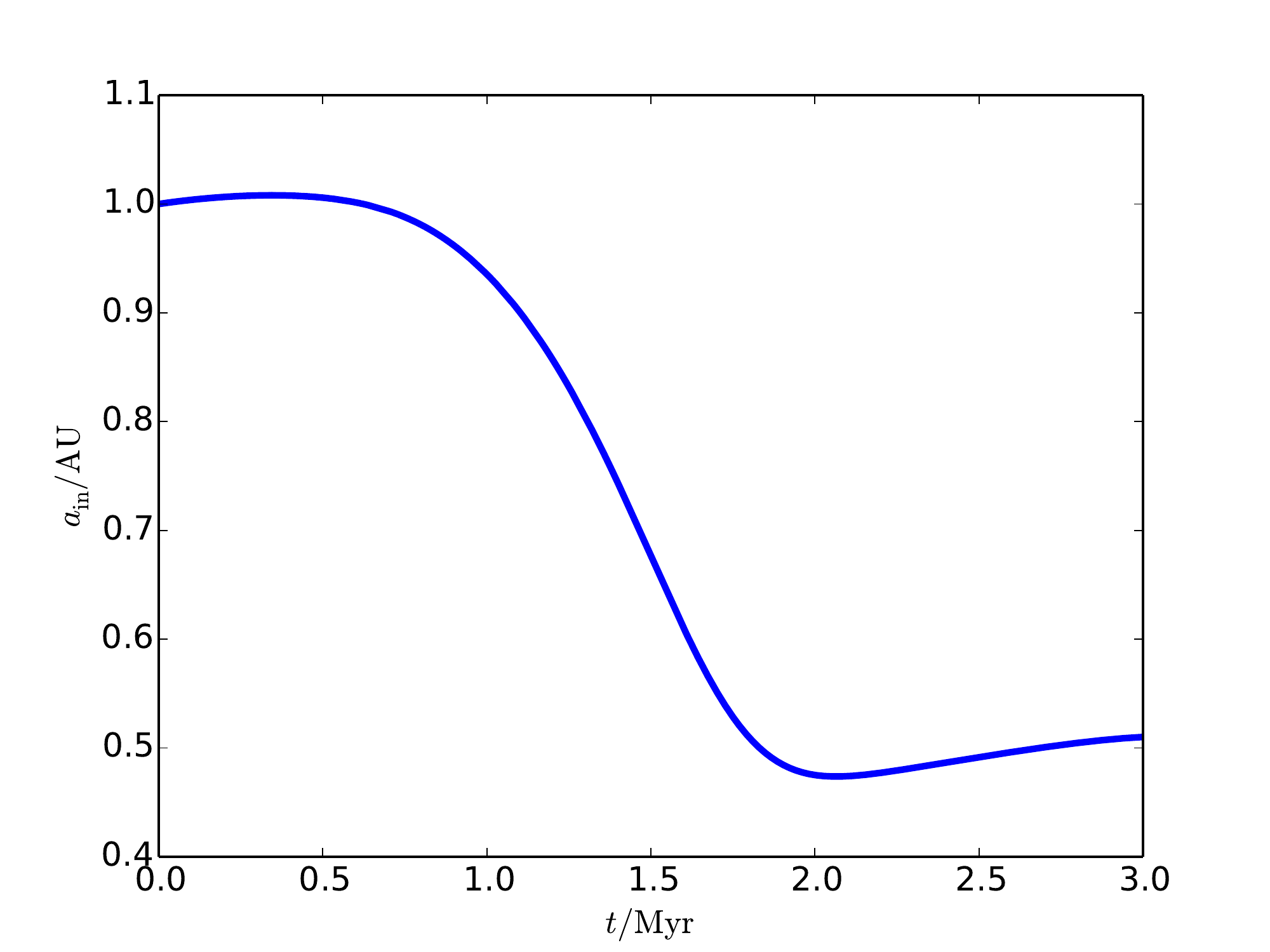} 
\label{fig:ex_eta_carinae_ain}
\end{figure}

The long-term evolution of the progenitor candidate of Eta Carinae shows another interesting feature in Fig.\,\ref{fig:ex_eta_carinae_ein}~and~\ref{fig:ex_eta_carinae_i}, i.e. the Lidov-Kozai cycles are quenched. Here the precession due to the distortion and rotation of the stars dominates over the precession caused by the Lidov-Kozai mechanism. As a result, the amplitudes of the cycles in  inner eccentricity and mutual inclination are reduced. After approximately 1.5Myr, the evolution of the system is completely dominated by tides, i.e. the system circularizes and the inner semi-major decreases accordingly (Fig.\,\ref{fig:ex_eta_carinae_ain}). After circularization of the inner binary has been achieved ($\gtrsim 2$Gyr), the inner semi-major axis increases again due to the stellar winds from the inner binary. The evolution of the Eta Carinae progenitor (Fig.\,\ref{fig:ex_eta_carinae_ein_zoom}-\ref{fig:ex_eta_carinae_ain}) illustrates that both three-body dynamics and stellar evolution matter, and neither can be neglected.

After about 3Myr, the primary star fills its Roche lobe and initiates a mass transfer phase (Fig.\,\ref{fig:ex_eta_carinae_mr}). The inner semi-major axis is about 0.5AU (109\Rsolar), and the massive primary has increased in size to 51\Rsolar.
Even though, the inner semi-major axis (and Roche lobe) of the primary are $\sim10$\% smaller around 2Gyr, there is no RLOF yet as the radius of the primary star is $\sim$50\% smaller. 
At RLOF, the masses of the inner stars have reduced from the initial 110\Msolar\,and 30\Msolar\,to 75\Msolar\,and\,29\Msolar, and both stars are still on the MS. The mass transfer phase proceeds in an unstable manner (Sect.\,\ref{sec:si_stability}), and a common-envelope develops that leads to a merger of the inner stars (Sect.\,\ref{sec:si_ce}). A new star is formed, that is still on the MS, with a mass of 104\Msolar. We assume this merger proceeds conservatively, and therefore the outer orbit is not affected, such that $a_{\rm out} \approx 32$AU, $e_{\rm out} \approx 0.2$ and $P\approx 15.7$yr. Prior to the merger, the outer semi-major axis has increased from the initial value of 25 AU to 32AU due to the stellar winds.

The resulting binary is similar to the current Eta Carinae system in mass and orbital period.  It is not an exact match, as the evolution of this specific triple is shown for illustrative purposes, and has not been fitted to match the currently observed system.  A progenitor study of Eta Carinae to improve the match is beyond the scope of this paper. 

We note that the current eccentricity of our remaining binary is low i.e. $e \approx 0.2$ compared to the observed $e=0.9$. 
In our simulations, the post-merger eccentricity is equal to the pre-merger outer eccentricity. During the evolution of the Eta Carinae progenitor, the outer eccentricity has remained roughly equal to its initial value of $e_{\rm out} = 0.2$. The outer eccentricity is not affected strongly by stellar evolution or Lidov-Kozai cycles. 
If we study the evolution of an alternative progenitor similar to the favourite model of \citet{Por16}, but with $e_{\rm out} = 0.9$, the system is dynamically unstable at birth. 
In order for the triple to be dynamically stable $e_{\rm out} \lesssim 0.81$ for the standard $i=60^{\circ}$, or up to $e_{\rm out} \lesssim 0.84$ for $i=90^{\circ}$. 
The evolutions of the dynamically stable systems with $e_{\rm out} \lesssim 0.7$ show similar behaviour as our initial system (Tbl.\,\ref{tbl:ex}), and the merger leads to a similar binary as in the case of our initial system. For dynamically stable systems with higher outer eccentricities, the merger time decreases strongly, and the inner system does not reach circularization before the merger takes place. The merger product is more massive, as less mass is lost in stellar winds. 
In the simulation of \citep{Por16}, the outer orbit has an eccentricity $e_{\rm out}=0.2$ initially, but becomes highly eccentric in the merger phase due to asymmetric mass loss.


\subsection{MIEK-mechanism}
\label{sec:ex_miek}
In this section, we illustrate the dynamical effect of mass loss on a triple system from the point of view of transitions in dynamical regimes, e.g. the regime without Lidov-Kozai cycles,  with regular or with eccentric Lidov-Kozai behaviour. 
Here, we focus on the transition from the regular Lidov-Kozai regime to the eccentric regime, i.e. where the octupole term is significant. This transition has been labelled 'mass-loss induced eccentric Kozai' or MIEK (Sect.\,\ref{sec:bg_ts_ml}). 
The canonical example of MIEK-evolution is a triple with the initial conditions as given by Tbl.\,\ref{tbl:ex} \citep{Sha13, Mic14}. 
To reproduce the experiment of \citet{Sha13}, we simulate the evolution of this triple with \code~including three-body dynamics and wind mass losses, however, without stellar evolution in radius, luminosity, or stellar core mass etc. 

\begin{figure}[h!]
\caption{\csentence{Radius and mass evolution} The evolution of the radii (dashed line) and mass (dash-dotted line) as a function of time 
for the stars in a triple that transitions from a region with regular to eccentric Lidov-Kozai behaviour i.e. MIEK. The initial conditions of the triple are given in Tbl.\,\ref{tbl:ex}, based on \citet{Sha13} and \citet{Mic14}.
The primary star is shown in blue, the secondary in green and the tertiary in red.  The Roche lobe of the primary and secondary are overplotted (blue solid line and green solid line respectively). 
The Roche lobe of the tertiary is 6500-8000\Rsolar. 
After about 55.5Myr, the primary star fills its Roche lobe. 
}
\centering
\includegraphics[width=0.95\columnwidth]{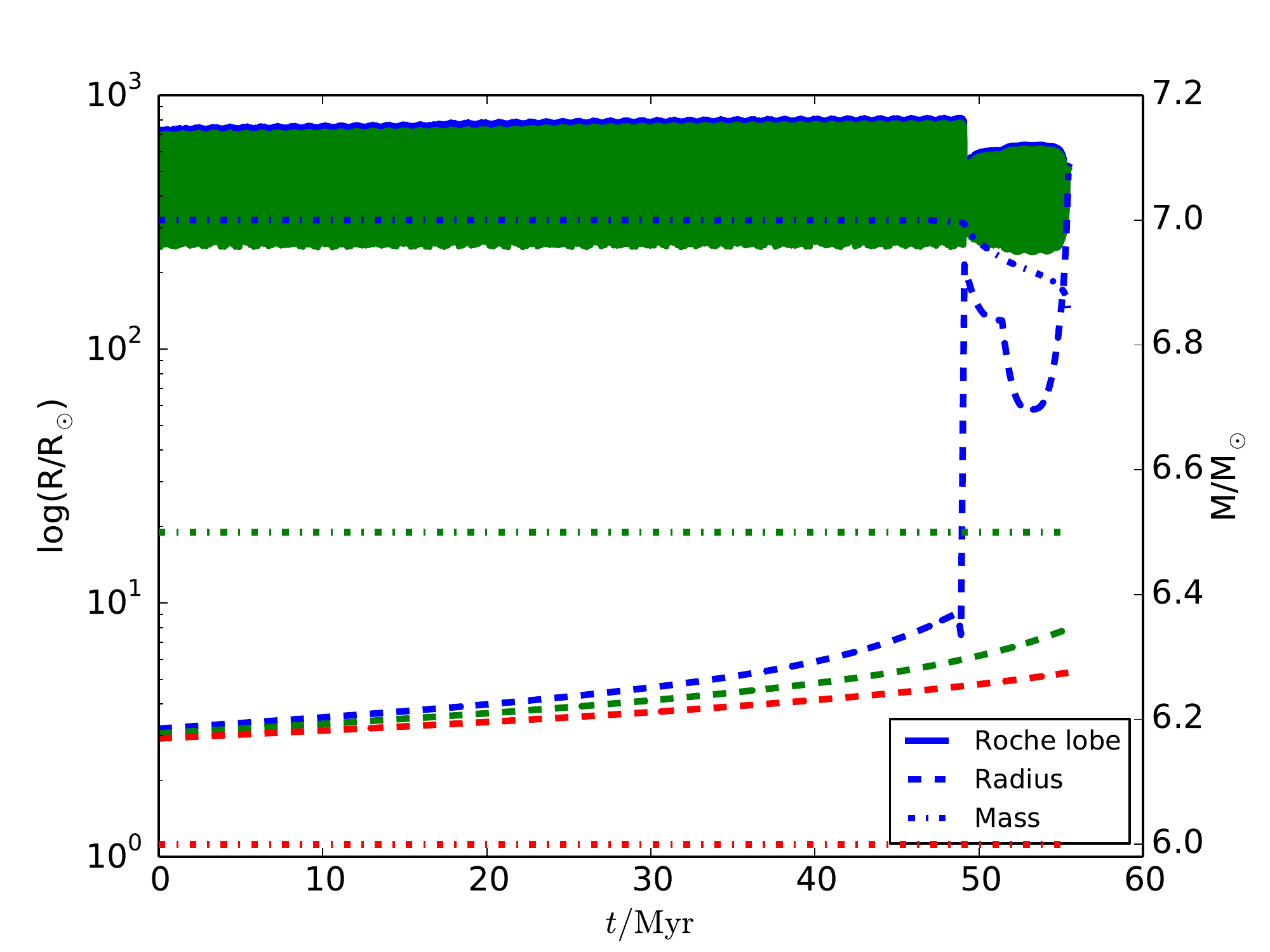}
\label{fig:ex_miek_mr}
\end{figure}

\begin{figure}[h!]
\caption{\csentence{Inner eccentricity evolution} 
The evolution of the inner eccentricity $e_{\rm in}$ as a function of time for a triple that transitions from a region with regular to eccentric Lidov-Kozai behaviour i.e. MIEK.  The initial conditions of the triple are given in Tbl.\,\ref{tbl:ex}, based on \citet{Sha13} and \citet{Mic14}.
For this figure, the stars are not allowed to evolve in \code, except for wind mass losses. If stellar evolution is taken into account fully, RLOF initiates at 55.5Myr, before the transition to MIEK can develop. }
\centering
\includegraphics[width=\columnwidth]{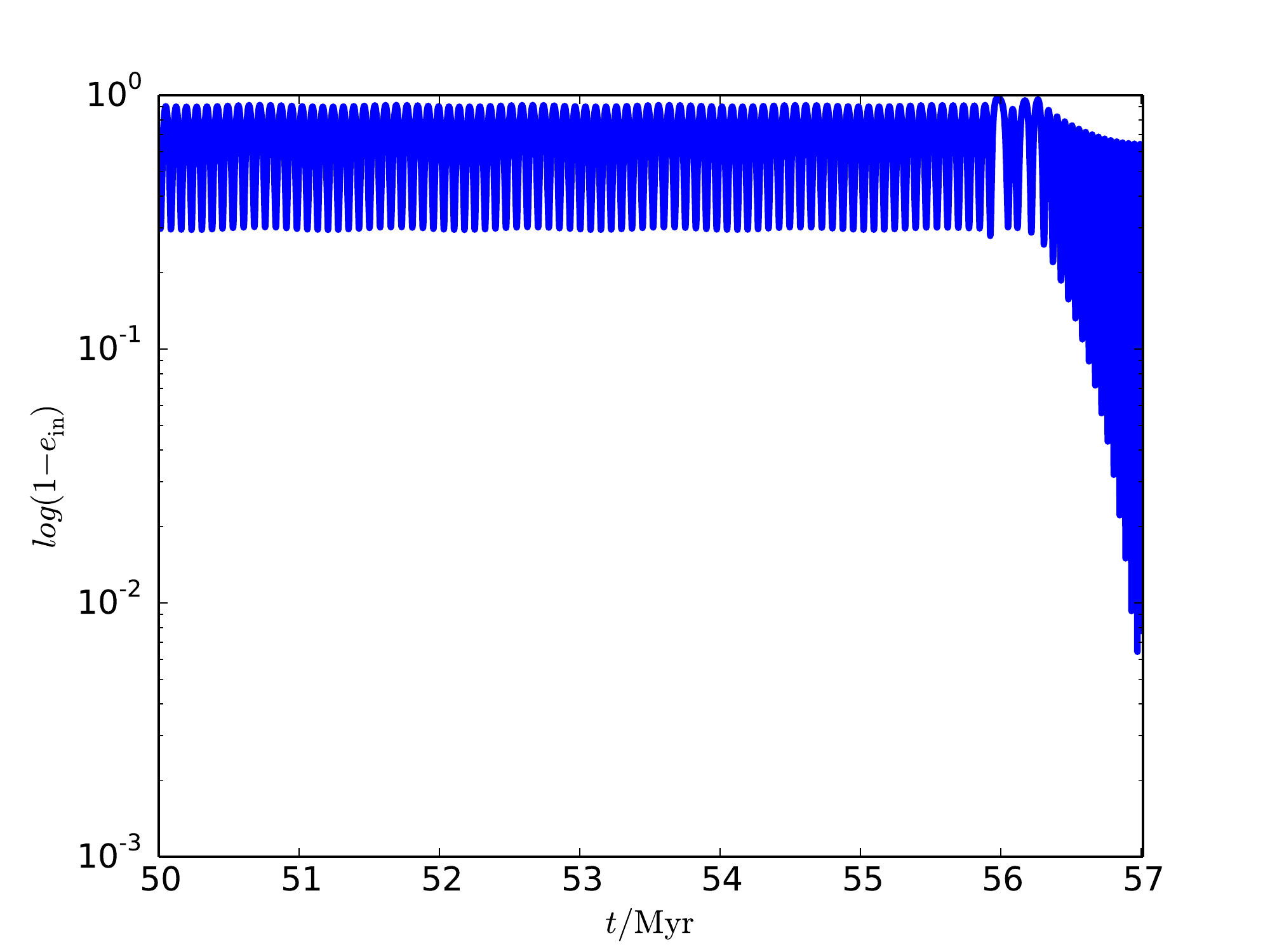}
\label{fig:ex_miek_ein}
\end{figure}

\begin{figure}[h!]
\caption{\csentence{Mutual inclination evolution} The evolution of the mutual inclination $i$ as a function of time for the same triple as in Fig.\,\ref{fig:ex_miek_ein}. The triple transitions from a region with regular to eccentric Lidov-Kozai behaviour at 56Myr.}
\centering
\includegraphics[width=\columnwidth]{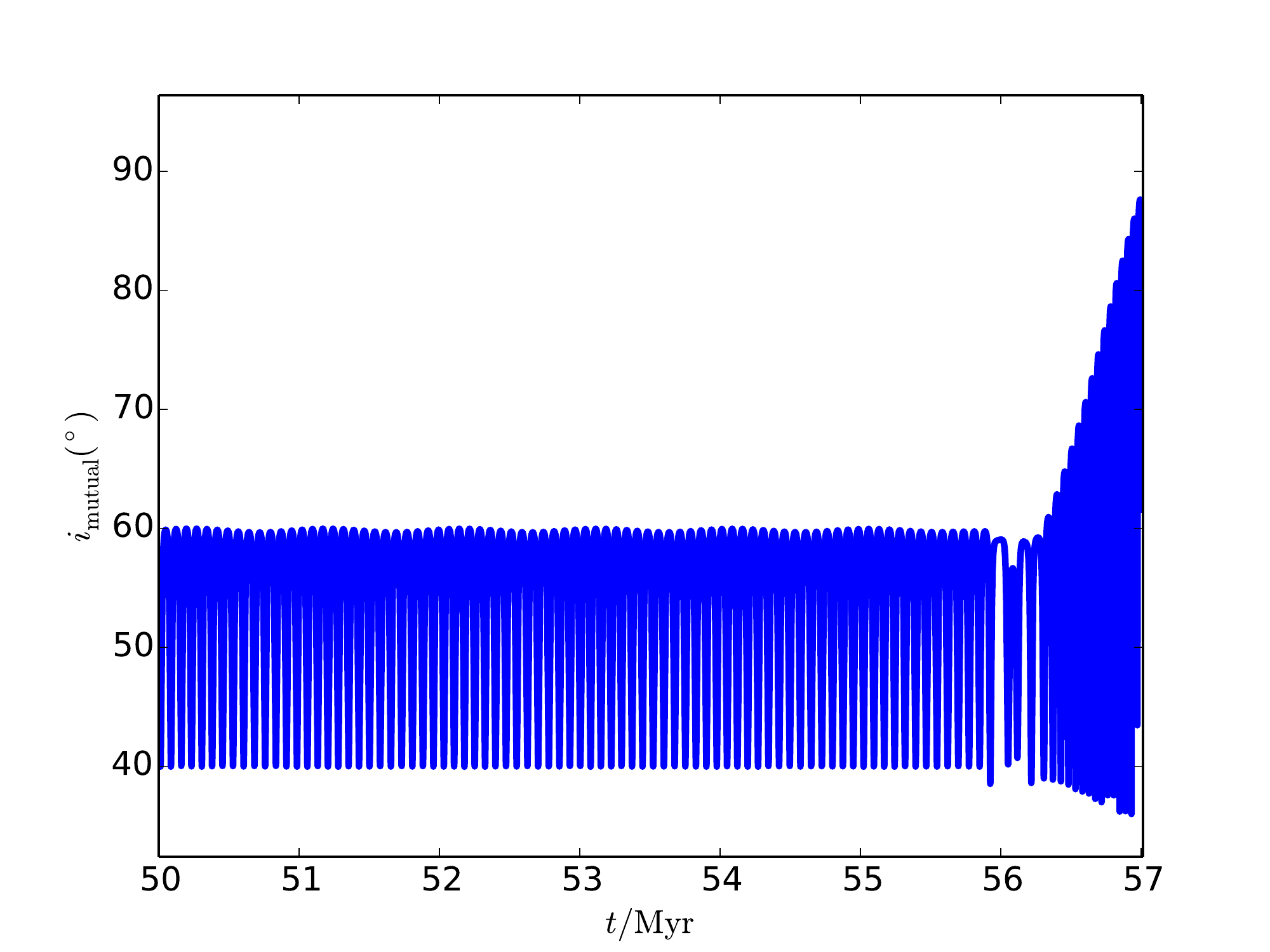}
\label{fig:ex_miek_i}
\end{figure}

Starting from the birth of the triple system, its orbit is susceptible to Lidov-Kozai cycles (Fig.\,\ref{fig:ex_miek_ein}~and~\ref{fig:ex_miek_i}). The timescale of the cycles is approximately 0.1Myr. The cycles are in the regular regime, i.e. $\epsilon_{\rm oct} = 0.002$. As time passes, the stars evolve. The primary star evolves off the MS at 49Myr, and after 55Myr it reaches the AGB with a mass of 6.9\Msolar\,(Fig.\,\ref{fig:ex_miek_mr}). Subsequently, it quickly loses a few solar masses in stellar winds, before it becomes an oxygen-neon white dwarf of 1.3\Msolar\,at 56Myr. The outer orbit widens to about $a_{\rm out}\sim350$AU due to the wind mass losses.

The wind mass loss allows the triple to transition to the eccentric Lidov-Kozai regime at about 56Myr, i.e. $\epsilon_{\rm oct} = 0.045$ at this time.  
The system is driven into extremely high eccentricities, and also the amplitude of the Lidov-Kozai cycle in inclination increases. 
The evolution of the system as shown in Figs.\,\ref{fig:ex_miek_ein}~and~\ref{fig:ex_miek_i} is qualitatively similar to that found by \citet{Sha13} based on N-body calculations
and \citet{Mic14} based on the secular approach.  
In these studies, stellar winds are implemented ad-hoc with a constant mass loss rate for a fixed time interval starting at a fixed time.
Moreover, the system is followed for multiple Myrs after the mass loss event in both papers, such that the inclination rises above 90$^{\circ}$, and the inner and outer orbit become retrograde to each other. 
In our case the simulation is stopped before such a flip in inclination develops, as RLOF is initiated in the inner binary 
 when the inner eccentricity is high.


However, if we fully include stellar evolution, as in the standard version of \code, the triple is not driven into the octupole regime. On the AGB, the radius of a 7\Msolar-star can reach values as large as $\sim$1000\Rsolar\,(Fig.\,\ref{fig:ex_miek_mr}), and therefore RLOF initiates before the MIEK-mechanism takes place. 
Even if the inner binary would be an isolated binary, RLOF would occur for initial separations of $a<15$AU. 
For triples, RLOF can occur for larger initial (inner) separations, as the Lidov-Kozai cycles can drive the inner eccentricity to higher values.
For wider inner binaries i.e. $a_{\rm in} > 16$AU, the MIEK-mechanism does not occur either, as the triple is dynamically unstable. 
This example indicates that the parameter space for the MIEK-mechanism to occur is smaller than previously thought, and so it may occur less frequently. 
Moreover, this example demonstrates the importance of taking into account stellar evolution when studying the evolution of triples.

For the canonical triple with $a_{\rm in} =10$AU and $a_{\rm out}=250$AU, RLOF occurs at 
55.5Myr, just a few 0.1Myr after the primary star arrives on the AGB. In that time, the radius of the primary  increased by a factor $\sim$3, and tides can no longer be neglected. 
The tidal forces act to circularize and synchronize the inner system, such that $e_{\rm in}=0$ at RLOF. 
The eccentric Kozai-mechanism does not play a role at this point, i.e. $\epsilon_{\rm oct} = 0.0008$. 
The mass of the core has not had enough time to grow to the same size as in the example without RLOF, i.e. the core mass is 1.25\Msolar\,instead of 1.3\Msolar.  
Stellar winds have reduced the mass of the primary star to 6.8\Msolar. As the primary has a convective envelope and is more massive than the secondary, a CE-phase develops. 
We envision three scenarios based on the different models for CE-evolution (Sect.\,\ref{sec:ce}~and~\ref{sec:si_ce}). First, the CE-phase leads to a merger of the inner binary, when the inner orbit shrinks strongly, as for the $\alpha$-model of CE-evolution with $\alpha\lambda_{\rm ce} = 0.25 $ (Sect.\,\ref{sec:ce}). Second, the CE-phase leads to strong shrinkage of the orbit, but not enough for the inner stars to merge. In this scenario, the envelope of the donor star is completely removed from the system, and the outer orbit widens to about 350AU, under the assumption that the mass removal affects the outer orbit as a fast wind. Assuming $\alpha\lambda_{\rm ce} = 2 $ (Eq.\,\ref{eq:alpha-ce}, Sect.\,\ref{sec:ce}), $a_{\rm in}\sim0.33$AU and $\epsilon_{\rm oct} = 0.0009$ after the CE-phase. In this scenario, the triples does not enter the octupole regime, and the MIEK-mechanism does not manifest. Lastly, the CE-phase does not lead to a strong shrinkage of the inner orbit, as for the $\gamma$-model of CE-evolution with $\gamma=1.75 $ (Eq.\,\ref{eq:gamma-ce}, Sect.\,\ref{sec:ce}). The inner semimajor-axis even increases from 6.0 to 7.3AU. In this scenario, $\epsilon_{\rm oct} = 0.02$, such that the perturbations from the octupole level become significant. In this last scenario, the triple undergoes the MIEK mechanism, despite \textit{and} because of the mass transfer phase.

\section{Discussion and conclusion}
\label{sec:disc}

In this paper, we discuss the principle complexities of the evolution of hierarchical triple star systems. Hierarchical triples are fairly common and potentially long-lived, which allows for their evolution to be affected by (secular) three-body dynamics, stellar evolution and their mutual influences. 
We present an overview of single star evolution and binary evolution with a focus on those aspects that are relevant for triple evolution. Subsequently, we describe the processes that are unique to systems with multiplicities of higher order than for binaries. 

In some cases, the evolution of a hierarchical triple can be adequately described by the evolution of the inner and outer binary separately. In other cases, the presence of the outer star significantly alters the evolution of the inner binary.
Several examples of the latter are given in detail. These examples also show the richness of the regime in which both three-body dynamics and stellar evolution play a role simultaneously. Moreover, the examples demonstrate the importance of coupling three-body dynamics with stellar evolution. 

Additionally, we present heuristic recipes for the principle processes of triple evolution. These descriptions are incorporated in a public source code \code\,for simulating the evolution of hierarchical, coeval, dynamically stable stellar triples.  
We discuss the underlying (sometimes simplifying) assumptions of the heuristic recipes. Some recipes are exact or adequate (e.g. gravitational wave emission, wind mass loss or Lidov-Kozai cycles), and others are admittedly crude (e.g. mass transfer). 
The recipes are based on simple assumptions and should be seen as a starting point for discussion and further study. 
When more sophisticated
models become available of processes that influence triple evolution,
these can be included in \code, and subsequently the effect on the triple populations can be studied. For now, the accuracy levels of the heuristic recipes are sufficient to initiate the systematic exploration of triple evolution (e.g. populations, evolutionary pathways), while taking into account three-body dynamics and stellar evolution consistently. 
We note that simulating through a phase of stable mass transfer in an eccentric inner orbit is currently beyond the scope of the project. However, 
appropriate methodology for eccentric mass transfer \citep[e.g.][]{Sep07b, Sep09,Dou16b} has been developed that we aim to implement at a later stage.

The triple evolution code \code\,is based on the secular approach to solve for the dynamics of the triple system.
It has been shown that this approach is in good agreement with N-body simulations of systems in which the secular approximations are valid \citep{Nao13, Ham13, Mic14}.
The advantage of the secular approach is that the computational time is orders of magnitudes shorter than for an N-body simulation.
The secular approach, however, is not valid when the evolutionary processes occur on timescales shorter than the dynamical timescale of the system. In these cases, we either stop the simulation (e.g. during a dynamical instability) or simulate the process as an instantaneous event (such as a common-envelope phase). 
Lastly, the secular approximation becomes inaccurate when the triple hierarchy is weaker \citep[e.g.][]{Ant12, Kat12, Ant14b, Bod14, Luo16}. In this case, the timescale of the perturbation from the outer star onto the inner binary during its periastron passage, is comparable to the dynamical timescale of the inner binary.
This can result in extremely high eccentricities and collisions between the stars in the inner binary. With the secular approach, as in \code, these occurrences are probably underestimated in systems with moderate hierarchies \citep[see also][]{Nao16}.

\code\,is written in the Astrophysics Multipurpose
Software Environment, or \texttt{AMUSE} \citep{Por09, Por13}, which is based on Python.  
\texttt{AMUSE} including \code\,can be downloaded for free at amusecode.org and github.com/amusecode/amuse.
Due to the nature
of \texttt{AMUSE}, the triple code can be easily extended to include a detailed stellar evolution code or a direct N-body code. Regarding the latter, this is interesting in the context of triples with moderate hierarchies where the orbit-averaged technique breaks down (as discussed above).
Furthermore, it is relevant for triples that become dynamically unstable during and as a consequence of their evolution. For example, \citet{Per12} show triples that become dynamically unstable due to their internal wind mass losses, are responsible for the majority of stellar collisions in the Galactic field. Consequently, the majority of stellar collisions do not take place between two MS stars, but involve an evolved star of giant-dimensions. 
Another interesting prospect is the inclusion of triples in simulations
of cluster evolution, where triples are often not taken into account 
e.g. in the initial population, through dynamical formation nor a consistent treatment of the evolution of triple star systems. 
However, dynamical encounters involving
triples are common, reaching or even exceeding the encounter rate involving solely single or binary stars, in particular in low- to moderate-density
star clusters \citep{Lei11, Lei13}. Therefore, the evolution of triples might not only be important for the formation and destruction of compact or exotic binaries, but also for the dynamical evolution of clusters in general.


\begin{backmatter}

\section*{Competing interests}
  The authors declare that they have no competing interests.

\section*{Author's contributions}
ST wrote the draft of this paper and the public source code \code. 
AH derived the equations for the orbital evolution of a triple during a supernova explosion as given in Appendix\,\ref{sec:app_sn}. AH also composed the ODE solver routine from updating the routine presented in \citet{Ham13}. 
SPZ assisted with the construction of the heuristic recipes for triple evolution. All of the authors contributed corrections and improvements on the draft of the manuscript. All authors read and approved the final manuscript.

\section*{Acknowledgements}
  We are grateful to Arjen van Elteren for his advice regarding particle sets within the \texttt{AMUSE}-framework. 
  This work was supported by the
Netherlands Research Council (NWO grant numbers 612.071.305 [LGM]
and 639.073.803 [VICI]), the Netherlands Research School for
Astronomy (NOVA), the Interuniversity Attraction Poles
Programme (initiated by the Belgian Science Policy Office, IAP P7/08
CHARM) and by the European Union’s Horizon 2020 research and
innovation programme (grant agreement No 671564, COMPAT project).


\bibliographystyle{bmc-mathphys} 
\bibliography{bmc_article}      
\nocite{label2}


\normalsize
\onecolumn
\section{\large Appendix}
\label{sec:app}
\subsection{Derivation of the orbital evolution of a triple during a supernova explosion}
\label{sec:app_sn}
\subsubsection{Definitions and assumptions}
Assume a hierarchical triple with inner binary masses $m_1$ and $m_2$, and tertiary mass $m_3$. The corresponding position and velocity vectors of the bodies, with respect to an arbitrary inertial reference frame, are denoted with $\ve{r}_i$ and $\ve{v}_i$, respectively. Let primed quantities denote quantities after the SN. It is assumed that body 1 collapses in a SN in such a way that the mass of body 1 changes instantaneously from $m_1$ to $m_1' = m_1 - \Delta m_1$ and that body 1 receives a kick velocity $\ve{v}_\mathrm{k}$. Therefore  the new velocity of body 1 is $\ve{v}_1' = \ve{v}_1 + \ve{v}_\mathrm{k}$. The position of body 1 is assumed not to change, i.e. $\ve{r}'_1 = \ve{r}_1$. Similarly, changes are not assumed to occur in the masses and instantaneous positions and velocities of the other bodies, i.e. $m'_i = m_i$, $\ve{r}'_i = \ve{r}_i$ and $\ve{v}'_i = \ve{v}_i$ for $i \in \{2,3\}$.

The pre-SN inner and outer relative separation vectors are given by
\begin{align}
\ve{r}_\mathrm{in} &= \ve{r}_1 - \ve{r}_2 \\
\ve{r}_\mathrm{out} &= \ve{r}_\mathrm{cm,in} - \ve{r}_3,
\end{align}
where 
\begin{align}
 \ve{r}_\mathrm{cm,in} \equiv \frac{m_1 \ve{r}_1 + m_2 \ve{r}_2}{m_1+m_2}
\end{align}
is the inner binary centre of mass position vector. Similarly, the pre-SN inner and outer relative velocity vectors are given by
\begin{align}
\ve{v}_\mathrm{in} &= \ve{v}_1 - \ve{v}_2 \\
\ve{v}_\mathrm{out} &= \ve{v}_\mathrm{cm,in} - \ve{v}_3,
\end{align}
where 
\begin{align}
 \ve{v}_\mathrm{cm,in} \equiv \frac{m_1 \ve{v}_1 + m_2 \ve{v}_2}{m_1+m_2}
\end{align}
is the inner binary centre of mass velocity vector. With our assumptions, the corresponding post-SN quantities are given by
\begin{align}
\ve{r}'_\mathrm{in} &= \ve{r}'_1 - \ve{r}'_2 = \ve{r}_\mathrm{in} \\
\ve{r}'_\mathrm{out} &= \ve{r}'_\mathrm{cm,in} - \ve{r}'_3  = \ve{r}_\mathrm{out} - \frac{\Delta m_1}{m_1+m_2-\Delta m_1} \frac{m_2}{m_1+m_2} \ve{r}_\mathrm{in},
\end{align}
where we used that the change of $\ve{r}_\mathrm{cm,in}$ can be written as
\begin{align}
\ve{r}_\mathrm{cm,in}' = \ve{r}_\mathrm{cm,in} -  \frac{\Delta m_1}{m_1+m_2-\Delta m_1} \frac{m_2}{m_1+m_2} \ve{r}_\mathrm{in}.
\end{align}
The corresponding equations for the velocities are
\begin{align}
\ve{v}'_\mathrm{in} &= \ve{v}'_1 - \ve{v}'_2 = \ve{v}_\mathrm{in} + \ve{v}_\mathrm{k} \\
\ve{v}'_\mathrm{out} &= \ve{v}_\mathrm{out} + \ve{v}_\mathrm{sys},
\end{align}
where the systemic velocity $\ve{v}_\mathrm{sys}$ is given by
\begin{align}
\ve{v}_\mathrm{sys} &\equiv \ve{v}_\mathrm{cm,in}' - \ve{v}_\mathrm{cm,in} =  - \frac{\Delta m_1}{m_1+m_2-\Delta m_1}  \cdot  \left [ \frac{m_2}{m_1+m_2} \ve{v}_\mathrm{in} + \left(1 - \frac{m_1}{\Delta m_1} \right ) \ve{v}_\mathrm{k} \right ].
\end{align}

\subsubsection{Orbital elements of the inner binary}
To derive the change of the semimajor-axis in the inner orbit, we use the following standard relation between the semimajor-axis, specific binding energy and the relative position and velocity vectors,
\begin{align}
E_\mathrm{in} = - \frac{G(m_1+m_2)}{2a_\mathrm{in}} = \frac{1}{2} v_\mathrm{in}^2 - \frac{G(m_1+m_2)}{r_\mathrm{in}}.
\end{align}
After the SN, the inner binary specific energy is given by
\begin{align}
\nonumber E'_\mathrm{in} &= - \frac{G(m_1+m_2-\Delta m_1)}{2a'_\mathrm{in}} = \frac{1}{2} \ve{v}_\mathrm{in}'^2 - \frac{G(m_1+m_2-\Delta m_1)}{r'_\mathrm{in}} \\
\nonumber &= \frac{1}{2} \left [v_\mathrm{in}^2 + v_\mathrm{k}^2 + 2 \left( \ve{v}_\mathrm{in} \cdot \ve{v}_\mathrm{k} \right ) \right ]- \frac{G(m_1+m_2-\Delta m_1)}{r_\mathrm{in}}  - \frac{G(m_1+m_2)}{2a_\mathrm{in}} + \frac{1}{2} \left [ v_\mathrm{k}^2 + 2 \left( \ve{v}_\mathrm{in} \cdot \ve{v}_\mathrm{k} \right ) \right ] + \frac{G \Delta m_1}{r_\mathrm{in}}.
\end{align}
This relation can be rewritten as
\begin{align}
\frac{a'_\mathrm{in}}{a_\mathrm{in}} = \left (1-\frac{\Delta m_1}{m_1+m_2} \right ) \left (1- \frac{2a_\mathrm{in}}{r_\mathrm{in}} \frac{\Delta m_1}{m_1+m_2} - 2 \frac{\left( \ve{v}_\mathrm{in} \cdot \ve{v}_\mathrm{k} \right )} {v_\mathrm{c,in}^2} - \frac{v_\mathrm{k}^2}{v_\mathrm{c,in}^2} \right )^{-1},\
\label{eq_app:a_in}
\end{align}
where $v_\mathrm{c,in}$ is the relative inner orbital speed for a circular orbit, i.e.
\begin{align}
v_\mathrm{c,in} \equiv \sqrt{ \frac{ G(m_1+m_2)}{a_\mathrm{in}} }.
\end{align}
Eq.\,\ref{eq_app:a_in} is equivalent to Eq.\,\ref{eq:a_sn_bin} and to Eq.\,(7) of \citet{Pij12}.

To find the post-SN eccentricity, we use the following standard relation for the specific angular momentum of the inner orbit,
\begin{align}
G(m_1+m_2) a_\mathrm{in} \left(1-e_\mathrm{in}^2\right) &= || \ve{r}_\mathrm{in} \times \ve{v}_\mathrm{in} ||^2 \\
&= r_\mathrm{in}^2 v_\mathrm{in}^2 - \left(\ve{r}_\mathrm{in} \cdot \ve{v}_\mathrm{in} \right )^2.
\end{align}
The post-SN inner orbit specific angular momentum is given by
\begin{align}
\nonumber G(m_1+m_2 &  -\Delta m_1 ) a'_\mathrm{in}  \left(1-e_\mathrm{in}'^2\right) = || \ve{r}'_\mathrm{in} \times \ve{v}'_\mathrm{in} ||^2 \\
&= r_\mathrm{in}'^2 v_\mathrm{in}'^2 - \left(\ve{r}_\mathrm{in}' \cdot \ve{v}_\mathrm{in}' \right )^2 \\
&= r_\mathrm{in}^2 \left [v_\mathrm{in}^2 + 2 \left( \ve{v}_\mathrm{in} \cdot \ve{v}_\mathrm{k} \right ) + v_\mathrm{k}^2 \right ] - \left [ \left ( \ve{r}_\mathrm{in} \cdot \ve{v}_\mathrm{in} \right ) + \left ( \ve{r}_\mathrm{in} \cdot \ve{v}_\mathrm{k} \right ) \right ]^2 \\
&= G(m_1+m_2) a_\mathrm{in} \left (1 - e_\mathrm{in}^2 \right ) +  r_\mathrm{in}^2 \left [2 \left( \ve{v}_\mathrm{in} \cdot \ve{v}_\mathrm{k} \right ) + v_\mathrm{k}^2 \right ] - 2 \left ( \ve{r}_\mathrm{in} \cdot \ve{v}_\mathrm{in} \right ) \left ( \ve{r}_\mathrm{in} \cdot \ve{v}_\mathrm{k} \right ) - \left ( \ve{r}_\mathrm{in} \cdot \ve{v}_\mathrm{k} \right )^2.
\end{align}
Inserting the relation for the post-SN inner semimajor-axis (Eq.\,\ref{eq_app:a_in}), we find the following expression for the post-SN eccentricity,
\begin{align}
\nonumber 1-e_\mathrm{in}'^{2} &=  \left (\frac{m_1+m_2}{m_1+m_2-\Delta m_1} \right )^2 \left (1- \frac{2a_\mathrm{in}}{r_\mathrm{in}} \frac{\Delta m_1}{m_1+m_2} - 2 \frac{\left( \ve{v}_\mathrm{in} \cdot \ve{v}_\mathrm{k} \right )} {v_\mathrm{c,in}^2} - \frac{v_\mathrm{k}^2}{v_\mathrm{c,in}^2} \right ) \bigg\{ \left(1-e_\mathrm{in}^2\right)  \\
& \quad
+ \frac{1}{G(m_1+m_2)a_\mathrm{in}} \left [ r_\mathrm{in}^2 \left (2 \left( \ve{v}_\mathrm{in} \cdot \ve{v}_\mathrm{k} \right ) + v_\mathrm{k}^2 \right ) - 2 \left ( \ve{r}_\mathrm{in}\cdot \ve{v}_\mathrm{in} \right) \left( \ve{r}_\mathrm{in} \cdot \ve{v}_\mathrm{k} \right ) - \left ( \ve{r}_\mathrm{in} \cdot \ve{v}_\mathrm{k} \right )^2 \right ] 
\bigg\} .
\label{eq_app:e_in}
\end{align}
The terms in the second line of Eq.\,\ref{eq_app:e_in} are missing in Eq.\,8a of Pijloo et al. (2012).

\subsubsection{Orbital elements of the outer binary}
Using the same method as above, we derive equations for the post-SN outer binary semimajor-axis and eccentricity. The outer orbit specific energy is given by
\begin{align}
E_\mathrm{out} &= - \frac{G(m_1+m_2+m_3)}{2 a_\mathrm{out}} = \frac{1}{2} \ve{v}_\mathrm{out}^2 - \frac{G(m_1+m_2+m_3)}{r_\mathrm{out}},
\end{align}
and changes according to
\begin{align}
\nonumber E'_\mathrm{out} &= - \frac{G(m_1+m_2+m_3-\Delta m_1)}{2 a'_\mathrm{out}} = \frac{1}{2} \ve{v}_\mathrm{out}'^2 - \frac{G(m_1+m_2+m_3-\Delta m_1)}{r_\mathrm{out}'} \\
\nonumber &= \left [ \frac{1}{2} v_\mathrm{out}^2 - \frac{G(m_1+m_2+m_3)}{r_\mathrm{out}} \right ] + \frac{G(m_1+m_2+m_3)}{r_\mathrm{out}} + \left( \ve{v}_\mathrm{out} \cdot \ve{v}_\mathrm{sys} \right ) + v_\mathrm{sys}^2 - \frac{G(m_1+m_2+m_3-\Delta m_1)}{r_\mathrm{out}'} \\
&= - \frac{G(m_1+m_2+m_3)}{2 a_\mathrm{out}} + \left( \ve{v}_\mathrm{out} \cdot \ve{v}_\mathrm{sys} \right ) + v_\mathrm{sys}^2 + \frac{G(m_1+m_2+m_3)}{r_\mathrm{out}} \left (1 - \frac{r_\mathrm{out}}{r_\mathrm{out}'} \frac{m_1+m_2+m_3-\Delta m_1}{m_1+m_2+m_3} \right ).
\end{align}
Rewriting this relation, we find
\begin{align}
\frac{a_\mathrm{out}'}{a_\mathrm{out}} = \left(1 - \frac{\Delta m_1}{m_1+m_2+m_3} \right) \left ( 1 - \frac{2a_\mathrm{out}}{r_\mathrm{out}'} \frac{\Delta m_1}{m_1+m_2+m_3} + 2 a_\mathrm{out} \frac{r_\mathrm{out}-r_\mathrm{out}'}{r_\mathrm{out} r_\mathrm{out}'} -  2 \frac{\left(\ve{v}_\mathrm{out} \cdot \ve{v}_\mathrm{sys} \right )}{v_\mathrm{c,out}^2} - \frac{v_\mathrm{sys}^2}{v_\mathrm{c,out}^2} \right )^{-1},
\label{eq_app:a_out}
\end{align}
where, analogously to the inner orbit, we defined
\begin{align}
v_\mathrm{c,out} \equiv \sqrt{ \frac{ G(m_1+m_2+m_3)}{a_\mathrm{out}} }.
\end{align}
Eq.\,\ref{eq_app:a_out} is equivalent to Eq.\,\ref{eq:a_sn_tri} and to Eq.\,26 of \citet{Pij12}.

The outer orbit specific angular momentum is given by
\begin{align}
G(m_1+m_2+m_3) a_\mathrm{out} \left(1-e_\mathrm{out}^2\right) = || \ve{r}_\mathrm{out} \times \ve{v}_\mathrm{out} ||^2 = r_\mathrm{out}^2 v_\mathrm{out}^2 - \left(\ve{r}_\mathrm{out} \cdot \ve{v}_\mathrm{out} \right )^2.
\end{align}
After the SN, it changes according to
\begin{align}
\nonumber G(m_1+m_2+m_3-\Delta m_1 ) a'_\mathrm{out}  \left(1-e_\mathrm{out}'^2\right) &= || \ve{r}'_\mathrm{out} \times \ve{v}'_\mathrm{out} ||^2 = r_\mathrm{out}'^2 v_\mathrm{out}'^2 - \left(\ve{r}_\mathrm{out}' \cdot \ve{v}_\mathrm{out}' \right )^2 \\
\nonumber &= \left [ r_\mathrm{out}^2 - 2 \alpha \left ( \ve{r}_\mathrm{in} \cdot \ve{r}_\mathrm{out} \right ) + \alpha^2 r_\mathrm{in}^2 \right ] \left (\ve{v}_\mathrm{out} + \ve{v}_\mathrm{sys} \right )^2 \\
&\quad - \left [ \left ( \ve{r}_\mathrm{out} \cdot \ve{v}_\mathrm{out} \right ) - \alpha \left( \ve{r}_\mathrm{in} \cdot \ve{v}_\mathrm{out} \right ) + \left ( \ve{r}_\mathrm{out} \cdot \ve{v}_\mathrm{sys} \right ) - \alpha \left ( \ve{r}_\mathrm{in} \cdot \ve{v}_\mathrm{sys} \right ) \right ]^2,
\end{align}
where, for notational convenience, we define the mass ratio
\begin{align}
\alpha \equiv \frac{\Delta m_1}{m_1+m_2-\Delta m_1} \frac{m_2}{m_1+m_2}.
\end{align}
Inserting the relation for the post-SN outer semimajor-axis (Eq.\,\ref{eq_app:a_out}), we find the following expression for the post-SN outer eccentricity,
\begin{align}
\nonumber 1-e_\mathrm{out}'^{2} &=
\bigg( 1 - \frac{2a_\mathrm{out}}{r'_\mathrm{out}}\frac{\Delta m_1}{m_1+m_2+m_3} + 2 a_\mathrm{out} \frac{r_\mathrm{out} - r'_\mathrm{out}}{r_\mathrm{out} r'_\mathrm{out}} - 2 \frac{\left(\ve{v}_\mathrm{out} \cdot \ve{v}_\mathrm{sys} \right )}{v_\mathrm{c,out}^2}  - \frac{v_\mathrm{sys}^2}{v_\mathrm{c,out}^2} \bigg)  \\
& \nonumber \quad 
\left (\frac{m_1+m_2+m_3}{m_1+m_2+m_3-\Delta m_1} \right )^2 \cdot 
\bigg[ \left(1-e_\mathrm{out}^2\right) 
+ \frac{1}{G(m_1+m_2+m_3)a_\mathrm{out}} \\
& \nonumber \quad
\big\{ r_\mathrm{out}^2 \left [2 \left(\ve{r}_\mathrm{out} \cdot \ve{v}_\mathrm{sys} \right ) + v_\mathrm{sys}^2\right ] + \left [-2 \alpha \left(\ve{r}_\mathrm{in}\cdot \ve{r}_\mathrm{out} \right ) + \alpha^2 \right] \left ( \ve{v}_\mathrm{out} + \ve{v}_\mathrm{sys} \right )^2 + 2 \left(\ve{r}_\mathrm{out} \cdot \ve{v}_\mathrm{out} \right ) \\
& \quad  \left [ \alpha \left(\ve{r}_\mathrm{in}\cdot \ve{v}_\mathrm{out} \right ) - \left(\ve{r}_\mathrm{out} \cdot \ve{v}_\mathrm{sys} \right ) + \alpha \left(\ve{r}_\mathrm{in} \cdot \ve{v}_\mathrm{sys} \right ) \right ]
 - \left [ - \alpha \left(\ve{r}_\mathrm{in} \cdot \ve{v}_\mathrm{out} \right ) + \left (\ve{r}_\mathrm{out} \cdot \ve{v}_\mathrm{sys} \right ) - \alpha \left(\ve{r}_\mathrm{in} \cdot \ve{v}_\mathrm{sys} \right ) \right ]^2 \big\} \bigg]
\label{eq_app:e_out}
\end{align}

The terms in the last two lines of Eq.\,\ref{eq_app:e_out} 
are missing in Eq.\,27 of \citet{Pij12}.

\end{backmatter}
\end{document}